\documentclass{ws-ijmpa}
\usepackage[super,compress]{cite}
\usepackage{multirow}
\usepackage{url}
\usepackage{bm}
\usepackage{ascmac}
\usepackage{amsmath}
\usepackage{amssymb}

\def\draftnote{\today\quad\currenttime\quad WSPC/INSTRUCTION FILE\qquad\jobname\quad\mbox{HUPD-1302}}%

\begin{document}

\markboth{K.-I. Ishikawa, D. Kimura, K. Shigaki, A. Tsuji}
{A Numerical Evaluation of Vacuum Polarization ...}

\catchline{}{}{}{}{}

\title{
A NUMERICAL EVALUATION OF VACUUM POLARIZATION 
TENSOR IN CONSTANT EXTERNAL MAGNETIC FIELDS}

\author{KEN-ICHI ISHIKAWA\footnote{ishikawa@theo.phys.sci.hiroshima-u.ac.jp},
        DAIJI KIMURA$^b$, KENTA SHIGAKI, ASAKO TSUJI}
\address{Graduate School of Science, Hiroshima University,\\
         Higashi-Hiroshima, Hiroshima 739-8526, Japan}
\address{$^b$General Education, Ube National College of Technology,\\
         Ube, Yamaguchi 755-8555, Japan}

\maketitle


\begin{abstract}
Hattori-Itakura have recently derived the full Landau-level summation 
form for the photon vacuum polarization tensor in constant external magnetic 
fields at the one-loop level.
The Landau-level summation form is essential when the photon momentum 
exceeds the threshold of the pair creation of charged particles in 
a magnetic field stronger than the squared mass of the charged particle.
The tensor has three different form factors
depending on the tensor direction with respect to the external magnetic field.
The renormalization is nontrivial because these form factors are expressed 
in terms of double or triple summation forms. 
We give a numerical UV subtraction method which can be applied
to numerically evaluate the form factors in constant external magnetic fields.
We numerically investigate the photon vacuum polarization tensor in the form 
of the Landau-level summation and estimate the systematic errors coming from
truncation of the Landau-level summation in a parameter region realized 
in heavy ion collision experiments.
We find that the error is practically controllable at an $O(10^{-2})$ level for
electrons and muons in strong magnetic fields expected in heavy ion collisions
in the experimentally feasible kinematic parameter regions.

\end{abstract}

\ccode{PACS Nos.: 12.20.-m,11.15.Bt}

\section{Introduction}	

Photon vacuum polarization is a fundamental tool to access the structure 
of quantum vacuum. Strong external electromagnetic fields could affect
the structure of the QED vacuum and cause various non-perturbative phenomena
such as pair production via the Schwinger mechanism, photon splitting, 
electron-positron pair production from a photon, and vacuum birefringence of a photon
{\it etc.}.
\cite{Adler:1971wn,Shabad:1972rg,Shabad:1984ASTRO,Tsai:1974fa,Tsai:1975iz,Ritus:1985,Dittrich:1998gt,Weise:2006jw,Baier:2009it,Baier:2007dw,Dobrich:2012sw,Dobrich:2012jd,Tsai:1974,Shabad:1975ik,Urrutia:1977xb,Melrose:1976dr,Melrose:1977,Schubert:2000yt,Schubert:2000kf,Gies:2001zm,Hattori:2012je,Karbstein:2011ja,Kohri:2001wx,Gies:2011he,Gies:2005bz,Fukushima:2011nu}
There have been many theoretical works to evaluate the vacuum polarization 
tensor in strong electromagnetic fields to investigate 
these phenomena
\cite{Adler:1971wn,Shabad:1972rg,Shabad:1984ASTRO,Tsai:1974fa,Tsai:1975iz,Ritus:1985,Dittrich:1998gt,Weise:2006jw,Baier:2009it,Baier:2007dw,Dobrich:2012sw,Dobrich:2012jd,Tsai:1974,Shabad:1975ik,Urrutia:1977xb,Melrose:1976dr,Melrose:1977,Schubert:2000yt,Schubert:2000kf,Gies:2001zm,Hattori:2012je,Karbstein:2011ja,Kohri:2001wx,Gies:2011he,Gies:2005bz,Fukushima:2011nu}.
In reality heavy ion collision experiments at RHIC and LHC could generate 
very strong magnetic fields of $eB \sim O(m_{\pi}^2)$~\cite{Skokov:2009qp,Voskresensky:1980},
which could affect the QCD phase structure via the various chiral magnetic effects\cite{CHIRALMAGREV,CHIRALMAG}.
These effects can be experimentally looked into only if the 
invariant masses and the transverse momenta are in the regions 
accessible by the detectors\cite{Yee:2013qma,Tuchin:2010gx,Tuchin:2010vs,Turbide:2005bz,Chatterjee:2005de,Layek:2006um,Kopeliovich:2007fv,Kopeliovich:2007sd}.
Before arguing about the effect of the strong magnetic field on the QCD phase structure,
it is preferable to directly verify the existence of such a strong magnetic field in 
the heavy ion collisions.
The photon propagation could be a detection tool for the existence of 
very strong magnetic fields expected in heavy ion collision experiments at
RHIC and LHC\cite{Skokov:2009qp,Voskresensky:1980,Adare:2009qk,Alessandro:2006yt}.
We aim at providing a quantitative assessment of the effects on the photon propagator
to allow evaluation of the experimental feasibility.

The photon vacuum polarization tensor in constant external magnetic fields at the one-loop
diagram is expressed in a double-integral form with respect to two proper-time variables
\cite{Tsai:1974fa,Baier:2009it,Baier:2007dw,Tsai:1974,Shabad:1975ik,Urrutia:1977xb,Melrose:1976dr,Schubert:2000yt,Schubert:2000kf,Gies:2001zm,Hattori:2012je,Karbstein:2011ja,Kohri:2001wx,Gies:2011he,Gies:2005bz}.
When the virtual photon momentum exceeds the threshold of the pair creation, 
it becomes difficult to evaluate in the integral form as the integrand induces a 
complicated singular behavior originating from the pair creation of charged particles trapped in
the magnetic field, where the charged particles are quantized in the Landau-level.
The Landau-level summation form for the vacuum polarization tensor
is analytically expressed in terms of double or triple series
based on summation on the Landau-level of virtual charged particles
\cite{Shabad:1975ik,Melrose:1977,Weise:2006jw,Baier:2007dw,Hattori:2012je}.
The Landau-level summation form is essential particularly in magnetic fields
stronger than so-called the ``critical field strength'' $eB_c = m^2$ with the charged 
particle mass $m$. Such a super-critical field strength is expected in the heavy ion collisions at
RHIC and LHC\cite{Skokov:2009qp,Voskresensky:1980}.
The analytic form could be an important basis to discuss 
the polarization tensor in realistic external fields\cite{Gies:2011he,Gies:2005bz}.

The Landau-level form has been explored in Refs.
\citen{Shabad:1975ik,Melrose:1977,Weise:2006jw,Baier:2007dw,Hattori:2012je}.
The analytic Landau-level summation form obtained by Shabad\cite{Shabad:1975ik} was
limited to the imaginary part and the real part was only given in a form of 
non-absolutely converging series. Instead he extracted the real part by replacing 
the imaginary part induced by the residue theorem at each pole of the integrand
to the one-loop integral for particle-antiparticle trapped in 
the Landau-level by inversely using the Cutkosky rule.
Since the imaginary part is free from the UV divergence, the finiteness of the 
full form is nontrivial.
Melrose and Stoneham\cite{Melrose:1977} have obtained an analytic Landau-level 
summation form similar to the one obtained by Shabad\cite{Shabad:1975ik}.
Baier and Katkov\cite{Baier:2007dw}
also explored the Landau-level summation form and have obtained the
imaginary part by which they discussed the pair creation of the
electron-positron by a photon in a strong magnetic field.
Hattori-Itakura\cite{Hattori:2012je} have recently obtained a similar full Landau-level
summation form aiming for the vacuum birefringence of a photon by strong magnetic fields.
Because the numerical property of the Landau-level summation form seems 
to be still missing in the literature, we numerically investigate the 
Landau-level summation form of the vacuum polarization tensor covering 
the kinetic region of virtual photon momentum responsible for
the heavy ion collision experiments
at RHIC\cite{Adare:2009qk} and LHC\cite{Alessandro:2006yt}.

In this paper we follow and extend the method developed by Hattori-Itakura\cite{Hattori:2012je}
to numerically investigate the Landau-level form of the vacuum polarization tensor.
They have discussed the renormalization and the UV-structure of the form factors 
contained in the vacuum polarization tensor expressed in the Landau-level summation form.
Their subtraction is, however, not suitable to evaluate the form factors 
numerically, because the subtraction is defined between
the upper limit of the series and the UV cut-off of the subtraction integral.
We modify their subtraction to become a preferable form for numerical
evaluation of the vacuum polarization tensor and numerically investigate 
the property of the convergence of the series.

This paper proceeds as follows.  
In the next section, we present master formula for 
the vacuum polarization tensor in constant external magnetic fields
written in the Schwinger's proper time integral. 
Our subtraction method and the Landau-level summation form are explained
in Section~\ref{sec:Subtraction}. 
The convergence and
the systematic errors from truncation of the Landau-level summation are 
discussed in Section~\ref{sec:Asymmptotic}.
The numerical results with kinematic parameters accessible in 
the heavy ion collision experiments are given in Section~\ref{sec:Results}.
We summarize the paper in the last section.

\section{Vacuum Polarization Tensor in Constant External Magnetic Fields}
The vacuum polarization tensor $\Pi^{\mu\nu}$ in external fields has 
the following tensor structure.
\begin{equation}
    \Pi^{\mu\nu}(k) = 
\left(P^{\mu\nu}-P_{\parallel}^{\mu\nu}-P_{\perp}^{\mu\nu}\right)
 N_0(k) + P_{\parallel}^{\mu\nu} N_1(k) + P_{\perp}^{\mu\nu} N_2(k)
\end{equation}
where $k^{\mu}$ is the photon four-momentum and 
the projection tensors are defined by
\begin{equation}
  P^{\mu\nu}=k^2 \eta^{\mu\nu}-k^{\mu}k^{\nu}, 
\quad
  P_{\parallel}^{\mu\nu}=k_{\parallel}^2 \eta_{\parallel}^{\mu\nu}-k_{\parallel}^{\mu}k_{\parallel}^{\nu},
\quad
  P_{\perp}^{\mu\nu}=k_{\perp}^2 \eta_{\perp}^{\mu\nu}-k_{\perp}^{\mu}k_{\perp}^{\nu}.
\end{equation}
We define the external magnetic field $\bm{B}$ is directed along the $z$-axis and $B_z = B >0$.
The photon four momentum $k^{\mu}$ and the metric $\eta^{\mu\nu}$ are 
classified according to the direction of $\bm{B}$ as follows.
\begin{equation}
  k^{\mu}_{\parallel} = (k^0,0,0,k^3)=(\omega,0,0,k_z),
\quad
  k^{\mu}_{\perp}     = (0,k^1,k^2,0)=(0,k_x,k_y,0),
\end{equation}
\begin{equation}
  \eta^{\mu\nu}_{\parallel} = \mathrm{diag}(1,0,0,-1),
\quad
  \eta^{\mu\nu}_{\perp}     = \mathrm{diag}(0,-1,-1,0),
\end{equation}
\begin{eqnarray}
k_{\parallel}^2 &=& (k^0)^2-(k^3)^2 = \omega^2-k_z^2,    \\
k_{\perp}^2 &=& -(k^1)^2-(k^2)^2 = -(k_x^2+k_y^2)=-\bm{k}_{\perp}^2.
\end{eqnarray}

We consider Dirac fermions with the unit charge $e > 0$ and mass $m$. 
The one-loop contribution to the form factors, $N_j$'s ($j=0,1,2$), 
is given by
\begin{equation}
N_j = -\frac{\alpha}{4\pi}\int_{-1}^{1}dv 
   \int_{0-i\varepsilon}^{\infty-i\varepsilon} dz 
\left[
\tilde{N}_j(z,v)
e^{-i\psi(z,v)\eta-i\phi(v;r,\mu)z}
-\frac{1-v^2}{z} e^{-i \frac{z}{\mu}}
\right],
\label{eq:Nmaster}
\end{equation}
\begin{eqnarray}
\tilde{N}_0(z,v)&=&\frac{\cos(vz)-v \cot(z)\sin(vz)}{\sin(z)},
\nonumber\\
\tilde{N}_1(z,v)&=&(1-v^2)\cot(z),
\label{eq:Ntilde}
\\
\tilde{N}_2(z,v)&=&2 \frac{\cos(vz)-\cos(z)}{\sin^3(z)},
\nonumber\\
\psi(z,v)&=&\frac{\cos(vz)-\cos(z)}{\sin(z)},
\label{eq:PhasePsi}
\\
\phi(v;r,\mu) &=&  \frac{1-(1-v^2)r}{\mu},
\label{eq:PhasePhi}
\end{eqnarray}
where we introduce dimensionless parameters $\mu$, $r$, and $\eta$
defined by
\begin{equation}
  \mu = \frac{eB}{m^2}, \quad
  r = \frac{k_{\parallel}^2}{4m^2}, \quad
  \eta = \frac{2 q}{\mu}, \quad\mbox{with}\quad
  q = \frac{\bm{k}_{\perp}^2}{4m^2}.
\end{equation}
The $z$ integration, originating from the Schwinger's proper time, should 
be carried out on a line slightly lower along the real axis in the complex 
plane to have the Feynman propagator boundary condition.

All we want to know is the value of $N_j$'s with arbitrary values
of $\mu, r, q$ responsible for the heavy ion collision experiments.
When $0<r<1$, the both exponential factors in the integrand
converge to zero on the lower quarter circle path with an infinite radius.
The integrand has no pole in the lower complex plane. 
Thus the $z$ integral path can be continued to the lower imaginary axis as $z=-i x$ via
the Cauchy's integral theorem. This yields the well converging double integration form 
for $N_j$'s. On the other hand, when $1<r$,
the $z$ integral should be evaluated via the residue theorem~\cite{Inagaki:2003yi,Inagaki:2003ac}. In this case
the line integral is a closed path on the quarter sector in the first quadrant of 
the complex plane, and a difficulty arises when evaluating the residue of the integrand.
The poles of $\tilde{N}_j$'s locate on $z=n\pi$ ($n=1,2,\cdots$),
where the phase factor $\psi$ in Eq.~(\ref{eq:PhasePsi}) also has poles.
This means that the residue at $z=n\pi$ ($n=1,2,\cdots$) is indefinite except 
for the $\eta=0$ cases.

When $\eta=0$ (equivalently $q=0$), the $z$ integration can be analytically performed
yielding the DiGamma functions. The form for $N_1$ has been obtained in Ref.~\citen{Karbstein:2011ja}.
Similar forms can be obtained for both $N_0$ and $N_2$.

The Hattori-Itakura formula opens the way to evaluate $N_j$'s for $\eta >0$ with $1<r$. 
They used a different set of the form factors defined by
\begin{equation}
  \chi_0= -N_0, \quad
  \chi_1= -(N_1-N_0), \quad
  \chi_2= -(N_2-N_0),
\end{equation}
where the zero-field counter term is contained only in $\chi_0$. 
For $\chi_1$ ($\chi_2$) the subtraction occurs between $\tilde{N}_1$ ($\tilde{N}_2$) and  $\tilde{N}_0$.
They analytically expand the integrand of 
Eqs.~(\ref{eq:Ntilde})-(\ref{eq:PhasePhi}) in terms of $C^{n}_{\ell}(\eta)$ defined by
\begin{equation}
    C^n_{\ell}(\eta)= e^{-\eta}\eta^n \frac{\ell!}{(\ell+n)!} \left(L^{n}_{\ell}(\eta)\right)^2,
\label{eq:DefCoefC}
\end{equation}
where $L^{n}_{\ell}(\eta)$ is the associated Laguerre polynomial,
and integrate both of the $v$ and $z$ integrals. 
The expansion yields a double series on $n$ and $\ell$, where
the indexes $n$ and $\ell$ correspond to 
the Landau-level of virtual fermions trapped in the external magnetic field in 
the one-loop diagram. 
The zero field counter term in $\chi_0$ still remains in the integral form. 
Numerical UV cancellation between the double series and the double integral is impossible.
Although $\chi_1$ ($\chi_2$) is completely expanded in the double series, the location of
the UV-divergence in the series is not aligned well between $\tilde{N}_1$ ($\tilde{N}_2$) and $\tilde{N}_0$.
Thus the UV cancellation is nontrivial even if they are renormalized.
We therefore need a well organized renormalization method suitable for the numerical evaluation.
We show our renormalization method in the next section.

\section{Subtraction method and the Landau-level sum form}
\label{sec:Subtraction}
We rearrange Eq.~(\ref{eq:Nmaster}) to the following form.
\begin{eqnarray}
N_j &=& -\frac{\alpha}{4\pi}\int_{-1}^{1}dv 
   \int_{0-i\varepsilon}^{\infty-i\varepsilon} dz 
\left[
\tilde{N}_j(z,v)
e^{-i\psi(z,v)\eta}
\left(e^{-i\phi(v;r,\mu)z}-e^{-i\frac{z}{\mu}}\right)
\right]
\nonumber\\
&&
 -\frac{\alpha}{4\pi}\int_{-1}^{1}dv 
   \int_{0-i\varepsilon}^{\infty-i\varepsilon} dz 
\left[
\left(\tilde{N}_j(z,v)e^{-i\psi(z,v)\eta}-
\frac{1-v^2}{z} 
\right)e^{-i \frac{z}{\mu}}
\right].
\label{eq:Narranged}
\end{eqnarray}
The first line is UV finite and we can entirely expand it according to the Hattori-Itakura's expansion method.
The second part of $N_j$ corresponds to $N_j$ at $r=0$, for which $z$ integral can be analytically 
continued to $z=-i x$ yielding suitable forms for the numerical integration. 
We note for the $N_2$ form factor that the singularity of $\tilde{N}_2$ is worse than others as it contains $1/\sin^3(z)$.
This yields another difficulty for the numerical evaluation. To tame the difficulty we replace $\tilde{N}_2$ with
\begin{equation}
\tilde{N}_2 = 2i\frac{1}{\sin^2(z)} \frac{\partial}{\partial\eta},
\end{equation}
for the first line of Eq.~(\ref{eq:Narranged}) before applying the Hattori-Itakura's expansion.

Applying the above prescription, we obtain the following form for the form factors.
\begin{eqnarray}
N_j &=& -\frac{\alpha}{4\pi}\sum_{n=0}^{\infty}C_n
\sum_{\ell=0}^{\infty} \Omega^n_{j,\ell}(r,\eta,\mu)
\nonumber\\
&&
 -\frac{\alpha}{4\pi}\int_{-1}^{1}dv 
   \int_{0}^{\infty} dx
\left[
\left(\overline{N}_j(x,v)e^{\overline{\psi}(x,v)\eta}-
\frac{1-v^2}{x} 
\right)e^{-\frac{x}{\mu}}
\right],
\label{eq:Nfinal}
\end{eqnarray}
where $C_n=(2-\delta_{n,0})$ and
\begin{eqnarray}
\Omega^n_{0,\ell}(r,\eta,\mu) &=& 
  \left[(1-\delta_{n,0})C^{n-1}_{\ell}(\eta)+(1+\delta_{n,0})C^{n+1}_{\ell-1}(\eta)\right]
  {\cal F}^n_{\ell}(r,\mu)
\nonumber\\
&&
 - (n / \eta) \left[ C^n_{\ell}(\eta) + C^n_{\ell-1}(\eta)\right]
   {\cal G}^n_{\ell}(r,\mu),
\\
\Omega^n_{1,\ell}(r,\eta,\mu) &=& 
  \left[C^n_{\ell}(\eta)+C^{n}_{\ell-1}(\eta)\right]
  \left( {\cal F}^n_{\ell}(r,\mu) - {\cal H}^n_{\ell}(r,\mu) \right),
\\
\Omega^n_{2,\ell}(r,\eta,\mu) &=& 
4\frac{d C^n_{\ell}}{d\eta}(\eta) {\cal R}^n_{\ell}(r,\mu),
\\
\overline{N}_j(x,v)&=&-i \tilde{N}_j(-i x,v), \quad\quad
\overline{\psi}(x,v)= -i \psi(-i x,v),
\end{eqnarray}
with $C^n_{-1}(\eta)=0$. 
The functions,  
${\cal F}^n_{\ell}$,
${\cal G}^n_{\ell}$,
${\cal H}^n_{\ell}$, and ${\cal R}^n_{\ell}$, are
\begin{eqnarray}
{\cal F}^n_{\ell}(r,\mu)&=& \mu\left[F^n_{\ell}(r,\mu)-F^n_{\ell}(0,\mu)\right],
\\
{\cal G}^n_{\ell}(r,\mu)&=& \mu\left[G^n_{\ell}(r,\mu)-G^n_{\ell}(0,\mu)\right],
\\
{\cal H}^n_{\ell}(r,\mu)&=& \mu\left[H^n_{\ell}(r,\mu)-H^n_{\ell}(0,\mu)\right],
\\
{\cal R}^n_{\ell}(r,\mu)&=& \int_{-1}^{1}dv
\left[ \Psi\left(\frac{ S^n_{\ell+1}(v;r,\mu)}{2\mu}\right)
      -\Psi\left(\frac{ S^n_{\ell+1}(v;0,\mu)}{2\mu}\right)\right],
\label{eq:Rfactor}
\\
F^n_{\ell}(r,\mu)&=&\int_{-1}^{1}dv \frac{1}{S^n_{\ell}(v;r,\mu)},
\label{eq:FuncF}\\
G^n_{\ell}(r,\mu)&=&\int_{-1}^{1}dv \frac{v}{S^n_{\ell}(v;r,\mu)},
\label{eq:FuncG}\\
H^n_{\ell}(r,\mu)&=&\int_{-1}^{1}dv \frac{v^2}{S^n_{\ell}(v;r,\mu)},
\label{eq:FuncH}\\
S^n_{\ell}(v;r,\mu)&=& r v^2 -(n\mu)v + 1-r + (2\ell + n)\mu - i \varepsilon,
\label{eq:SPropLN}
\end{eqnarray}
where the $z$ integration is performed and the Feynman's $i\varepsilon$ prescription is 
restored to identify the absorptive part of these functions for the $v$ integration.
The function $\Psi(z)$ is the DiGamma function. We follow the notation for 
$F^n_{\ell}$,
$G^n_{\ell}$, and $H^n_{\ell}$ given by Ref.~\citen{Hattori:2012je} and
the $v$ integration  can be done analytically as given in \ref{sec:apdxA} 
except for ${\cal R}^n_{\ell}$. 
The form of ${\cal R}^n_{\ell}$ is inspired from Ref.~\citen{Karbstein:2011ja} in which
the analytic form for $N_1$ with $q=0$ has been obtained.

The DiGamma function $\Psi(z)$ has poles at $z=0$ and negative integers. 
When $1<r$, the argument $S^n_{\ell+1}/(2\mu)$ of $\Psi(z)$ in Eq.~(\ref{eq:Rfactor}) 
could hit the singularities in integrating $v$. 
In order to extract the absorptive part of Eq.~(\ref{eq:Rfactor}) we employ the recurrence formula, 
$\Psi(z) = \Psi(z+1) - 1/z$, until the argument becomes a non-zero positive number as
\begin{equation}
  \Psi(z) = \Psi(z+1) - \frac{1}{z} = \cdots =
\Psi(z+K+1) - \sum_{k=0}^{K}\frac{1}{z+k},
\end{equation}
where $K$ is a nonnegative integer chosen to satisfy $z+K+1 > 0$. Thus Eq.~(\ref{eq:Rfactor}) becomes
\begin{eqnarray}
{\cal R}^n_{\ell}(r,\mu)&=& \int_{-1}^{1}dv
\left[ \Psi\left(\frac{S^n_{\ell+1+K+1}(v;r,\mu)}{2\mu}\right)
      -\Psi\left(\frac{S^n_{\ell+1+K+1}(v;0,\mu)}{2\mu}\right)\right]
\nonumber\\
&&  -2 \sum_{k\ge 0}^{K}{\cal F}^n_{\ell+1+k}(r,\mu),
\label{eq:Rfactor2}\\
K &=& 
\left\{
    \begin{array}{lcr}
        -\mathrm{Ceiling}[A^n_{\ell+1}] &\quad\quad\quad& \mbox{( $|n\mu/(2r)|<1$ and $A^n_{\ell+1} \le 0$)}\\
        -1                   && \mbox{(otherwise)}
    \end{array}\right.
,\\
A^n_{\ell}&=& \frac{1}{2\mu}\left[1-r + (2\ell+n)\mu-\frac{(n\mu)^2}{4r}\right].
\label{eq:Amin}
\end{eqnarray}
The absorptive part is extracted as the sum of ${\cal F}^n_{\ell}$
and the $v$ integral can be numerically evaluated. 
When $|(v^2-1)r/(2\mu)| < 0.01$ the integrand of
Eq.~(\ref{eq:Rfactor2}) is evaluated using 8th order Taylor expansion 
to avoid a loss of significant digits;
\begin{equation}
\Psi(z+dz)-\Psi(z)\simeq 
\Psi^{(1)}(z) dz + \Psi^{(2)}(z) \frac{(dz)^2}{2}
 + \cdots + \Psi^{(8)}(z) \frac{(dz)^8}{8!},
\end{equation}
where $\Psi^{(j)}(z)$ is the polygamma function of order $j$.
To reduce the cost of numerical integrations at each $\ell$ 
we can use the following recurrence formula for ${\cal R}^n_{\ell}$;
\begin{equation}
 {\cal R}^n_{\ell}(r,\mu) =  {\cal R}^n_{\ell-1}(r,\mu) + 2 {\cal F}^{n}_{\ell}(r,\mu).
\end{equation}

The form factors below the threshold ($r<1$) can be evaluated numerically for any $0<q$ 
by analytic continuation with $z=-i x$ in Eq.~(\ref{eq:Nmaster}).
The form factors in these regions have been investigated in Ref.~\citen{Kohri:2001wx}.
The values from the double integral are compared to our numerical estimates 
from the Landau-level summation Eq.~(\ref{eq:Nfinal}) to check the consistency.

With the vanishing transverse momentum ($q=0$), the $z$ integral in Eq.~(\ref{eq:Nmaster})
can be performed analytically.
Karbstein {\it et al.}\cite{Karbstein:2011ja}
have shown the analytic form for $N_1$ with $q=0$, which 
is the integral containing DiGamma functions similar to Eq.~(\ref{eq:Rfactor}).
This form is valid for any $r$. 
We obtain similar analytic expressions for $N_0$ and $N_2$ with $q=0$ as given 
in \ref{sec:apdxB} together with $N_1$ with $q=0$.
We can check the validity of the numerical values from 
Eq.~(\ref{eq:Nfinal}) with $q=0$ by comparing to the values from 
the DiGamma expressions, Eqs.~(\ref{eq:N0q0})-(\ref{eq:N2q0}) given in \ref{sec:apdxB}, 
in the case of $1<r$.
Before going to numerical evaluation, we discuss the convergence of the double sum
of Eq.~(\ref{eq:Nfinal}) by observing the asymptotic form in the next section.

\section{Asymptotic form of the double series}
\label{sec:Asymmptotic}

The asymptotic form for Eqs.~(\ref{eq:FuncF})-(\ref{eq:Rfactor})
in $1 \ll \ell$ is given by
\begin{eqnarray}
 {\cal F}^n_{\ell}(r,\mu)&\sim&\frac{r}{3\mu \ell^2} + O\left(\frac{1}{\ell^3}\right),
\\
 {\cal G}^n_{\ell}(r,\mu)&\sim&\frac{nr}{15\mu \ell^3} + O\left(\frac{1}{\ell^4}\right),
\\
 {\cal H}^n_{\ell}(r,\mu)&\sim&\frac{r}{15\mu \ell^2} + O\left(\frac{1}{\ell^3}\right),
\\
 {\cal R}^n_{\ell}(r,\mu)&\sim&-\frac{2r}{3\mu \ell} + \frac{5r(1+(n+1)\mu)-2r^2}{15\mu^2\ell^2}+ O\left(\frac{1}{\ell^3}\right).
\end{eqnarray}

When $\eta >0$ the coefficient function $C^n_{\ell}(\eta)$ and its derivative behave as
\begin{eqnarray}
C^n_{\ell}(\eta) &\sim& \frac{1}{\pi\sqrt{\eta\ell}}e^{-\frac{n+1}{4\ell}}\cos^2
\left(\Theta^n_{\ell}(\eta)\right),
\nonumber\\
\frac{d C^n_{\ell}}{d\eta}(\eta) &\sim& -\frac{1}{\pi \eta}e^{-\frac{n+1}{4\ell}}
\sin\left(2\Theta^n_{\ell}(\eta)\right),\\
\Theta^n_{\ell}(\eta)&=&2\sqrt{\eta\kappa^n_\ell}
-\frac{\pi}{2}\left(n+\frac{1}{2}\right),\\
\kappa^n_\ell&=&\ell+\frac{n+1}{2},
\end{eqnarray}
for $\eta < 4\kappa^n_\ell$ with $1 \ll \ell$.
This is followed by the asymptotic form for the Laguerre polynomials
\begin{equation}
  L^n_{\ell}(\eta)\sim 
\frac{(\ell+n)!}{\ell!}\frac{e^{\eta/2}}{\sqrt{\pi}}\left(\kappa^n_{\ell}\eta\right)^{-n/2-1/4}\cos
\Theta^n_\ell(\eta),
\end{equation}
based on Bessel function expansion\cite{OscillatoryRegion,StrongRatioAsymptotic}.

$\Omega_{0,\ell}^n$ and $\Omega_{1,\ell}^n$ are bounded by
\begin{equation}
 \Omega^n_{j,\ell} \leq \left|\Omega^n_{j,\ell}\right| \sim O\left(\frac{1}{\ell^{\frac{5}{2}}}\right)
\quad\quad\quad\mbox{(for $j=0$ and $1$)}.
\end{equation}
This is a slowly converging series at a fixed $n$.
For $\Omega_{2,\ell}^n$, however, it does not seem to be absolutely convergent
since $|\Omega^n_{2,\ell}|\sim O(1/\ell)$.
The cancellation due to the oscillatory behavior of $dC^n_{\ell}/d\eta$ or 
due to the sign mixture among terms with different $n$ could occur 
for the convergence.
The worst case is that the series for $\Omega_{2,\ell}^n$ is asymptotic.
We could not prove the convergence for $\Omega_{2,\ell}^n$ with $0<\eta$ case.

From Eq.~(\ref{eq:DefCoefC}) the coefficient function $C^n_{\ell}(\eta)$ and the derivative for $\eta=0$
become
\begin{eqnarray}
    C^n_{\ell}(0) &=& \delta_{n,0}, \\
\left.  \frac{n}{\eta}C^n_{\ell}(\eta)\right|_{\eta=0} &=& (\ell+1)\delta_{n,1}, \\
   \frac{d C^n_{\ell}}{d\eta}(0) &=& -(2\ell+1)\delta_{n,0}+(\ell+1)\delta_{n,1}.
\end{eqnarray}
Since these do not have damping factors for $1 \ll \ell$, 
the series convergence becomes critical. 
We check the convergence of the double series explicitly in the following.
For $\Omega_{0,\ell}^n$ and $\Omega_{1,\ell}^n$ with $\eta=0$, the double sum converges as follows.
\begin{eqnarray}
    \sum_{n=0}^{\infty}C_n\sum_{\ell=0}^{\infty} \Omega_{0,\ell}^{n} &=&
2\sum_{\ell=0}^{\infty}({\cal F}^{1}_{\ell}-(2\ell+1){\cal G}^{1}_{\ell})
\nonumber\\
&\sim& \sum_{\ell\gg 1}^{\infty} \left[\frac{2r}{5\mu\ell^2}+O\left(\frac{1}{\ell^3}\right) \right]< \infty,
\\
    \sum_{n=0}^{\infty}C_n\sum_{\ell=0}^{\infty} \Omega_{1,\ell}^{n} &=&
({\cal F}^{0}_{0}-{\cal H}^{0}_{0})+
\sum_{\ell=1}^{\infty} 2 ({\cal F}^{0}_{\ell}-{\cal H}^{0}_{\ell})
\nonumber\\
&\sim& \sum_{\ell\gg 1}^{\infty} \left[\frac{8r}{15\mu\ell^2}+O\left(\frac{1}{\ell^3}\right) \right]< \infty.
\end{eqnarray}

For $\Omega_{2,\ell}^n$ it becomes
\begin{eqnarray}
 \sum_{n=0}^{\infty}C_n\sum_{\ell=0}^{\infty} \Omega_{2,\ell}^{n} &=&
\sum_{\ell=0}^{\infty}\left[ -4(2\ell+1){\cal R}^0_{\ell}+8(\ell+1){\cal R}^1_{\ell}\right]
\nonumber\\
&\sim & \sum_{\ell \gg 1}^{\infty}O\left(\frac{1}{\ell^2}\right) < \infty,
\end{eqnarray}
where the linear and logarithmic divergences are canceled among $n=0$ and $n=1$ terms.
Thus the double series for $\Omega_{2,\ell}^n$ is not absolutely convergent and the result 
depends on the ordering of the summation. 
When $q=0$ ($\eta=0$) and $r<1$, we numerically observe a large discrepancy
caused by the conditional convergence property
between the Landau-level summation formula and the double integral formula.
Fortunately $N_2$ with $q=0$ does not contribute to the polarization tensor 
as it is multiplied by
the projection tensor $P_{\perp}^{\mu\nu}$ which is identical to zero.

So far we do not discuss the convergence of the summation on $n$ 
except for the case with $\eta=0$. 
To check the validity of the Landau-level summation form we
compare the value to those evaluated with the other forms
numerically instead of analytically.
The comparison is possible in the following two regions.
\begin{itemize}
\item[(A)] Double integral form in $r<1$.
\item[(B)] DiGamma form with $q=0$.
\end{itemize}
The double integral form is obtained by substituting $z=-i x$ in Eq.~(\ref{eq:Nmaster}).
The DiGamma form is given in \ref{sec:apdxB}. 
The integral is numerically evaluated using 
the double-exponential quadrature formula. We employ the program in 
Ref.~\citen{DEQUADRATURE} to evaluate the numerical integration not only for the
(A) and (B) above, but also for Eqs.~(\ref{eq:Rfactor2}) and (\ref{eq:Nfinal}).
The missing region for the validity check (A) and (B) is $1<r$ with $q\ne 0$.

Since the series coefficients $C^n_{\ell}$ and $dC^n_{\ell}/d\eta$ are 
independent from the choice of $r$ and only a finite set of $(n, \ell)$ 
induces the absorptive part in 
${\cal F}^{n}_{\ell}$, ${\cal G}^{n}_{\ell}$, 
${\cal H}^{n}_{\ell}$ and ${\cal R}^{n}_{\ell}$ for a finite $r$,
the double series does not change the asymptotic form irrespective of the choice of $r$.
Therefore we expect that
if we have the validity in the region of (A) $r<1$ with a truncated double series,
the same truncated series is also valid in the region of $1<r$.
The comparison in the region (B) provides a limited consistency check for the statement.
We make the above comparison numerically in the next section.

\section{Numerical Results}
\label{sec:Results}
We employ Fortran 90 language to evaluate Eq.~(\ref{eq:Nfinal}) in the double precision.
The double integral on $x$ and $v$ is evaluated by 
nesting the double-exponential quadrature formula subroutine of Ref.~\citen{DEQUADRATURE}.
In order to avoid the loss of significant digits in $\overline{N}_j(x,v)$ 
 ($\overline{\psi}(x,v)$) near $x=0$, we use the 9th order (11th order) Taylor expansion form 
for $x < 0.02$, respectively. To avoid an overflow of hyperbolic functions
in $\overline{N}_j(x,v)$ and $\overline{\psi}(x,v)$ for $1\ll x$,
we transform them to a well organized exponential form for $10<x$.

\begin{table}[t]
\tbl{Parameter combinations we investigated.
We use 
$m_e  =    0.5109989$ [MeV], 
$m_\mu=  105.6583668$ [MeV], 
$m_{\pi}=139.57018$ [MeV],
$0 <  k_{\parallel}^2 < 4^2 \mbox{\ [GeV$^2$]}$,
$0 < \bm{k}_{\perp}^2 < 3^2 \mbox{\ [GeV$^2$]}$.
We take 401 (41) sample points for both $r<1$ and $1<r$ region
at equal intervals.
}
{
\begin{tabular}{ccccc} \toprule
\multirow{2}{*}{Case}     &
\multirow{2}{*}{$[m,eB]$} &
\multirow{2}{*}{$\ell_{\mathrm{max}}$}
                                & \# of sample   & \# of sample   \\
          &               &     & points for $r$ & points for $q$ \\
\colrule
\multirow{2}{*}{[a-1]} & \multirow{2}{*}{$[m_{\mu},    10 m_{\pi}^2]$} & 1000 & 401 & 31 \\
      &                                                                & 2000, 4000, 8000 &  41 & 31 \\\relax
\multirow{2}{*}{[b-1]} & \multirow{2}{*}{$[m_{\mu},       m_{\pi}^2]$} & 1000 & 401 & 31 \\
      &                                                                & 2000, 4000, 8000 &  41 & 31 \\\relax
\multirow{2}{*}{[c-1]} & \multirow{2}{*}{$[m_{\mu},(1/10) m_{\pi}^2]$} & 1000 & 401 & 31 \\
      &                                                                & 2000, 4000, 8000 &  41 & 31 \\\relax
\multirow{2}{*}{[a-2]} & \multirow{2}{*}{$[    m_e,    10 m_{\pi}^2]$} & 1000 & 401 & 31 \\
      &                                                                & 2000, 4000, 8000 &  41 & 31 \\\relax
\multirow{2}{*}{[b-2]} & \multirow{2}{*}{$[    m_e,       m_{\pi}^2]$} & 1000 & 401 & 31 \\
      &                                                                & 2000, 4000, 8000 &  41 & 31 \\\relax
\multirow{2}{*}{[c-2]} & \multirow{2}{*}{$[    m_e,(1/10) m_{\pi}^2]$} & 1000 & 401 & 31 \\
      &                                                                & 2000, 4000, 8000 &  41 & 31 \\
\botrule
\end{tabular}
\label{tab:params}
}
\end{table}

\newcommand{\figscale}{2.5in}
\newcommand{\figscaleB}{2.5in}

The coefficient functions $C^n_{\ell}$ and $dC^n_{\ell}/d\eta$ are
computed using a three-term recurrence formula based on 
the Laguerre polynomials during the summation on $\ell$.
This means that we keep last several values of $L^n_{\ell-1}$ 
and $L^n_{\ell-2}$ {\it etc.} to compute $C^n_{\ell}$ and $dC^n_{\ell}/d\eta$ 
for $\Omega_{j,\ell}^{n}$ to avoid the full re-computation of $L^n_{\ell}$ 
at each $\ell$. The three-term recurrence formula and numerical method we employed 
is explained in \ref{sec:apdxC}.

The double series of Eq.~(\ref{eq:Nfinal}) must be truncated at a cutoff index
$(n_{\mathrm{max}},\ell_{\mathrm{max}})$ for the numerical evaluation.
The summation on $\ell$ is truncated at a $\ell_{\mathrm{max}}$ independent of $n$. While
the summation on $n$ is stopped when 
the partial sum $\delta N_j = \sum_{\ell=0}^{\ell_{\mathrm{max}}} \Omega^n_{j,\ell}$
becomes negligible compared to the current estimate of $N_j$
provided by the following condition;
\begin{equation}
\left(|\delta N_j| < 10^{-14} \quad \mbox{and}\quad |N_j| < 10^{-14}\right)
\quad \mbox{or}\quad |\Delta N_j|/|N_j|<10^{-14}
\end{equation}
in double precision arithmetic.

We show the combination of input parameters 
for the form factors in Table~\ref{tab:params}. 
We choose the magnetic field strength at $O(m_{\pi}^2)$ which is expected 
to exist in the heavy ion collisions at LHC\cite{Skokov:2009qp,Voskresensky:1980}.
The longitudinal and transverse momenta ranges we investigated are
$0 <  k_{\parallel}^2 < 4^2 \mbox{\ [GeV$^2$]}$ and 
$0 < \bm{k}_{\perp}^2 < 3^2 \mbox{\ [GeV$^2$]}$, respectively.

\begin{figure}[t]
\centerline{
\includegraphics[width=\figscale]{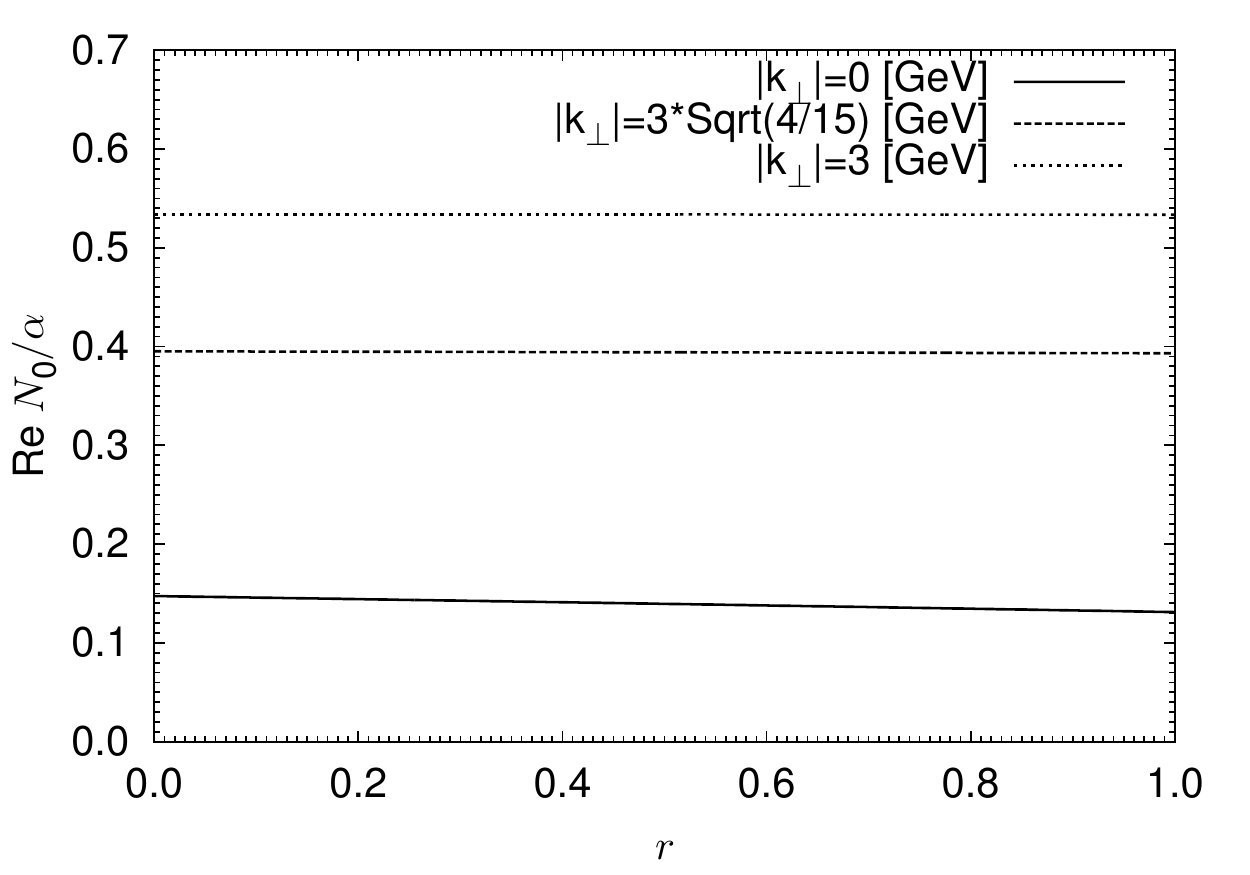}
\includegraphics[width=\figscale]{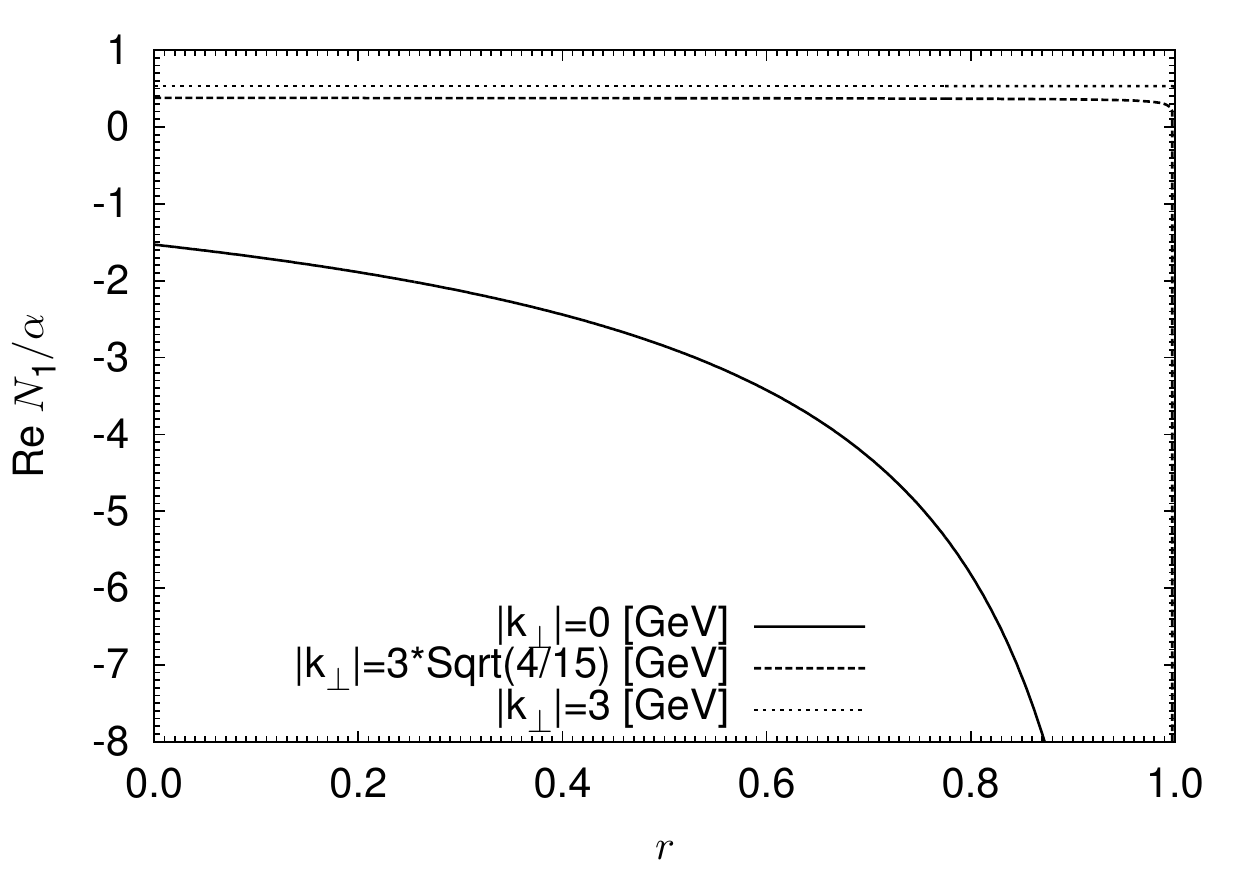}
}
\vspace*{-4pt}
\caption{Form factors $N_0$ (left) and $N_1$ (right) for muons (case [a-1])
in $r < 1$ with $\ell_{\mathrm{max}}=1000$.}
\label{fig:N0N1muon10mpiBelow}
\end{figure}

\begin{figure}[t]
\centerline{
\includegraphics[width=\figscale]{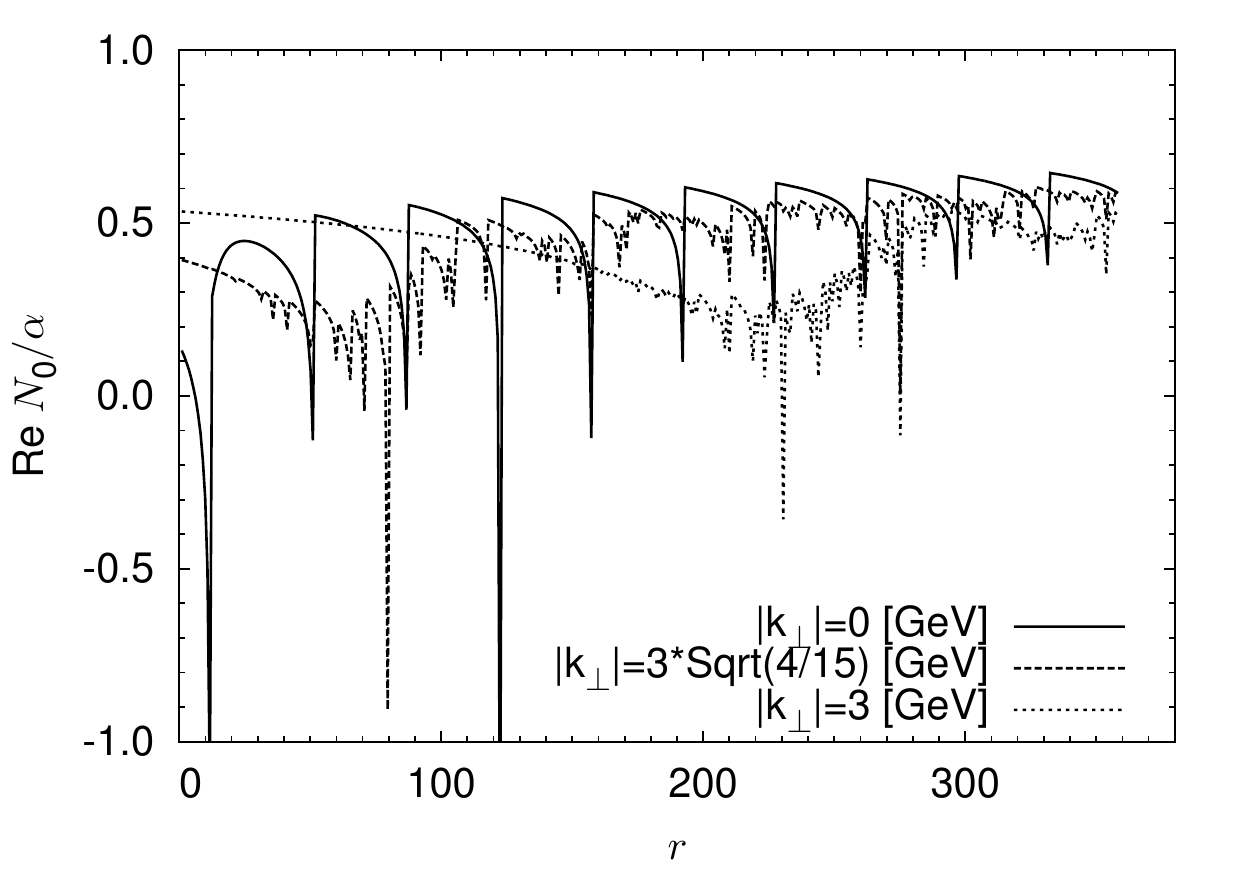}
\includegraphics[width=\figscale]{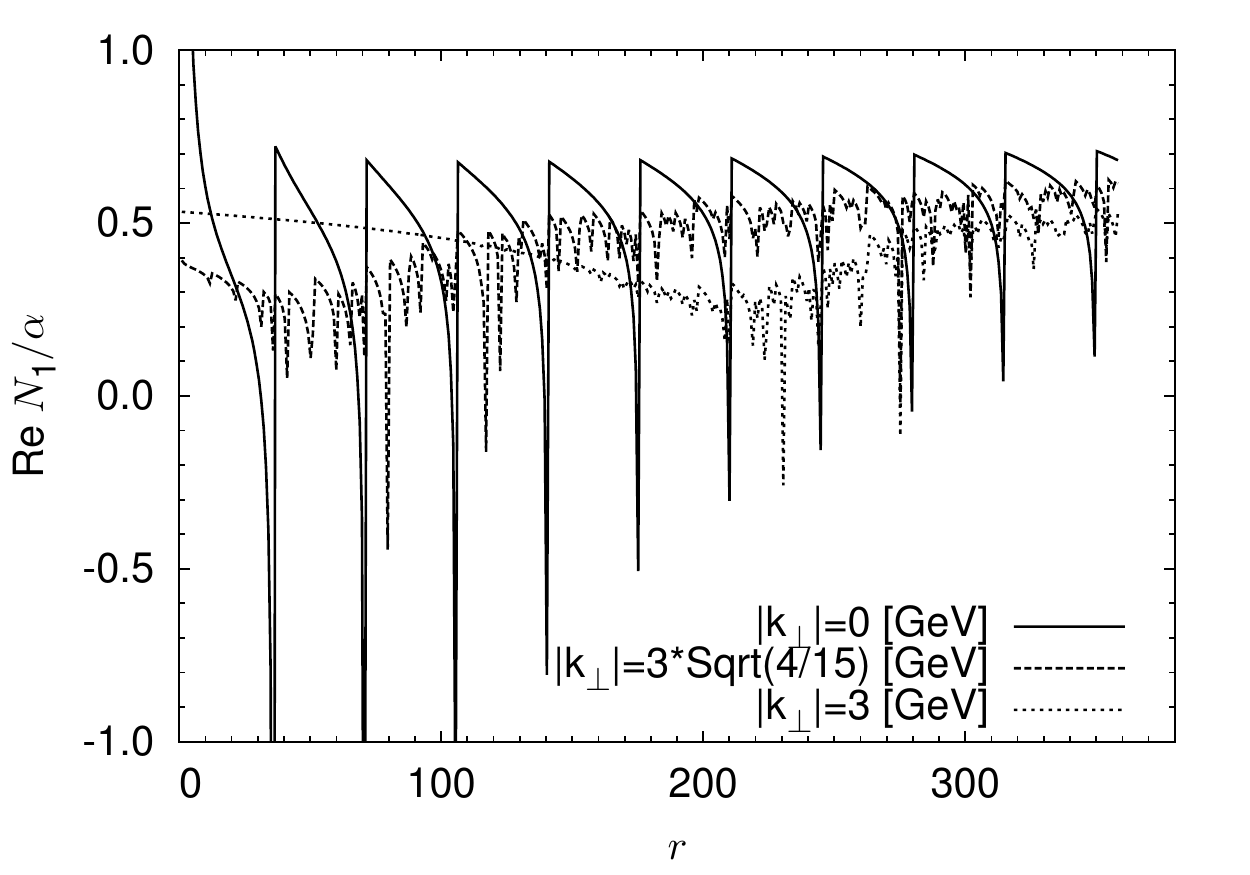}
}
\centerline{
\includegraphics[width=\figscale]{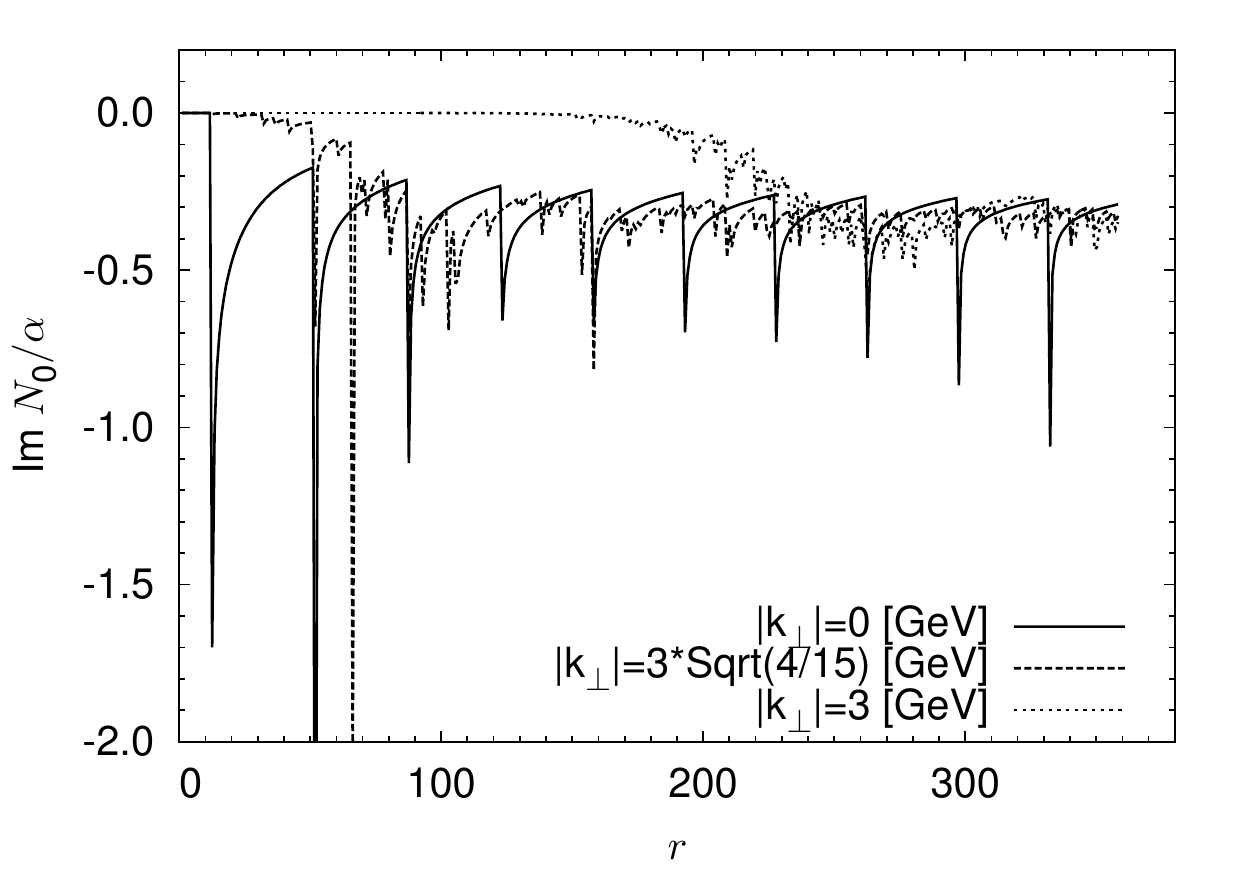}
\includegraphics[width=\figscale]{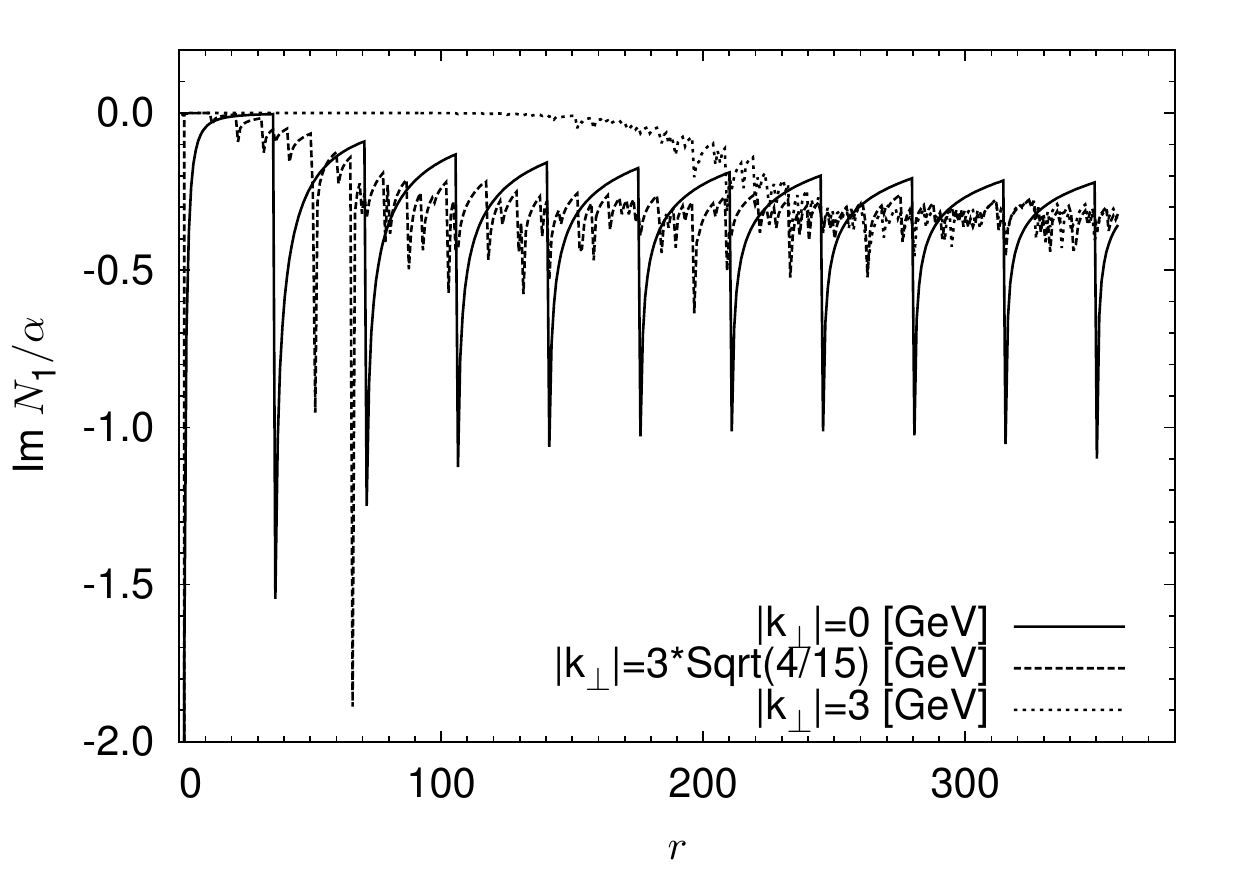}
}
\vspace*{-4pt}
\caption{Same as Fig.\ref{fig:N0N1muon10mpiBelow} 
 but for real (top) and imaginary (bottom) parts in $1<r$.}
\label{fig:N0N1muon10mpiAbove}
\end{figure}

\begin{figure}[th]
\centerline{
\includegraphics[width=\figscale]{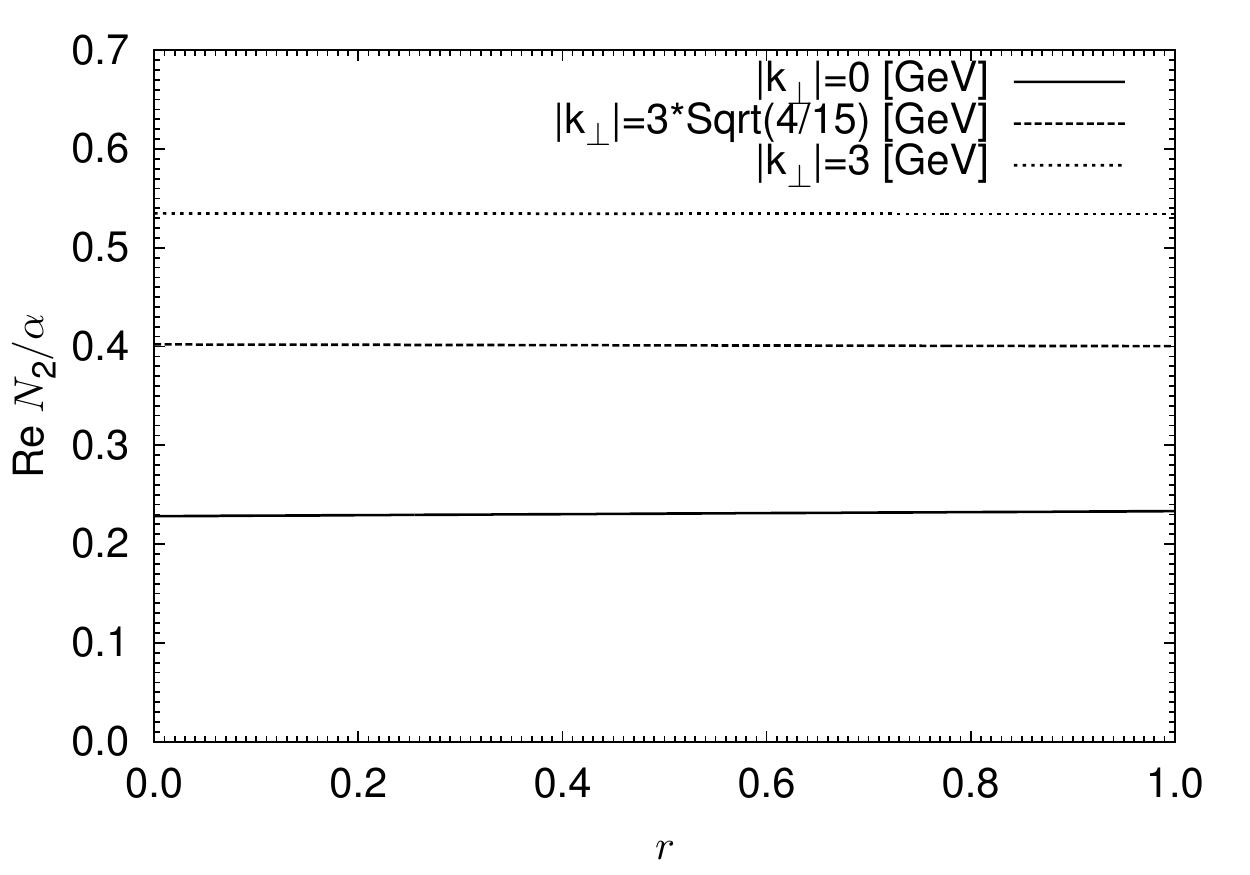}
}
\centerline{
\includegraphics[width=\figscale]{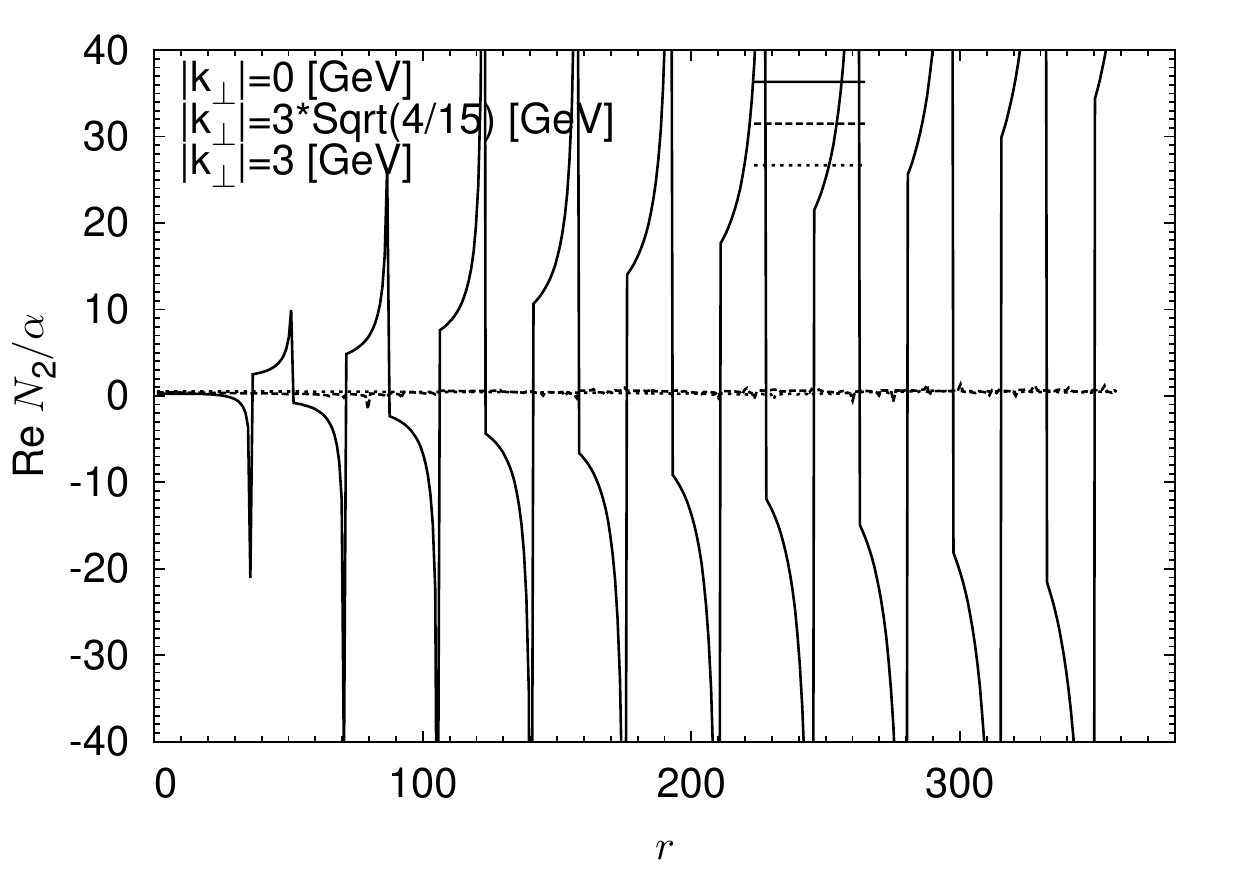}
\includegraphics[width=\figscale]{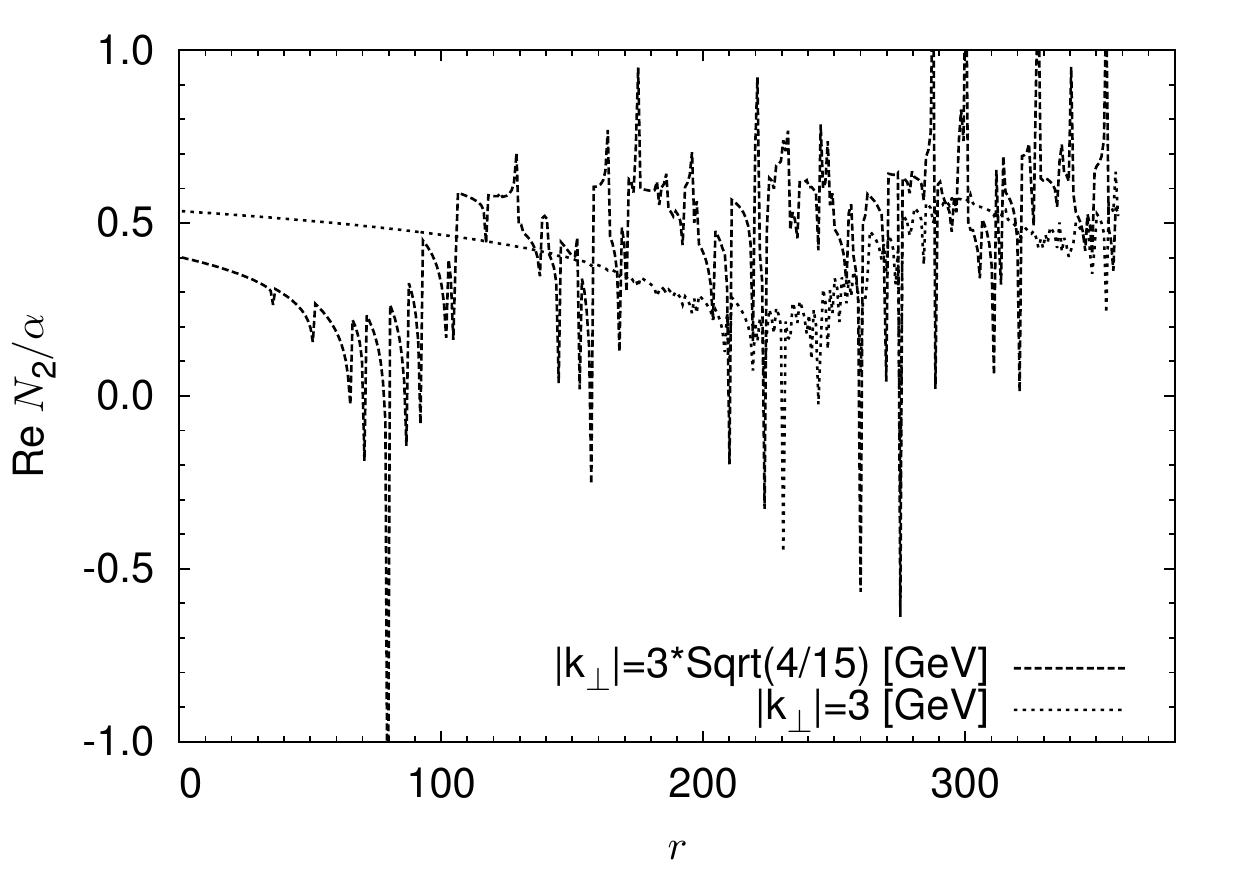}
}
\centerline{
\includegraphics[width=\figscale]{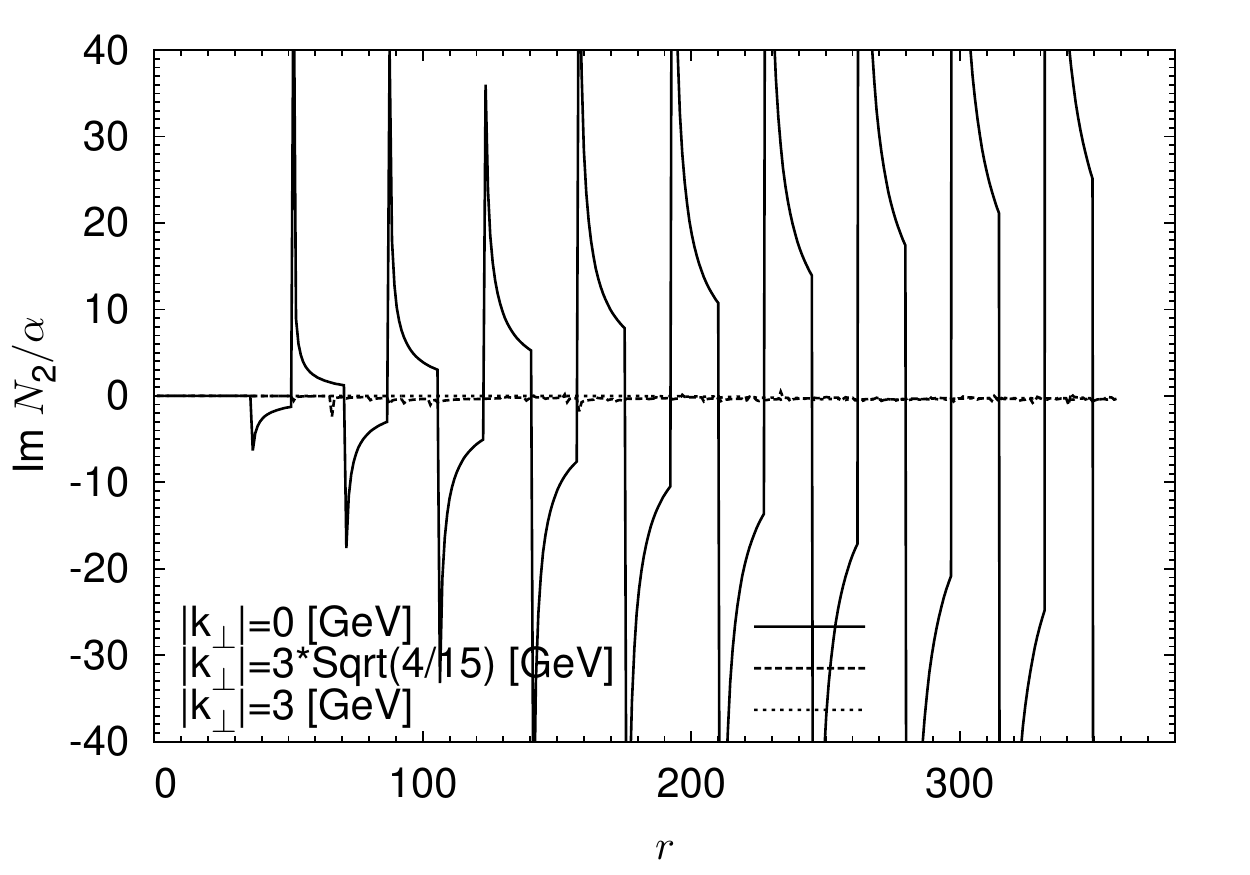}
\includegraphics[width=\figscale]{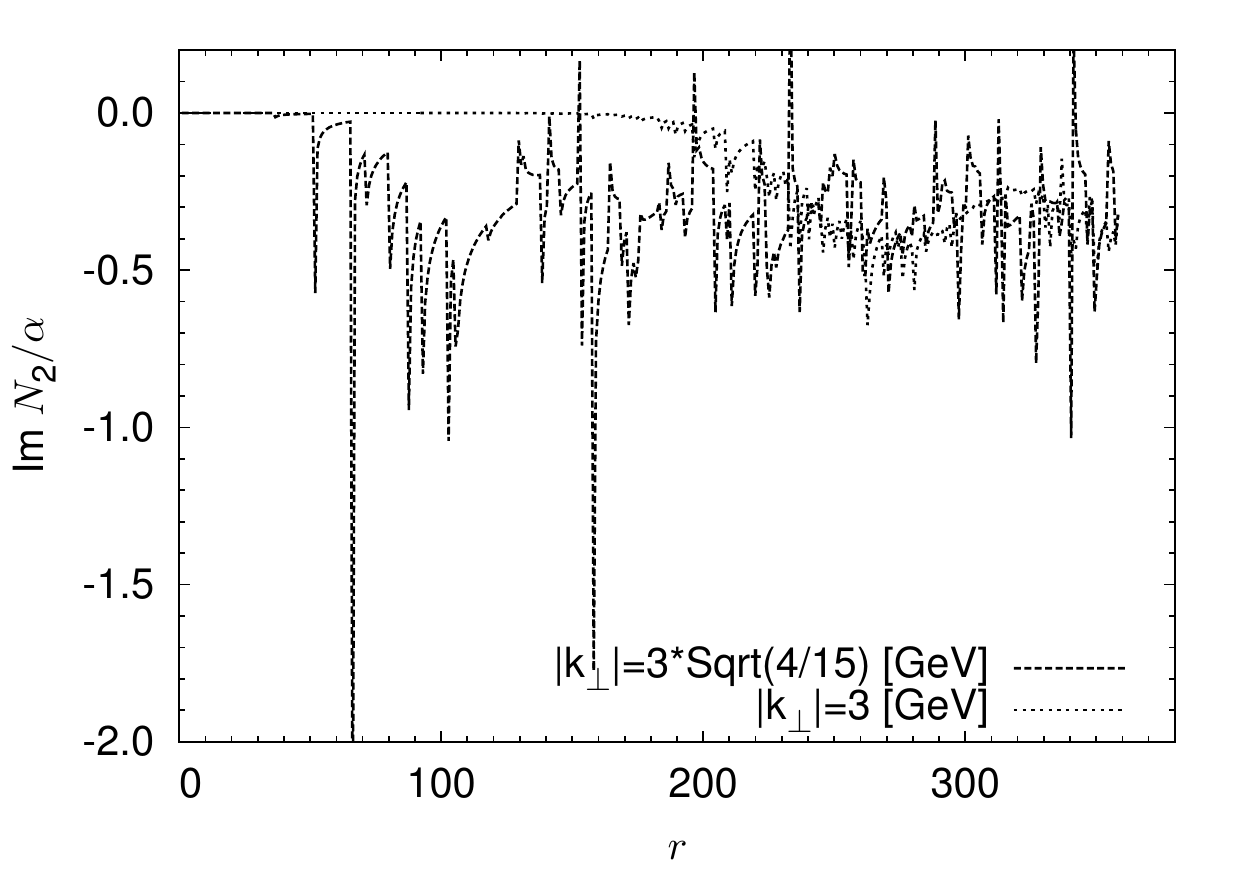}
}
\vspace*{-4pt}
\caption{Form factor $N_2$ for muons (case [a-1]) with
$\ell_{\mathrm{max}}=1000$
(top ($r<1$), middle (real part in $1<r$) and bottom (imaginary part in $1<r$).
Right panels in $1<r$ are magnification of left panels.}
\label{fig:N2muon10mpi}
\end{figure}

\begin{figure}[ht]
\centerline{
\includegraphics[width=\figscale]{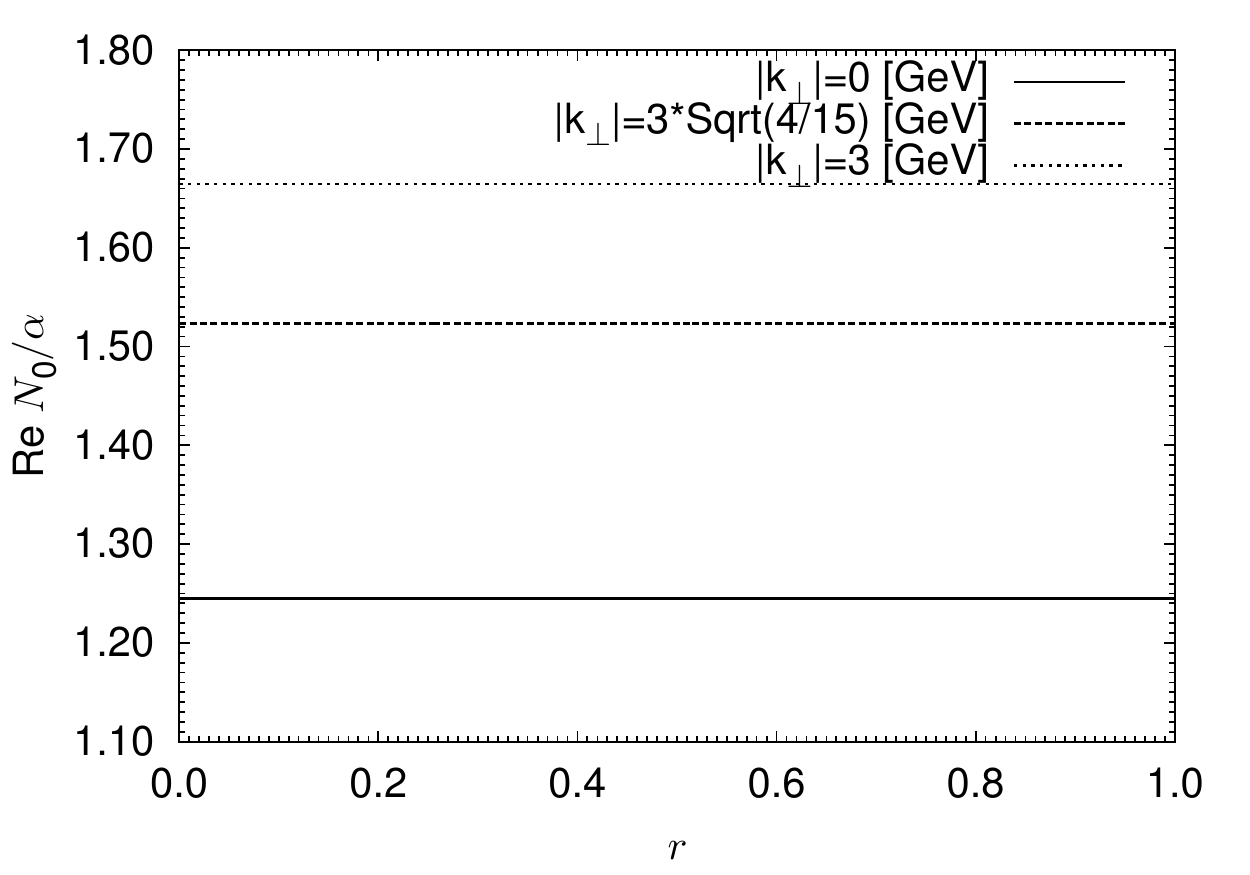}
\includegraphics[width=\figscale]{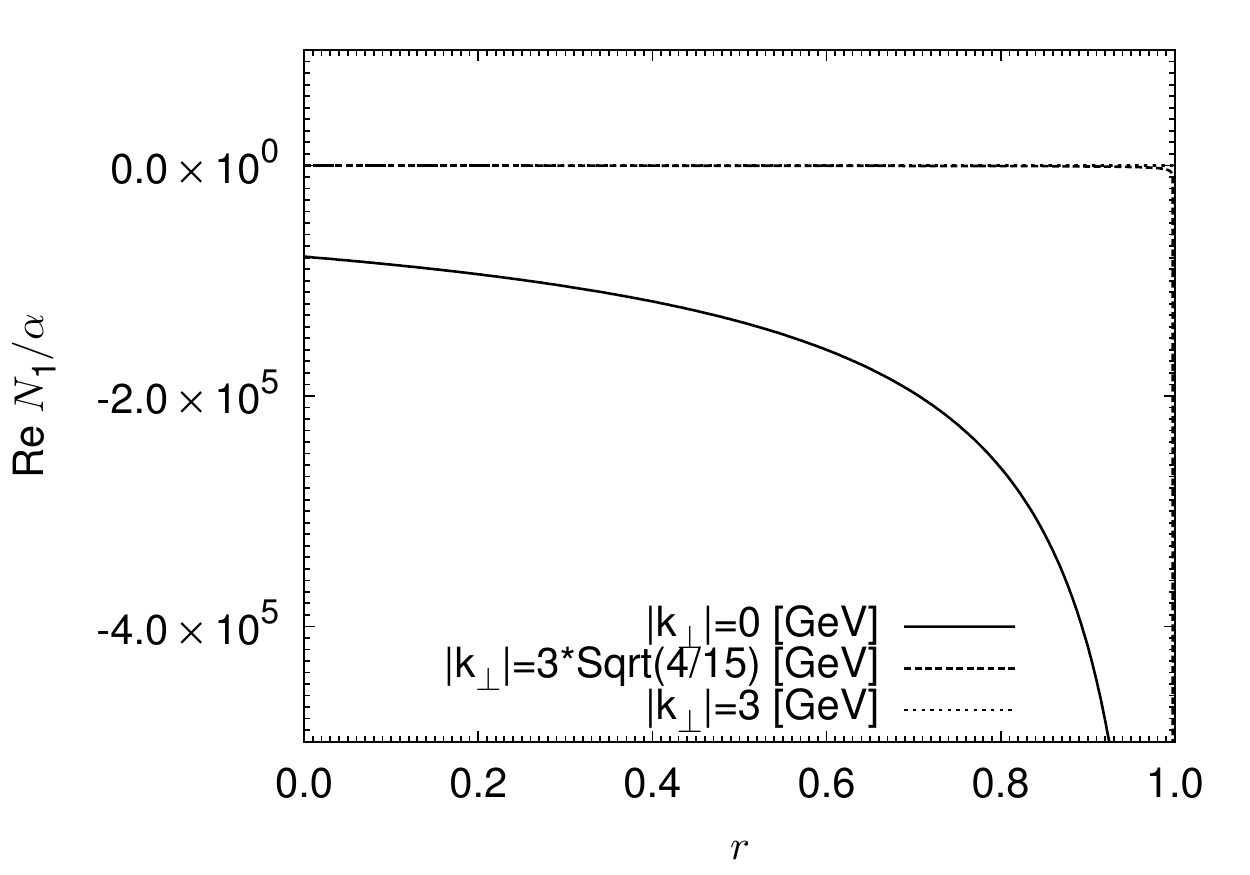}
}
\vspace*{-4pt}
\caption{Form factors $N_0$ (left) and $N_1$ (right) in $r<1$ for electrons (case [a-2])
with $\ell_{\mathrm{max}}=1000$.}
\label{fig:N0N1electron10mpiBelow}
\end{figure}

\begin{figure}[ht]
\centerline{
\includegraphics[width=\figscale]{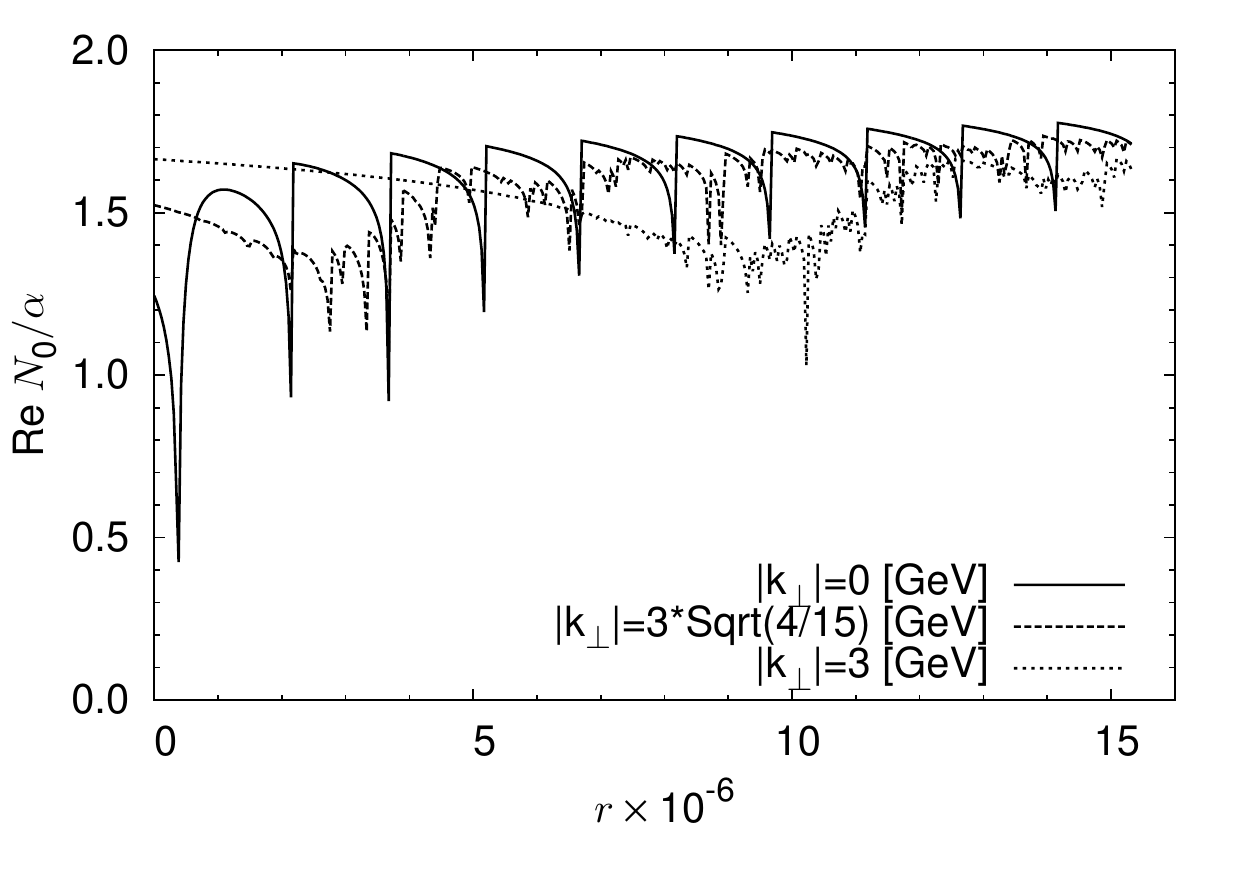}
\includegraphics[width=\figscale]{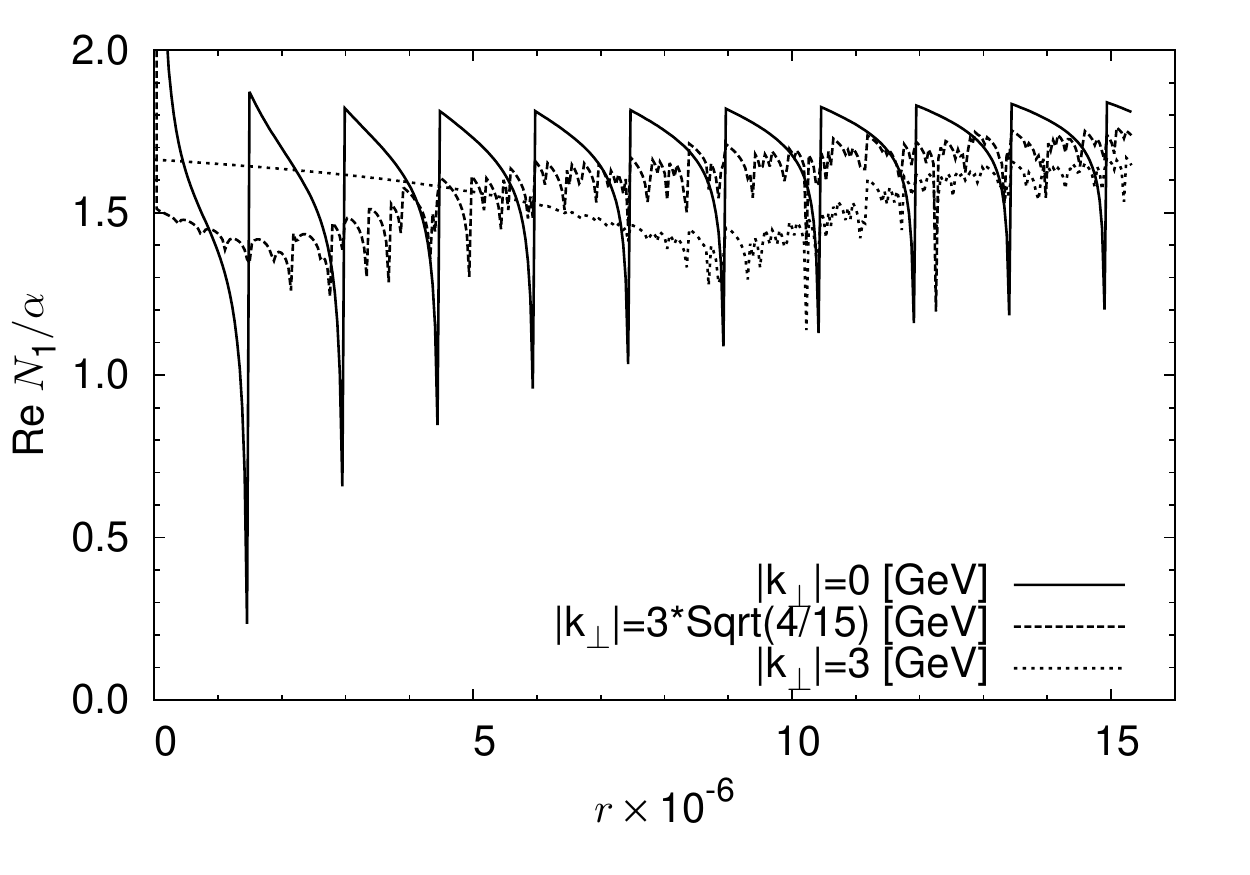}
}
\centerline{
\includegraphics[width=\figscale]{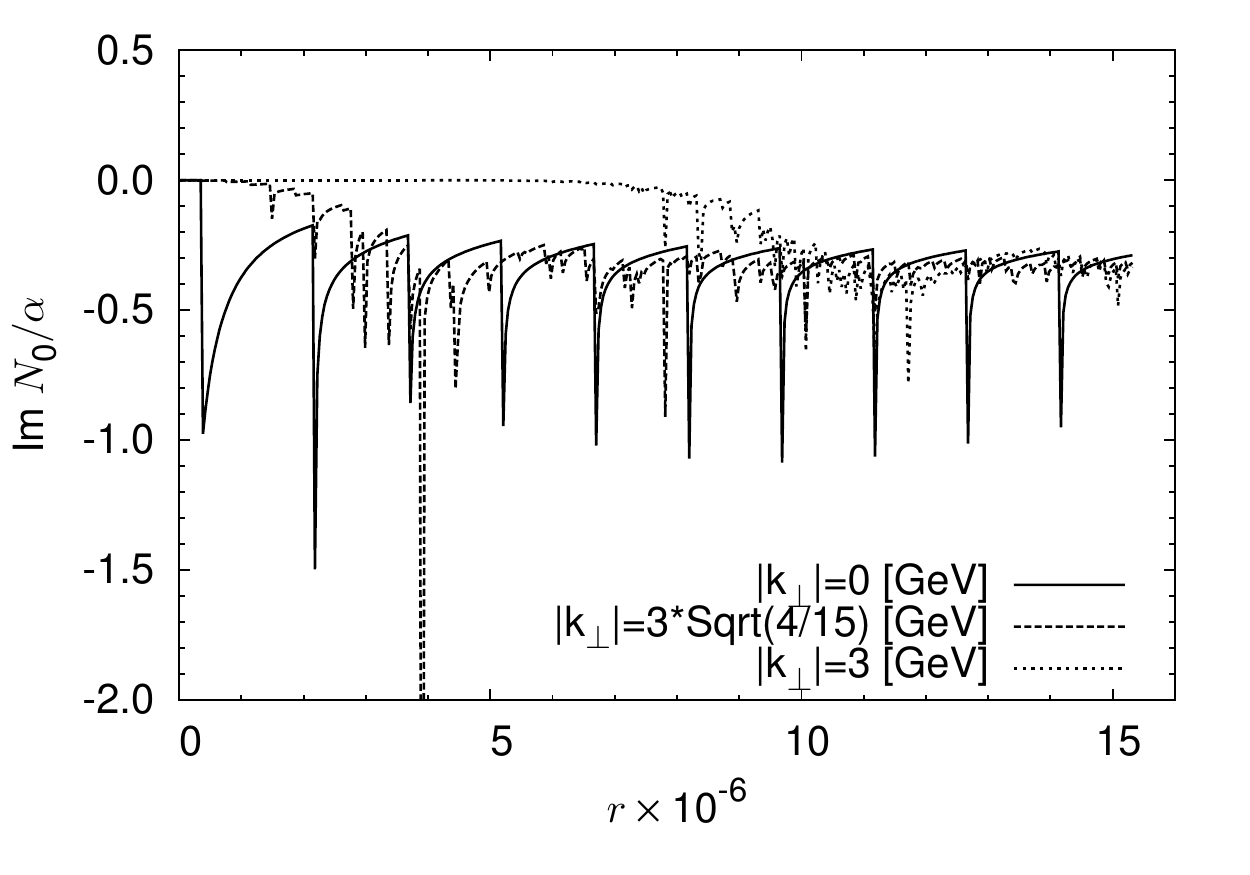}
\includegraphics[width=\figscale]{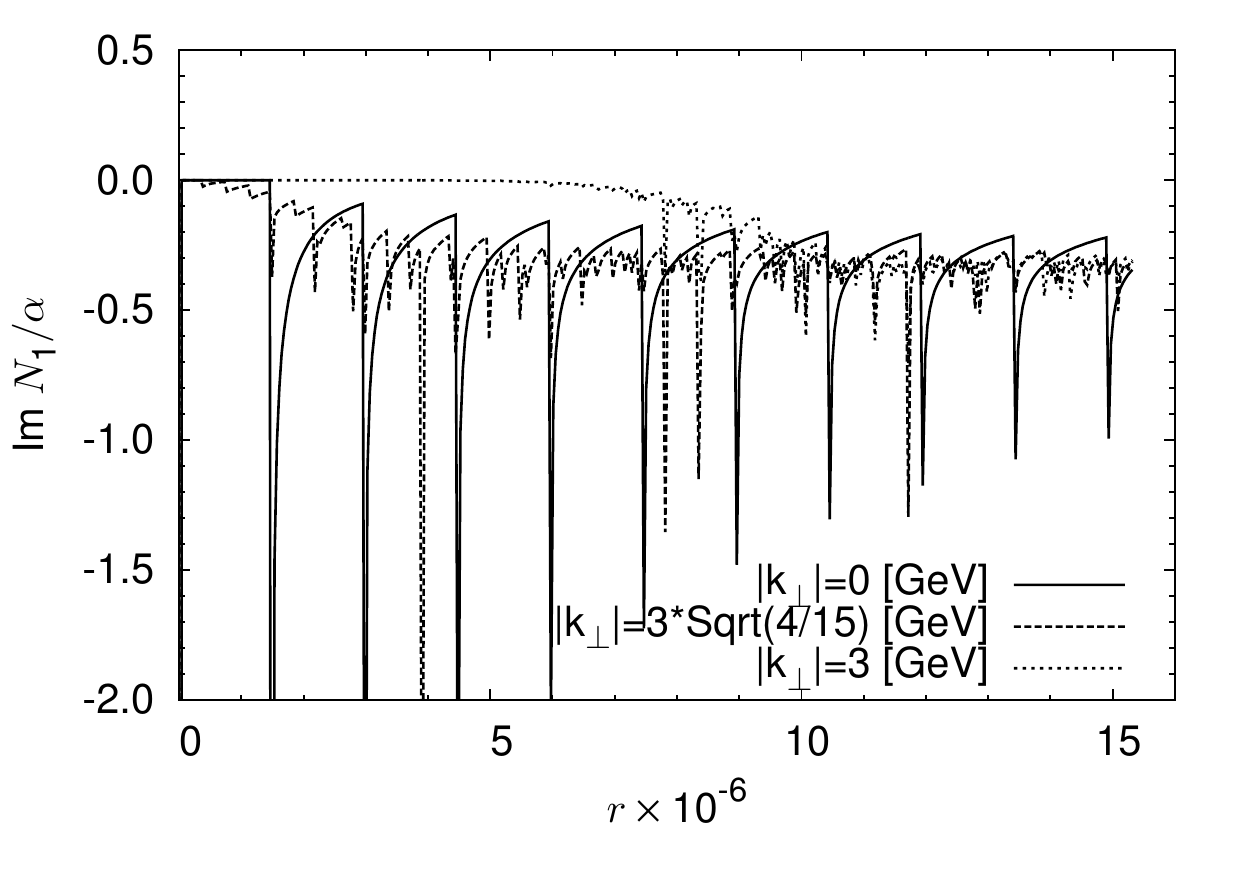}
}
\vspace*{-4pt}
\caption{Same as Fig.\ref{fig:N0N1electron10mpiBelow}
 but for real (top) and imaginary (bottom) parts in $1<r$.}
\label{fig:N0N1electron10mpiAbove}
\end{figure}

\begin{figure}[ht]
\centerline{
\includegraphics[width=\figscale]{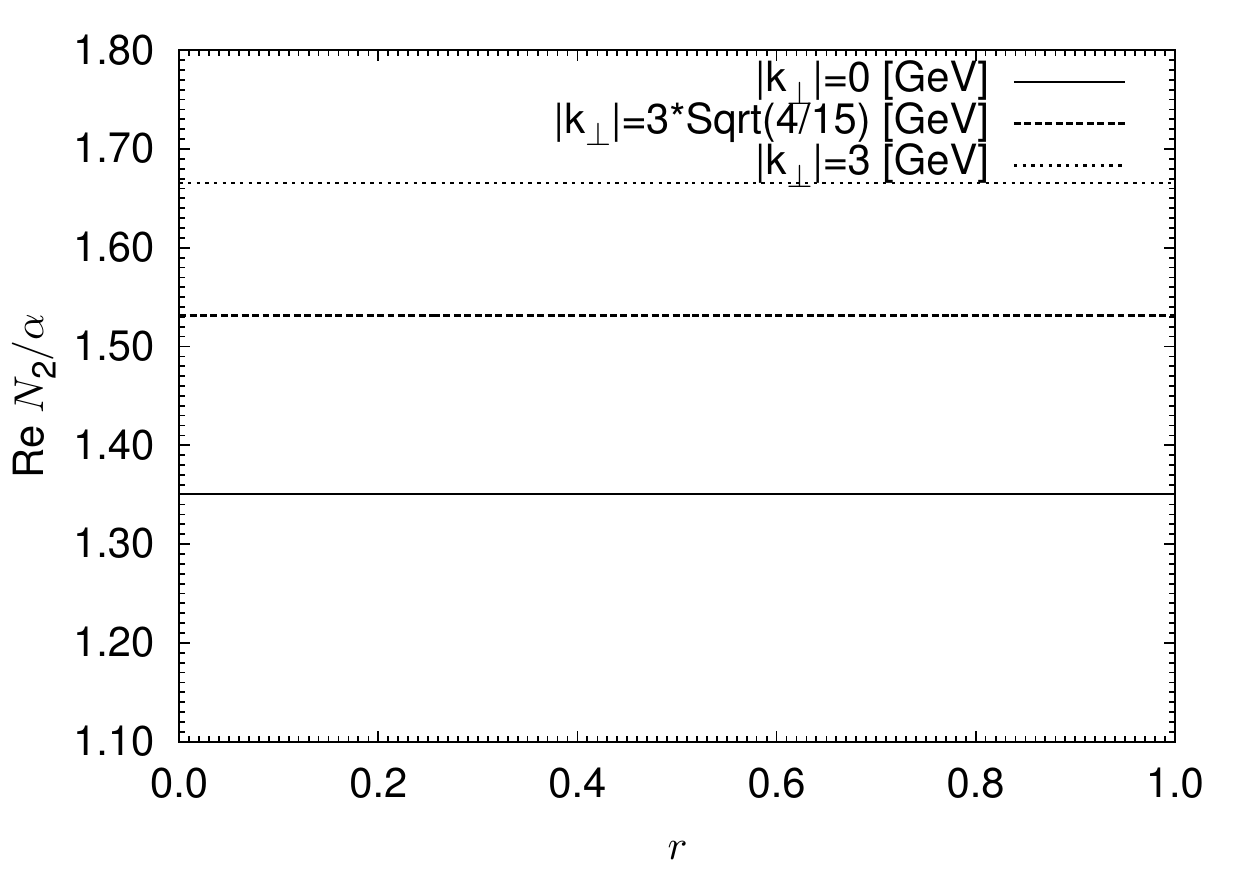}
}
\centerline{
\includegraphics[width=\figscale]{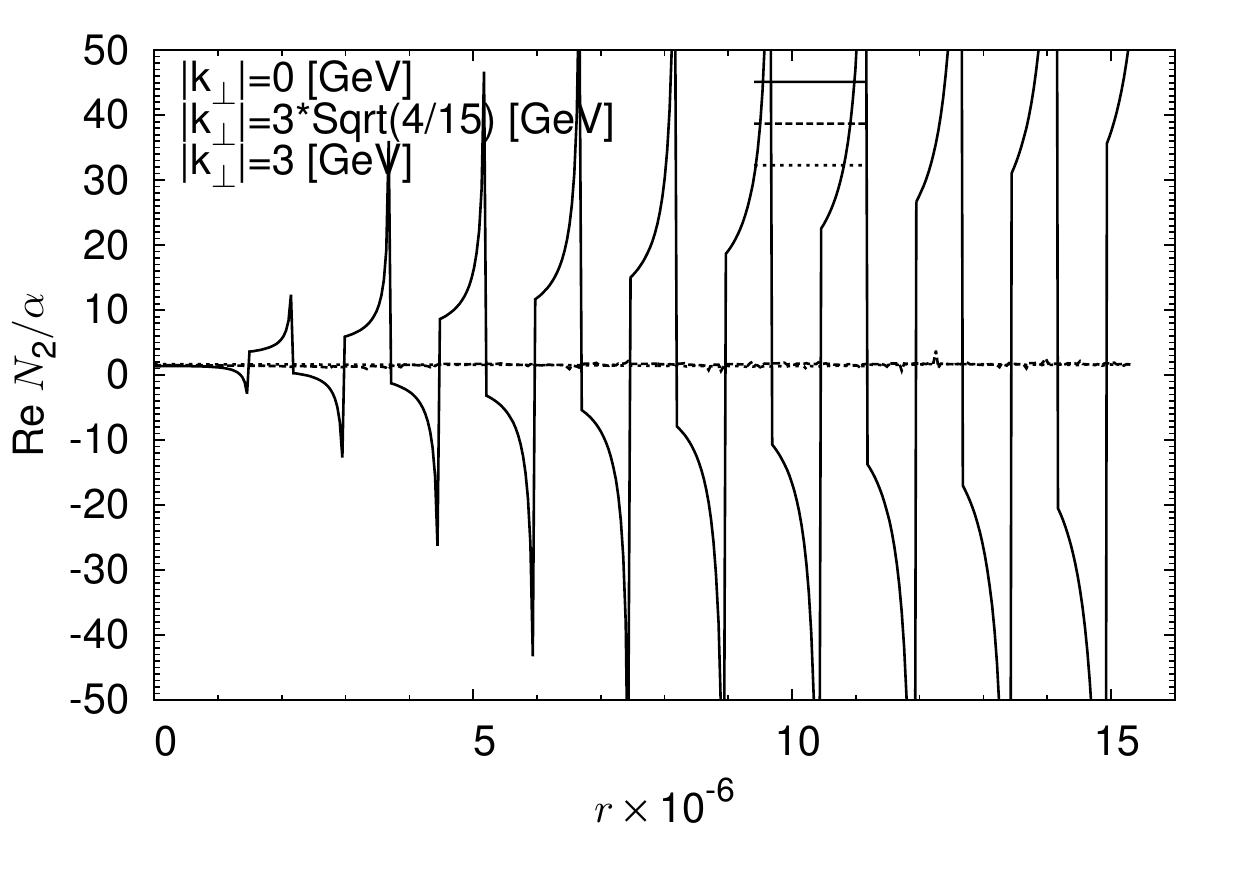}
\includegraphics[width=\figscale]{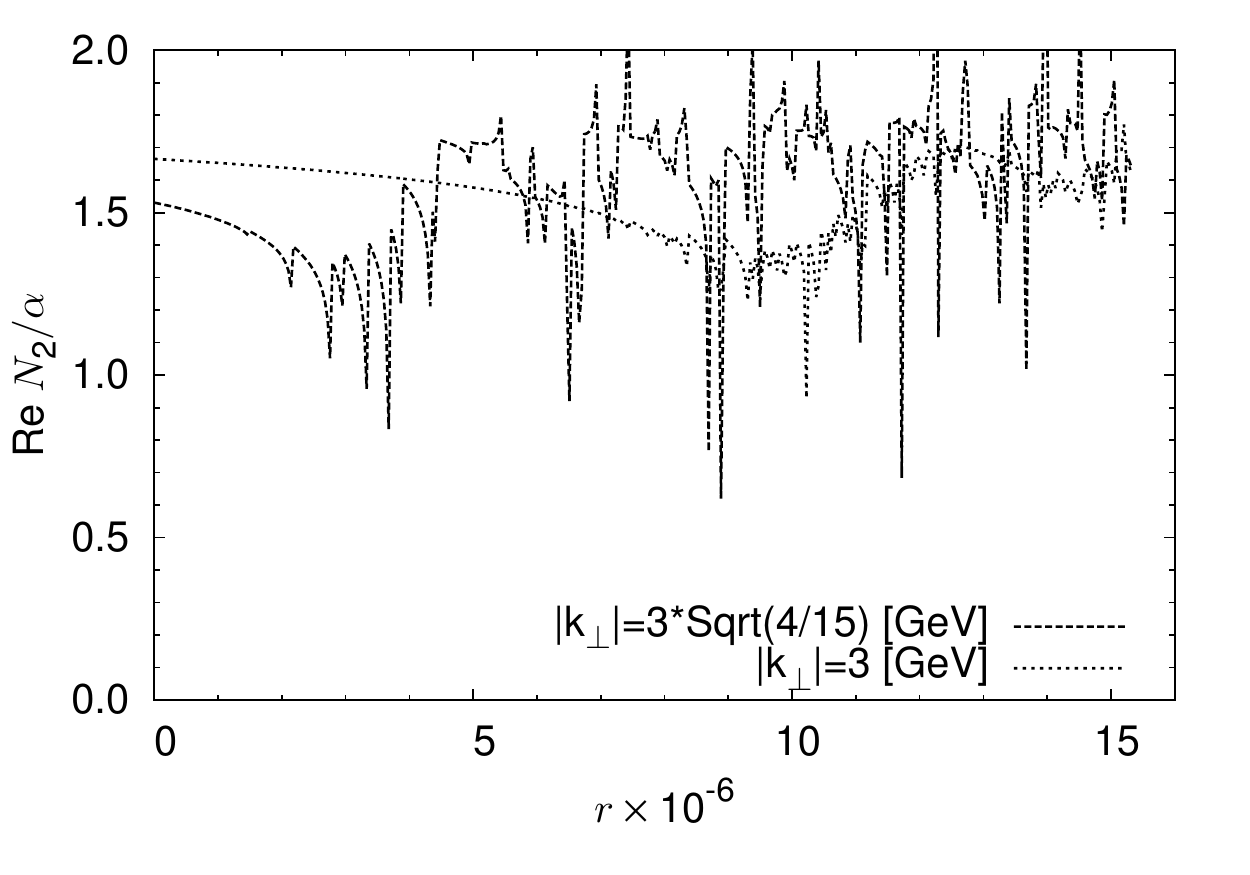}
}
\centerline{
\includegraphics[width=\figscale]{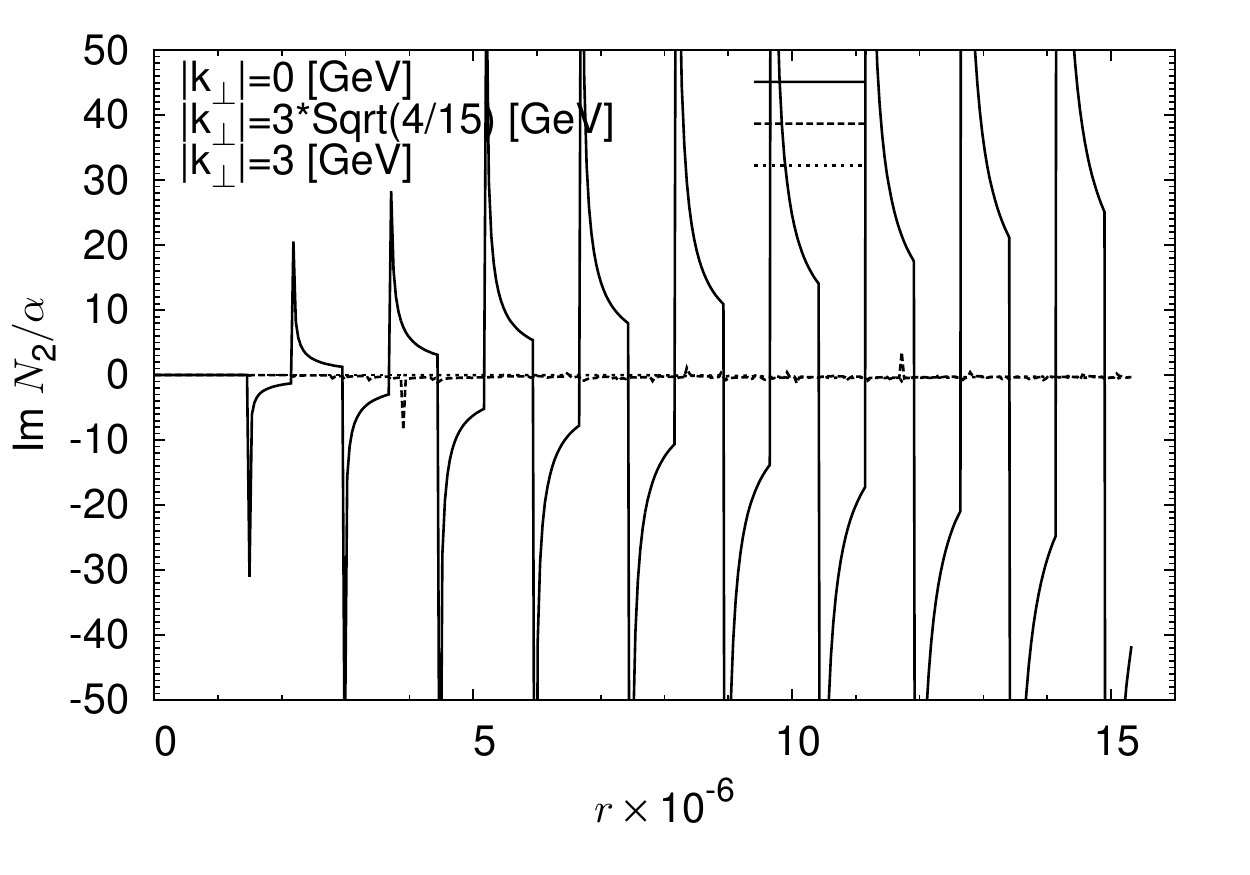}
\includegraphics[width=\figscale]{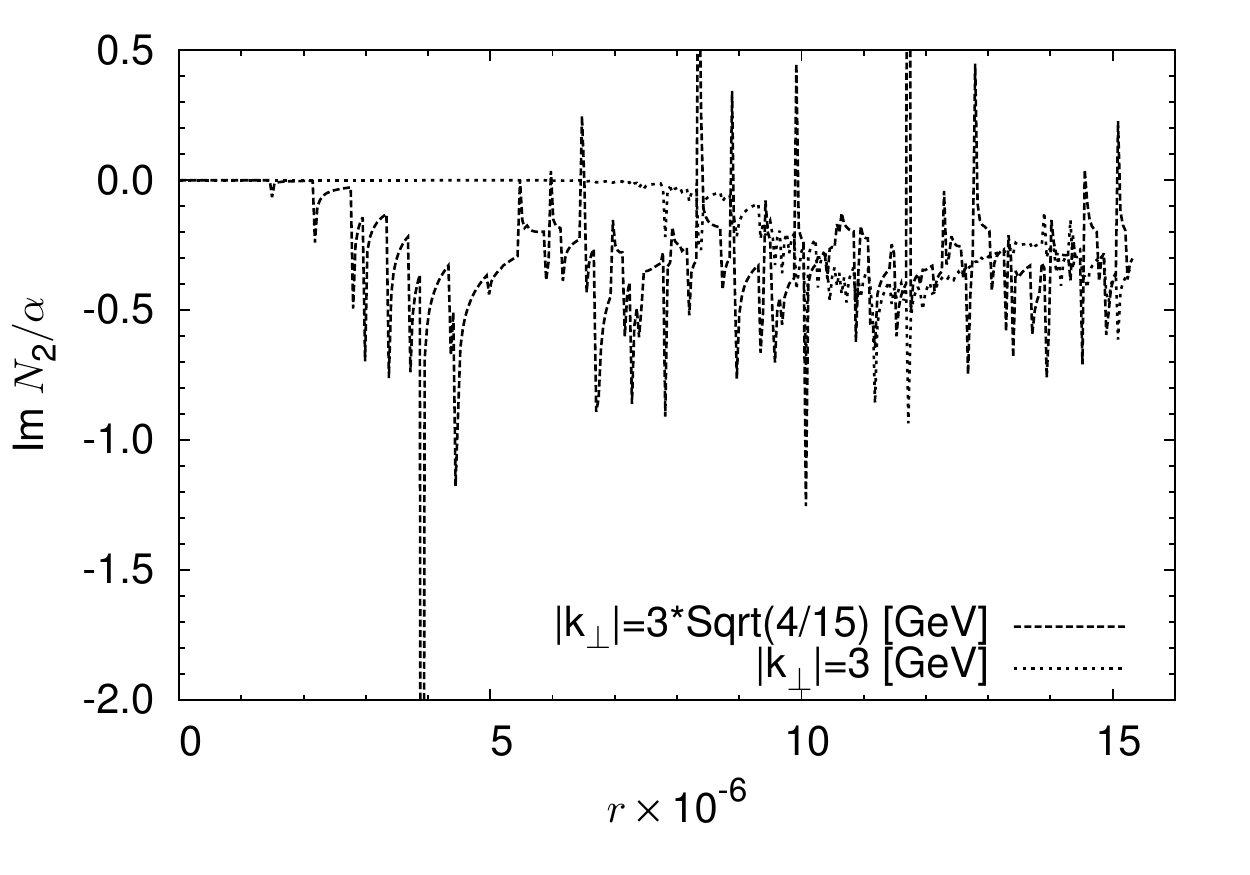}
}
\vspace*{-4pt}
\caption{Same as Fig.\ref{fig:N0N1electron10mpiBelow} but for $N_2$.}
\label{fig:N2electron10mpi}
\end{figure}

Figs.~\ref{fig:N0N1muon10mpiBelow}-\ref{fig:N2muon10mpi} show the form factors
$N_0$, $N_1$ and $N_2$ with $m=m_{\mu}$ and $eB=10m_{\pi}^2$ (case [a-1]).
The upper limit on $\ell$ is $\ell_{max}=1000$.
Figs.~\ref{fig:N0N1electron10mpiBelow}-\ref{fig:N2electron10mpi} are for electrons with $m=m_e$ 
with $eB=10m_{\pi}^2$ and $\ell_{max}=1000$ (case [a-2]).
Complicated threshold structures due to the Landau-levels are seen for $1<r$ 
in Figs.~\ref{fig:N0N1muon10mpiAbove}-\ref{fig:N2muon10mpi} 
and in Figs.~\ref{fig:N0N1electron10mpiAbove}-\ref{fig:N2electron10mpi}.
The pair creation trapped in the magnetic field occurs at each sharp peak (the cyclotronic resonance).~\cite{Shabad:1972rg}
The property of the singularity comes from the analytic property of the functions 
of Eqs.~(\ref{eq:FuncF})-(\ref{eq:FuncH}). 
The absorptive part is essentially from Eq.~(\ref{eq:AnlFuncF}) (see \ref{sec:apdxA}).
The real part diverges just below the threshold and finite just above the threshold.
The imaginary part is zero just below the threshold and diverges just above the threshold~\cite{Hattori:2012je,Shabad:1984ASTRO}.
The solid lines in the top panels of
Figs.~\ref{fig:N2muon10mpi} and \ref{fig:N2electron10mpi},
which correspond to $N_2$ with $q=0$,
have a different behavior compared to the other lines. 
This is because of the conditionally convergent property of $\Omega_{2,\ell}^{n}$
as explained in the last section. 
Thus the solid lines in the real part of $N_2$ in $1<r$ 
(middle and bottom left panels of Figs.~\ref{fig:N2muon10mpi} and \ref{fig:N2electron10mpi})
also contain the same systematic error. 
We compile other figures for the form factors with weaker magnetic fields 
in \ref{sec:apdxE} together with those at $eB=0$. 
As we decreasing the field strength the interval between the thresholds and 
the amplitude of the peaks decrease.
The form factors seem to approach the value with the vanishing field.


\begin{figure}[th]
\centerline{
\includegraphics[width=\figscaleB]{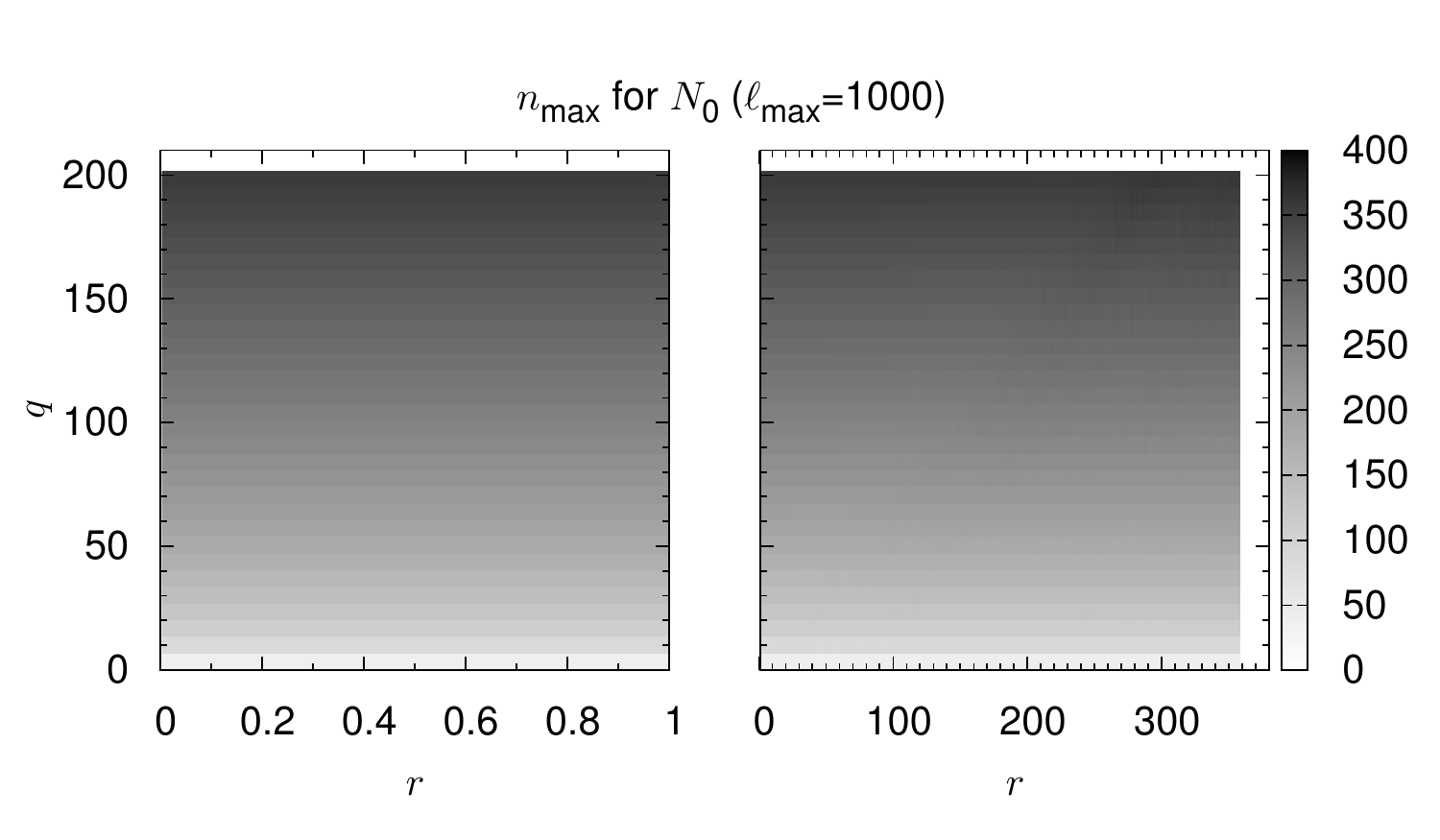}
\includegraphics[width=\figscaleB]{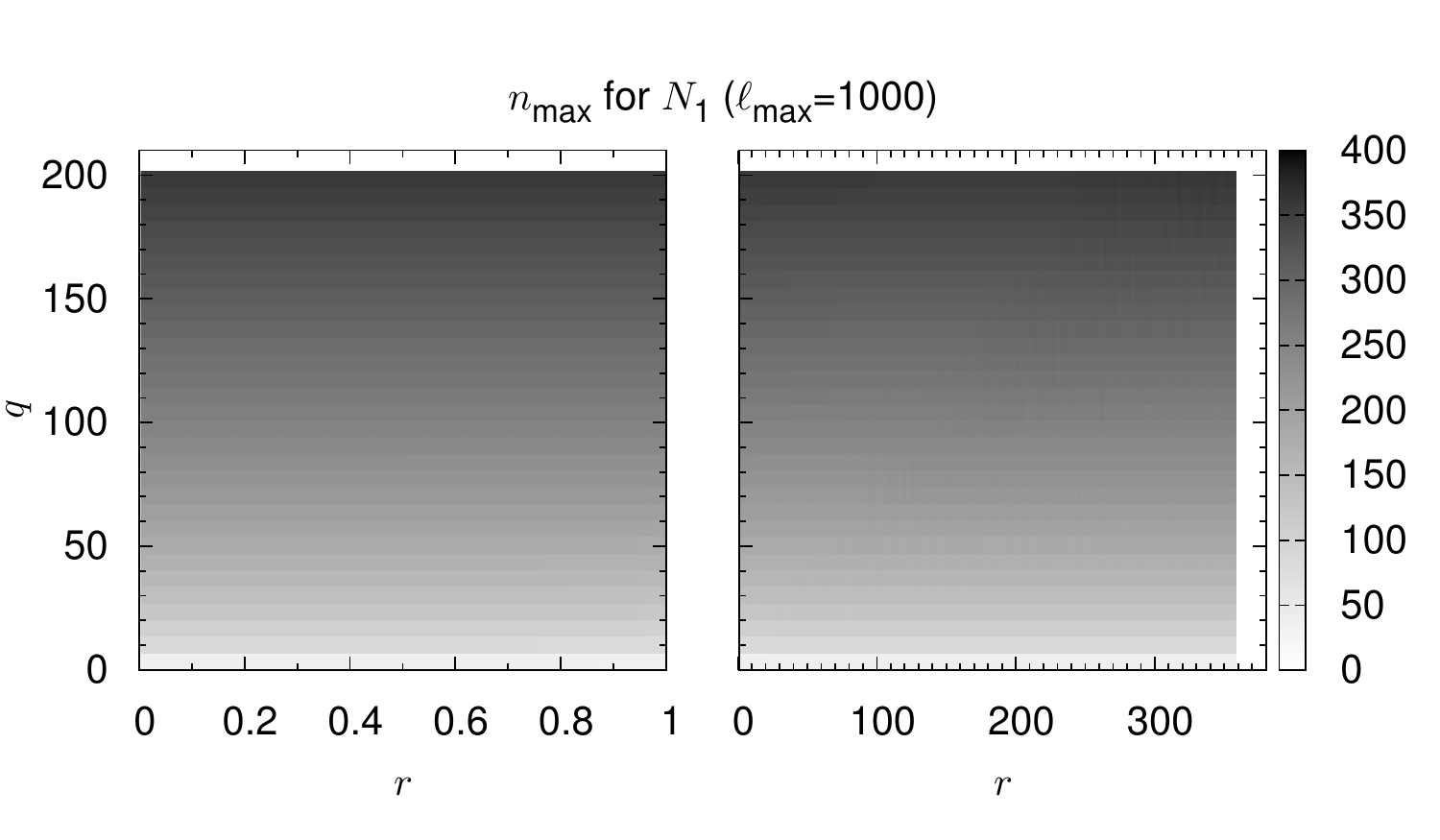}
}
\vspace*{-1em}
\centerline{
\includegraphics[width=\figscaleB]{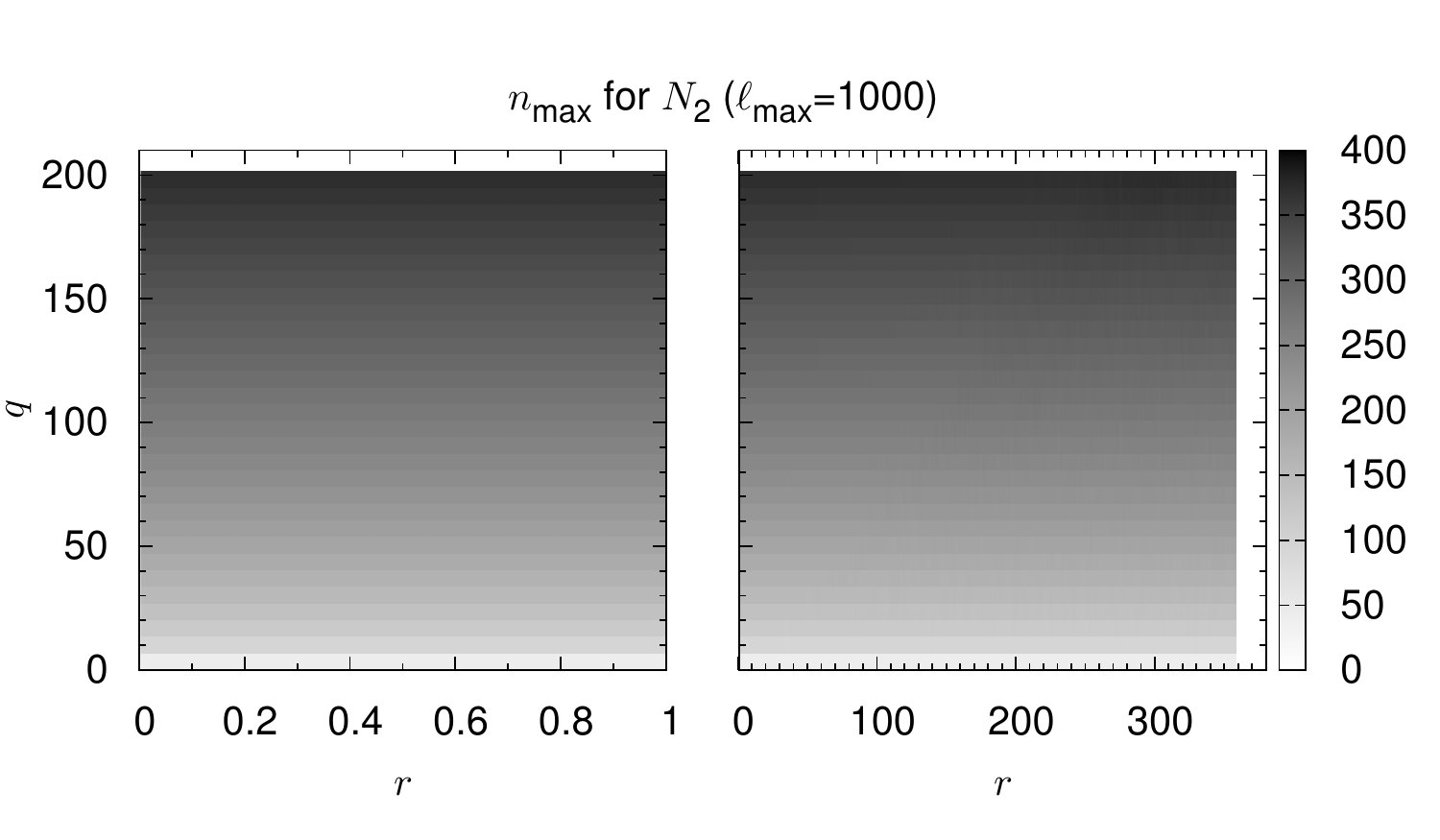}
}\vspace*{-1em}
\caption{$n_{\mathrm{max}}$ for muons (case [a-1]) with $\ell_{\mathrm{max}}=1000$.}
\label{fig:truncationindex}
\end{figure}

The truncation on the summation $n$ is monitored as shown in Fig.~\ref{fig:truncationindex}.
It requires 360--370 terms on $n$ for larger transverse momenta $q$.
We also observe that $n_{\mathrm{max}}$ depends linearly on $\sqrt{\ell_{\mathrm{max}}}$,
resulting in $n_{\mathrm{max}}=$ 920--940 at $\ell_{\mathrm{max}}=8000$.
We observe a similar behavior on $n_{\mathrm{max}}$ for electrons except for $N_1$.
An early truncation in  $r \lesssim 1$ for $N_1$ is seen as it is well approximated 
by the lowest Landau level approximation\cite{Dobrich:2012jd,Karbstein:2011ja,Fukushima:2011nu}.
As we decreasing $eB$ to $m_{\pi}^2/10$, $n_{\mathrm{max}}$ increases to 5460--5500
(with $\ell_{\mathrm{max}}=1000$) for both electrons and muons. 
To approach the zero field limit, we must accumulate more contributions 
from higher Landau levels. Verifying the zero field limit becomes numerically difficult.
The zero field limit for the imaginary parts with $q=0$ can be analytically taken as
shown in \ref{sec:apdxD}. 
The divergence at each threshold properly disappears in the case with $q=0$.
It should be noted that 
the truncation error involved in the figures with weaker fields in \ref{sec:apdxE}
could be rather large than those with $eB=10m_\pi^2$ as we explain in the following.

\begin{figure}[t]
\centerline{
\includegraphics[width=\figscale]{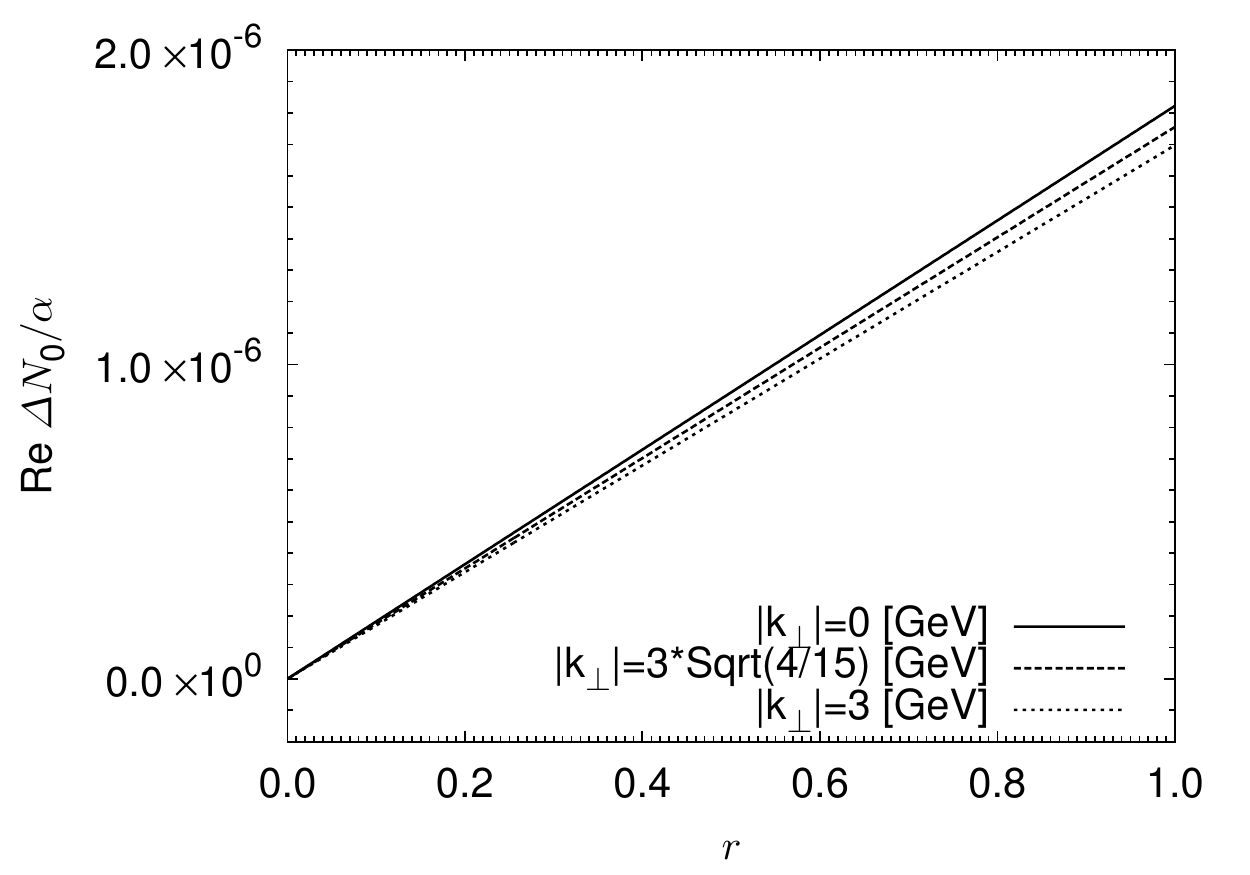}
\includegraphics[width=\figscale]{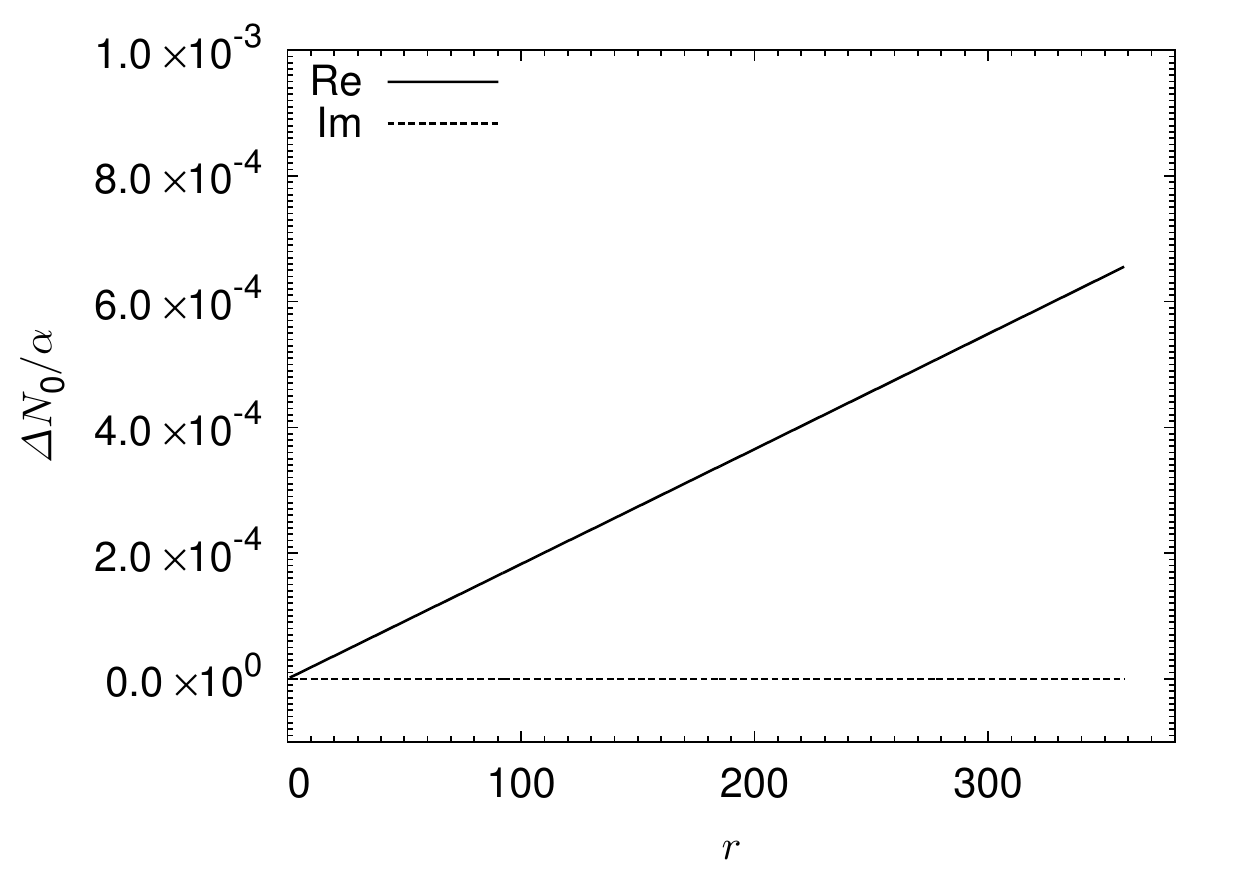}
}
\centerline{
\includegraphics[width=\figscale]{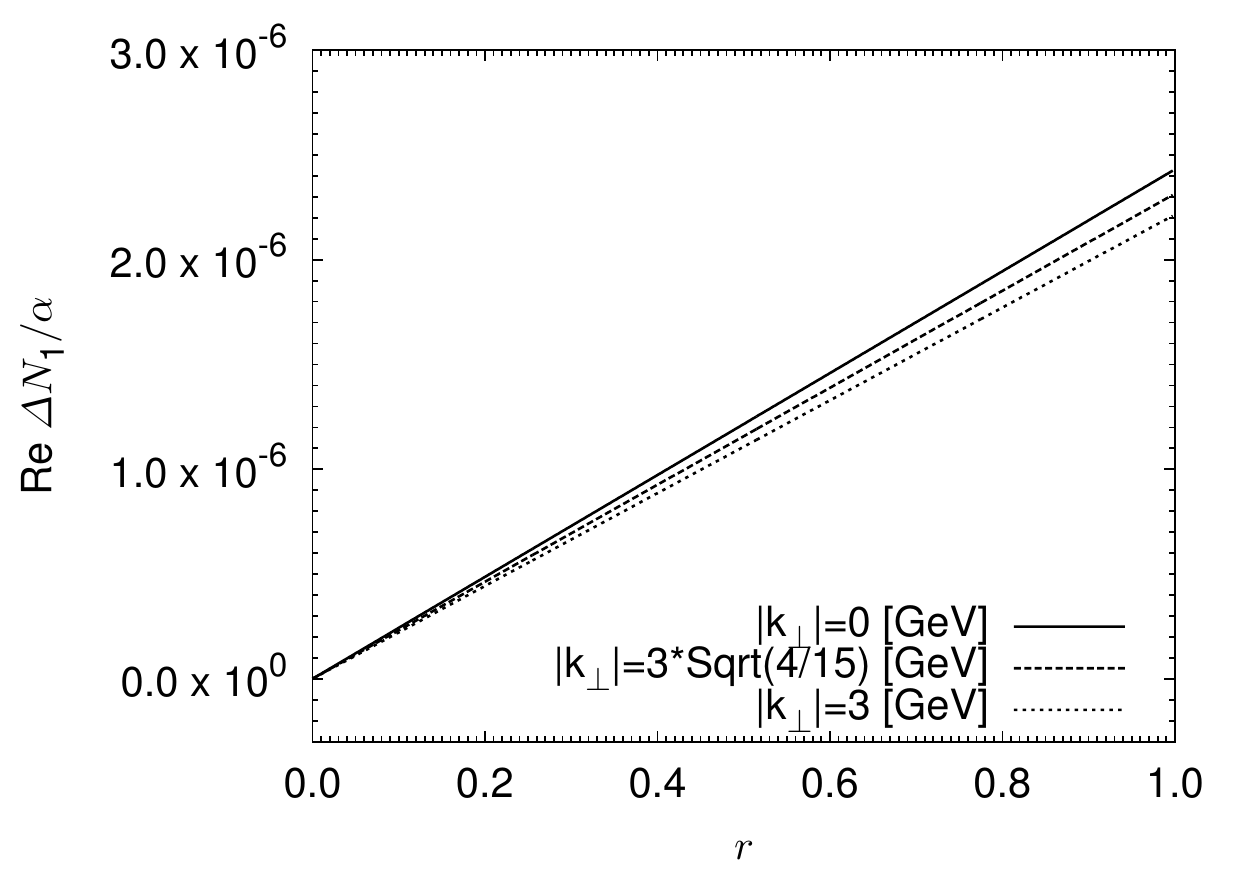}
\includegraphics[width=\figscale]{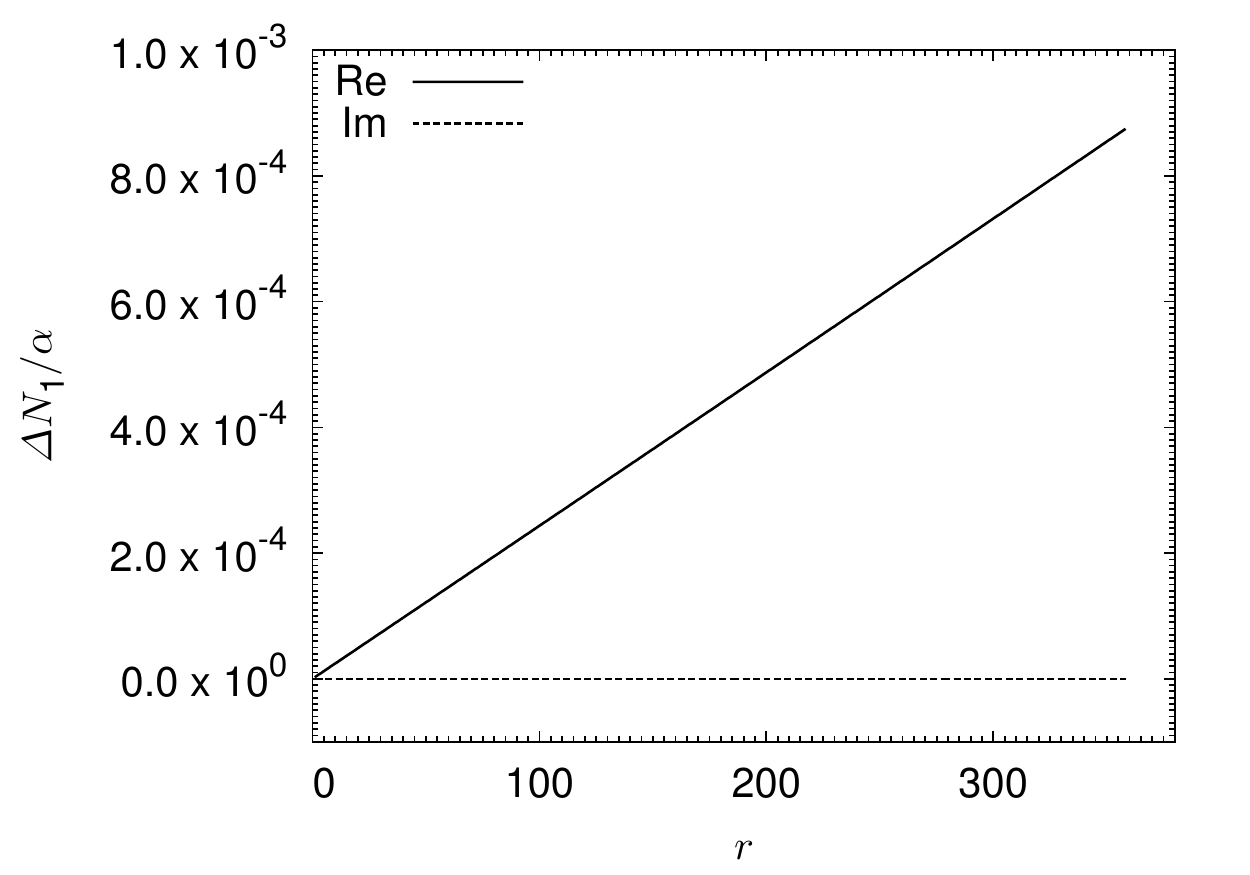}
}
\centerline{
\includegraphics[width=\figscale]{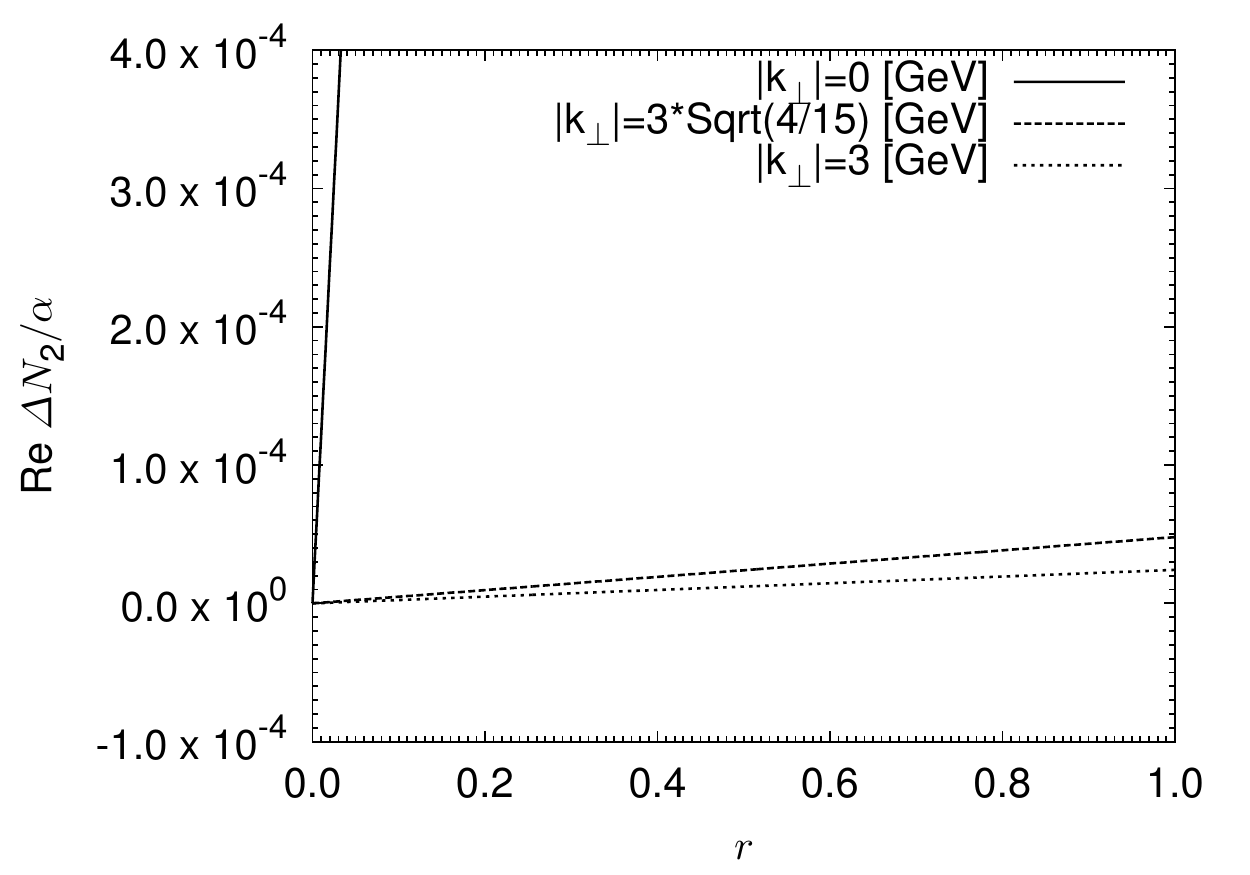}
\includegraphics[width=\figscale]{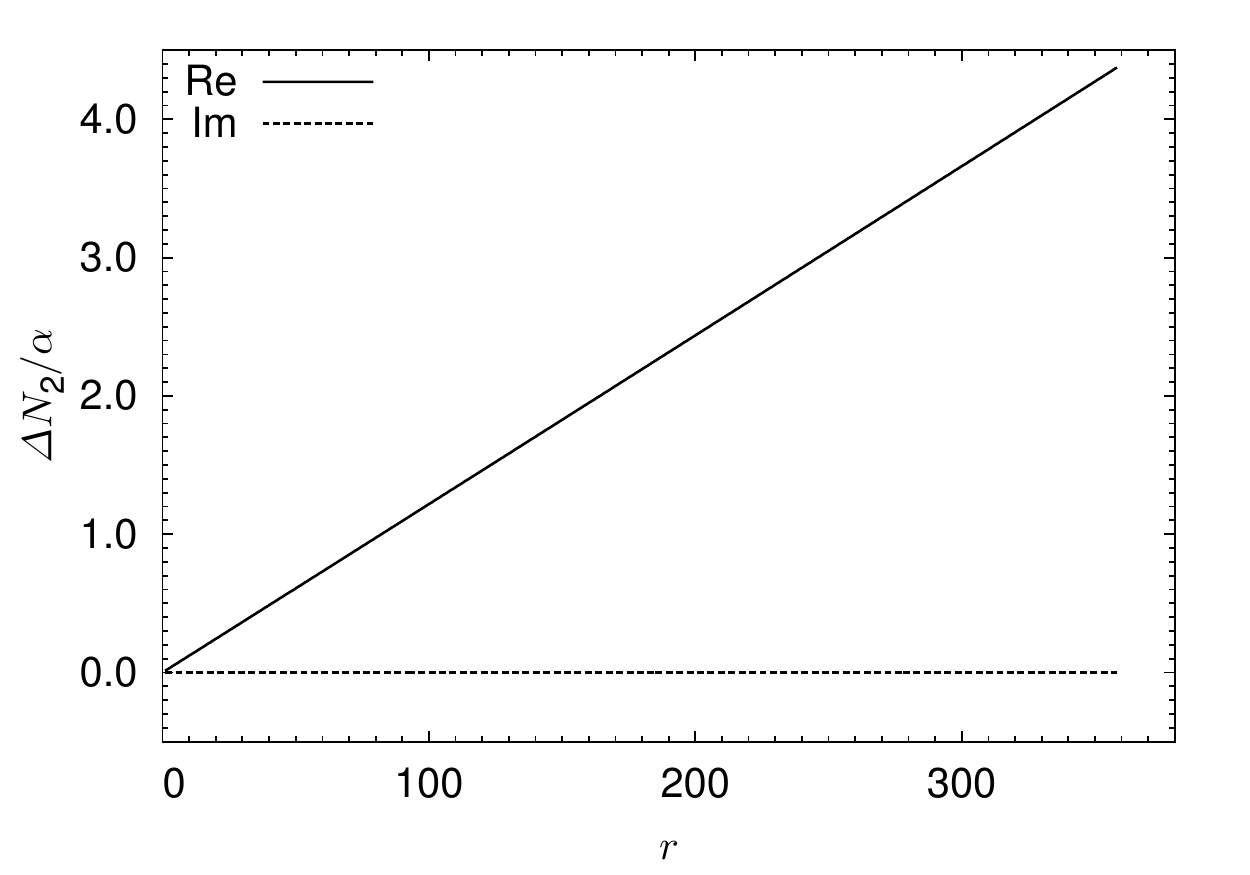}
}
\vspace*{-4pt}
\caption{Comparison (A) (left panels) and (B) (right panels) for 
muons (case [a-1]) with $\ell_{\mathrm{max}}=1000$}
\label{fig:compareAandB}
\end{figure}

The left panels in Fig.~\ref{fig:compareAandB} 
show  the discrepancy between the Landau-level form and the double-integral form 
as the consistency check (A) in $r<1$ for muons with $eB=10m_{\pi}^2$ and $\ell_{\mathrm{max}}=1000$ (case [a-1]).
The discrepancies of $\Delta N_0$ and $\Delta N_1$ is at $O(10^{-6})$ and decreases with increasing $q$.
This is practically satisfactory level.
While for $\Delta N_2$, the discrepancy with $q=0$ (solid line) has an $O(1)$ error and
it rapidly decreases to $O(10^{-4})$ with increasing $q$.
The truncation errors depend on $r$ linearly, which is consistent with our asymptotic analysis.
For electrons with $eB=10m_{\pi}^2$ (case [a-2]) we observe the same behavior in $r<1$
except for $N_1$.
Since the relative truncation error $|(\Delta N_1)/N_1|$ for electrons reaches 
the limit of double precision accuracy, we cannot extract the proper $r$ 
dependence for $\Delta N_1$ (case [a-2]).

\begin{figure}[t]
\centerline{
\includegraphics[width=\figscale]{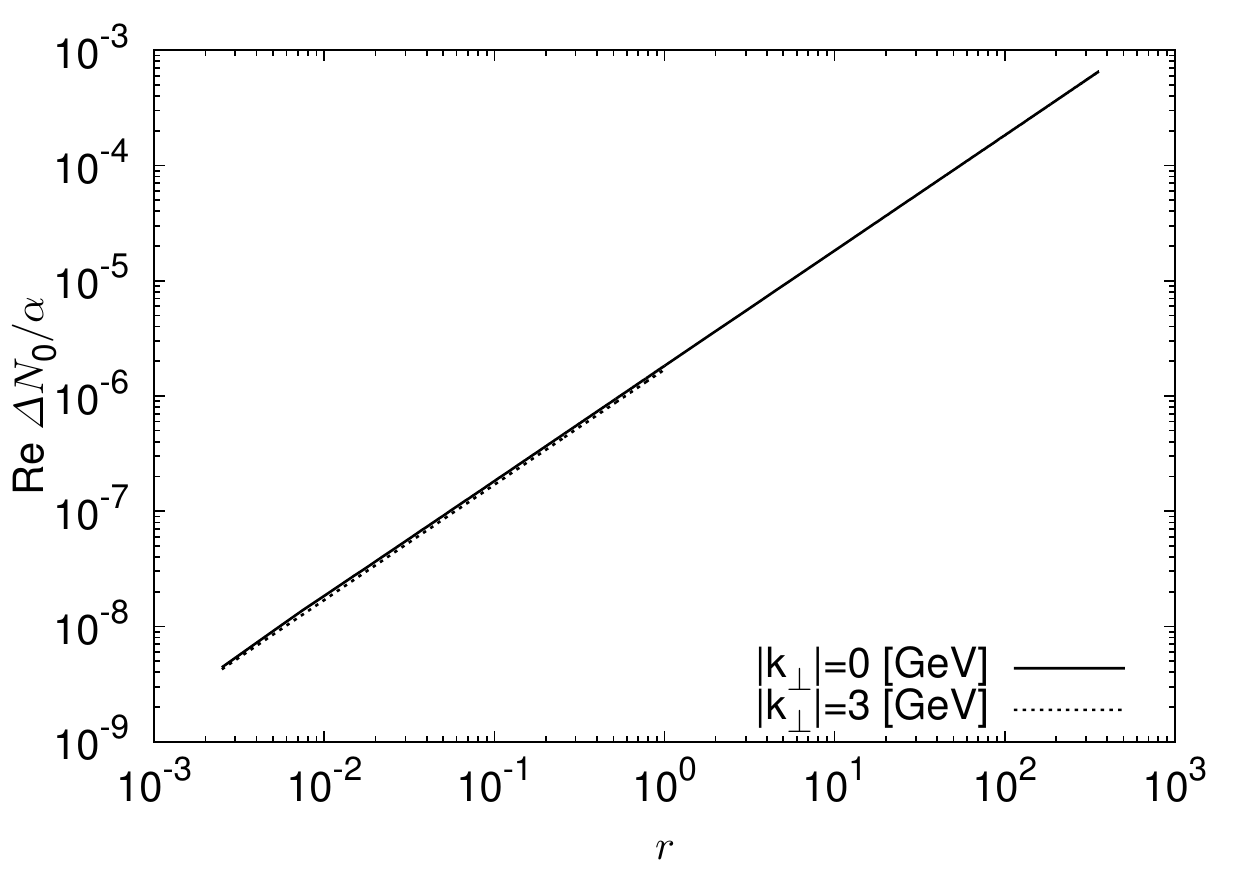}
\includegraphics[width=\figscale]{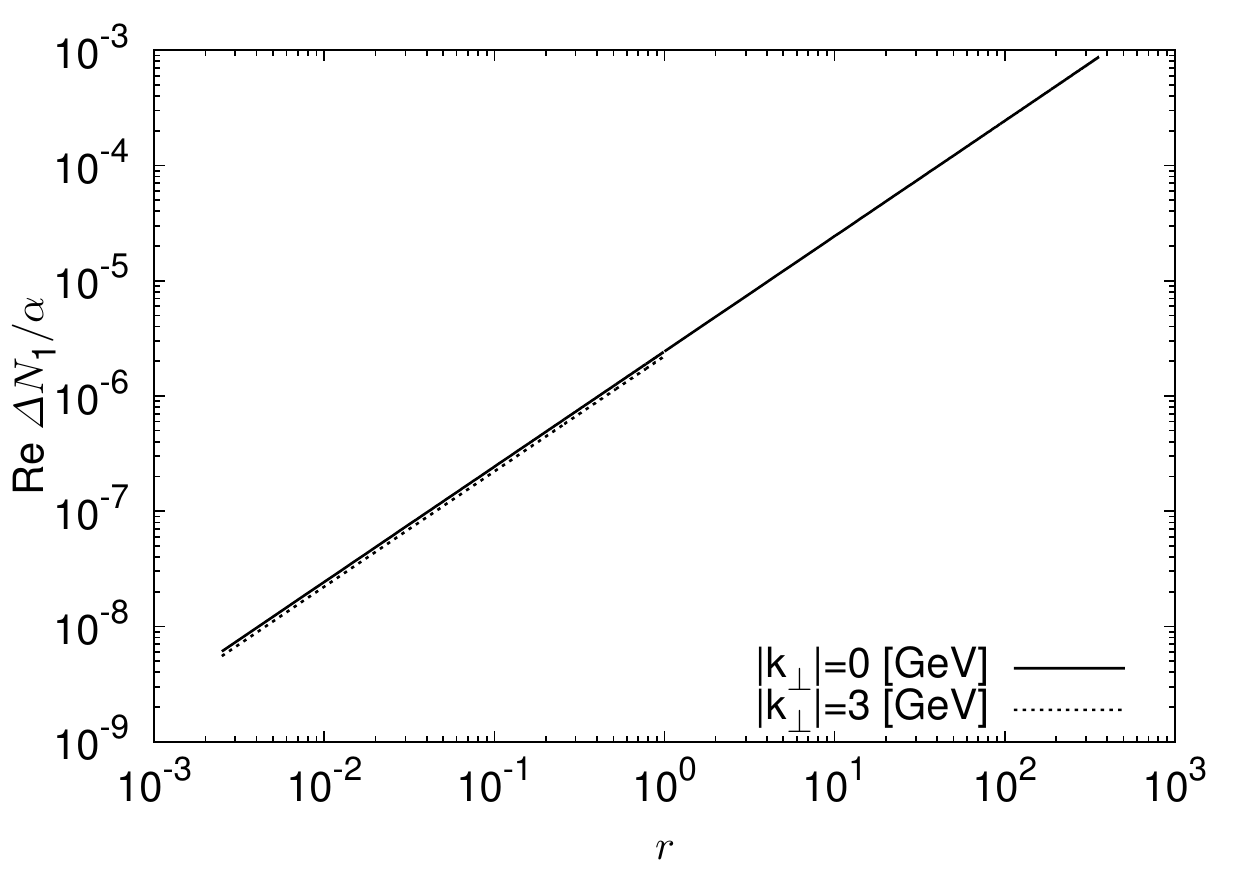}
}
\centerline{
\includegraphics[width=\figscale]{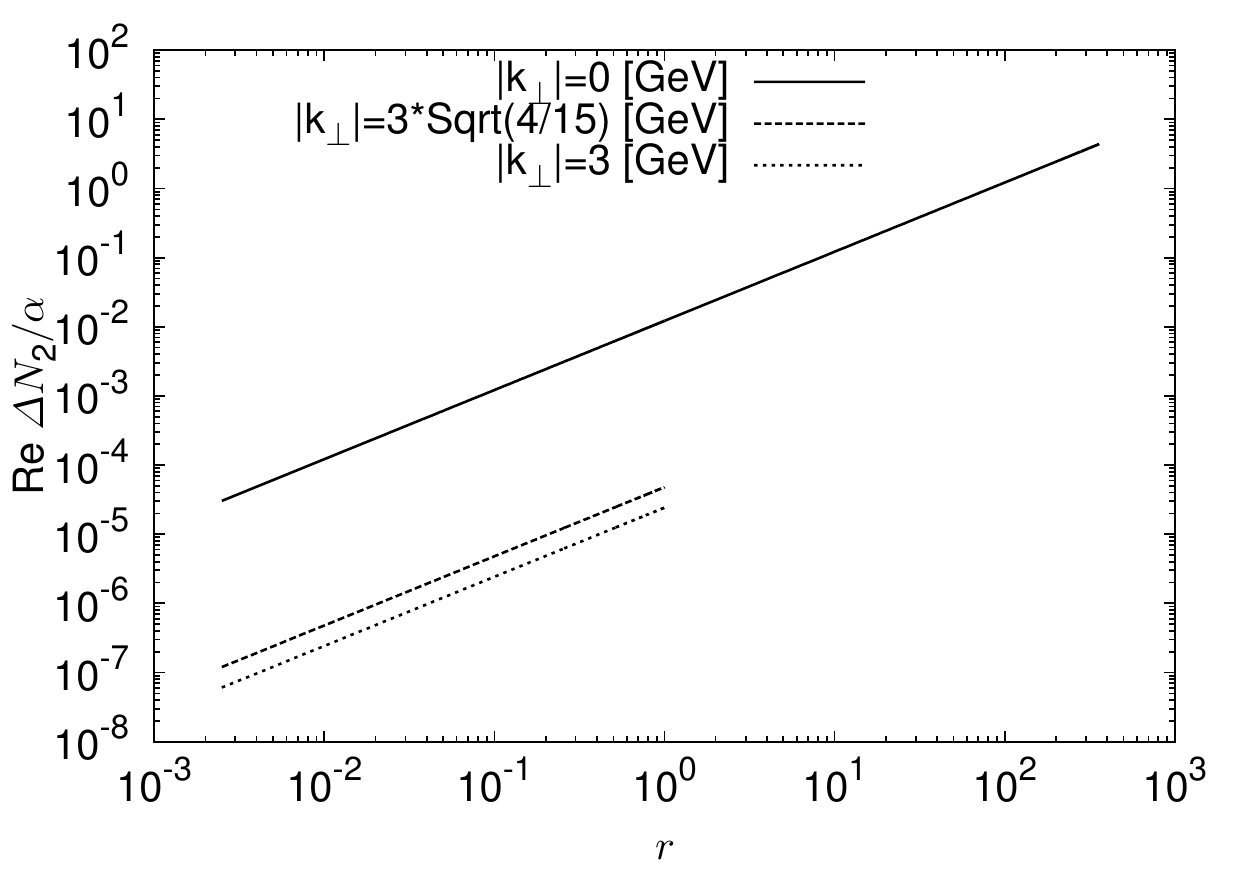}
}
\vspace*{-4pt}
\caption{Figs.~\ref{fig:compareAandB} are combined in log-log plots
(case [a-1] with $\ell_{\mathrm{max}}=1000$).}
\label{fig:compareC}
\end{figure}

The right panels in Fig.~\ref{fig:compareAandB} show the discrepancy between 
the Landau-level form and the DiGamma form in $1<r$ with $q=0$ as the consistency check (B)
for muons with $eB=10m_{\pi}^2$ and $\ell_{\mathrm{max}}=1000$ (case [a-1]). 
The imaginary parts perfectly coincide with each other for all form factors.
The real parts for $N_0$ and $N_1$ are linearly continued from 
the left panels and still remain below  $O(10^{-4})$ in the region we investigated. 
For $\Delta N_2$, however, it reaches $O(1)$.
If we extend the observation in the region with $r<1$ to $1<r$, 
we expect even with $q>0$ that $\Delta N_0$ and $\Delta N_1$ still remain at $O(10^{-4})$
and $\Delta N_2$ with $|\bm{k}_{\perp}| \gtrsim 3\sqrt{4/15}\sim 1.5$ [GeV] remains at $O(10^{-2})$ (see Fig.~\ref{fig:compareC}).


\begin{figure}[t]
\centerline{
\includegraphics[width=\figscale]{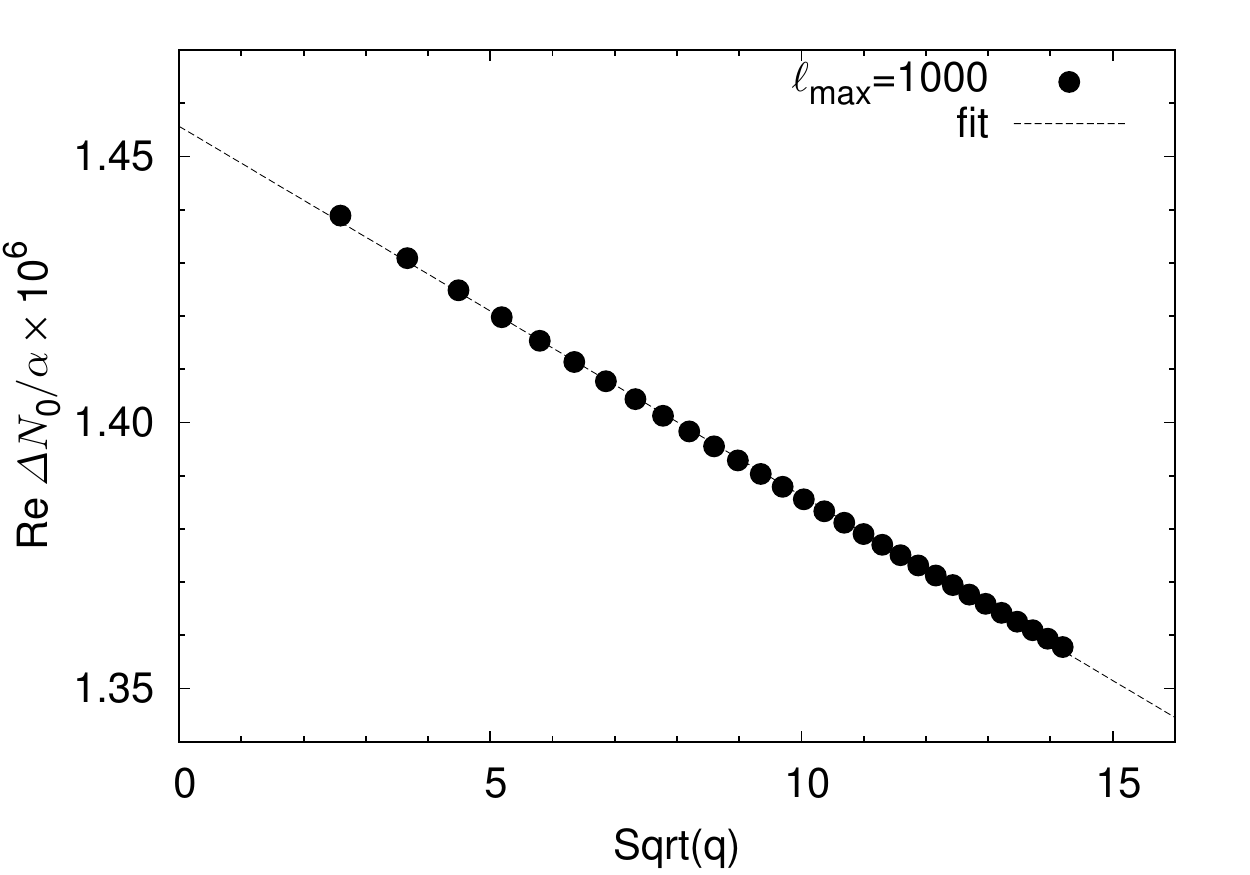}
\includegraphics[width=\figscale]{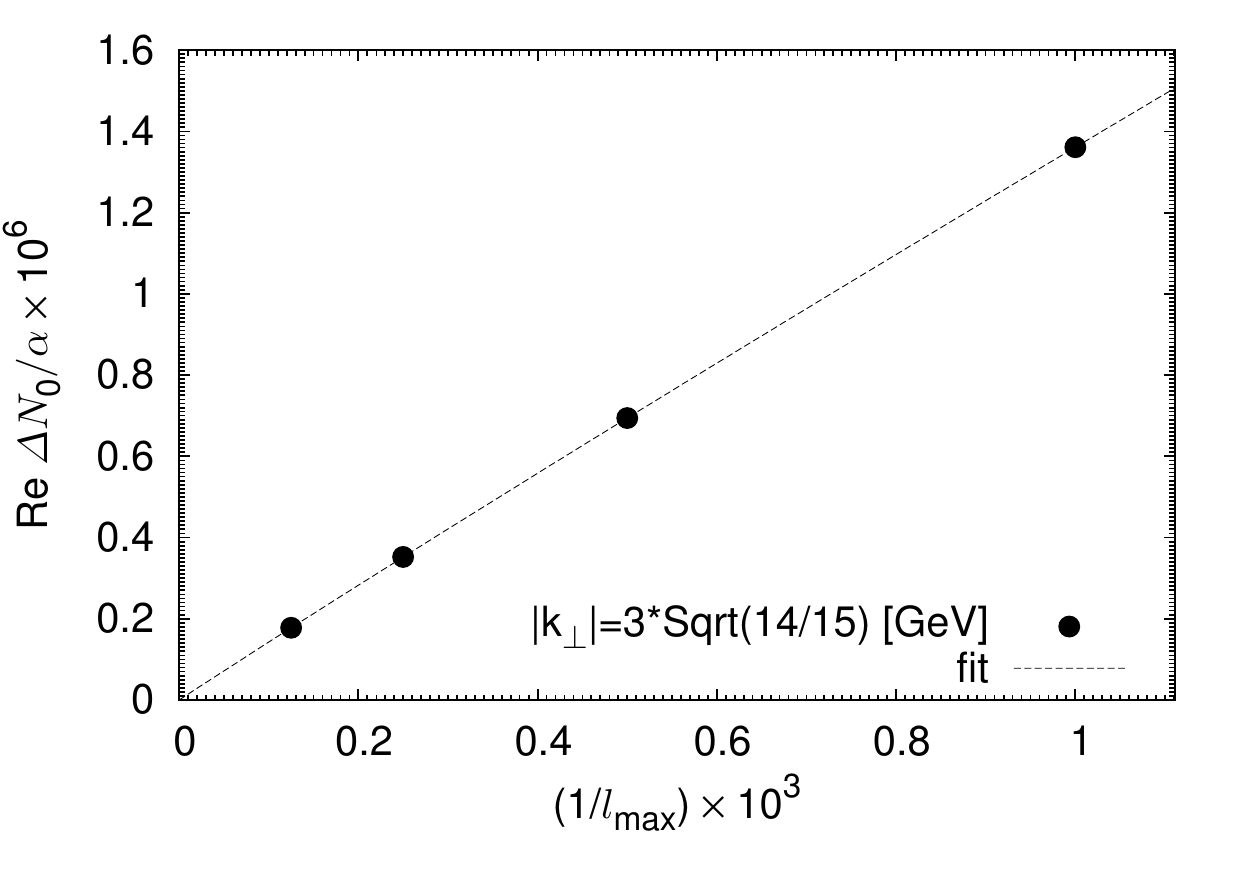}
}
\centerline{
\includegraphics[width=\figscale]{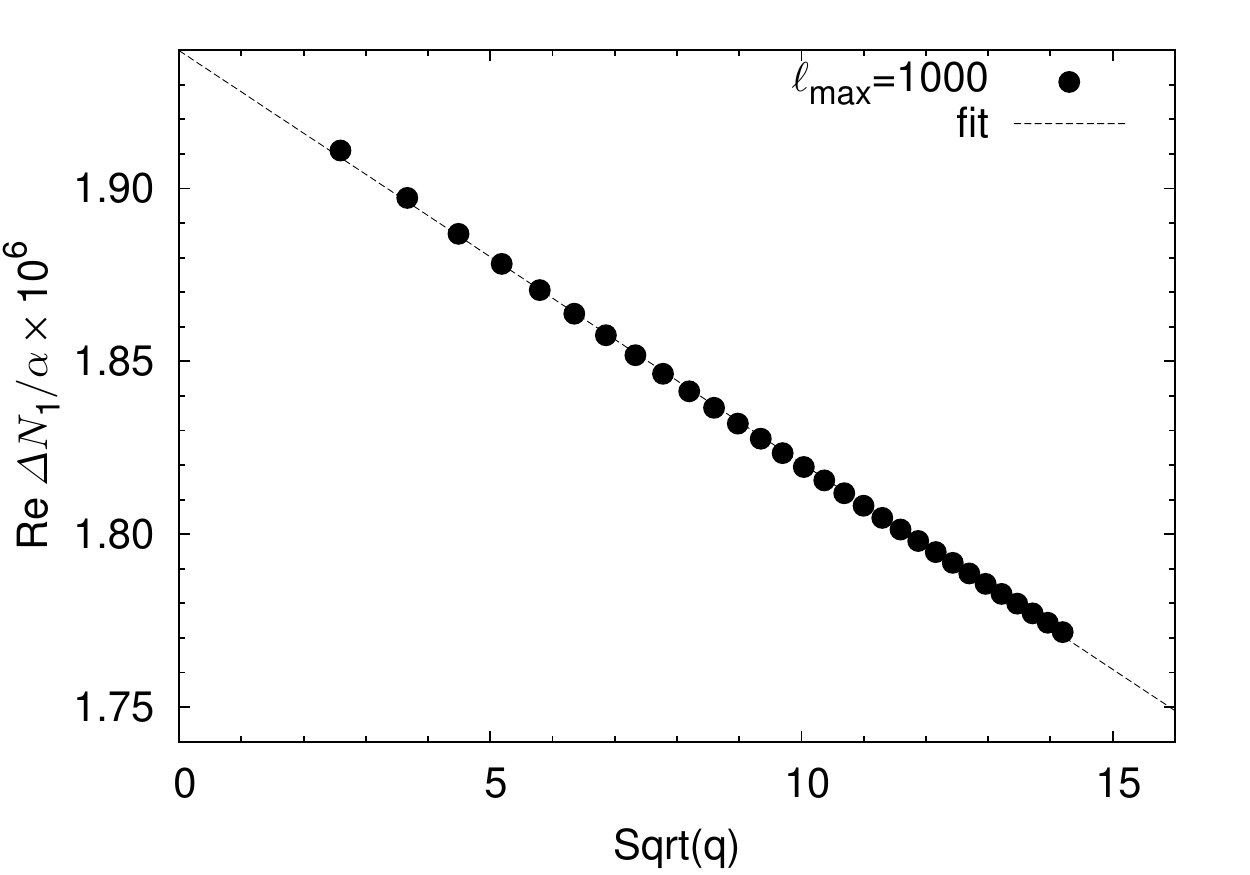}
\includegraphics[width=\figscale]{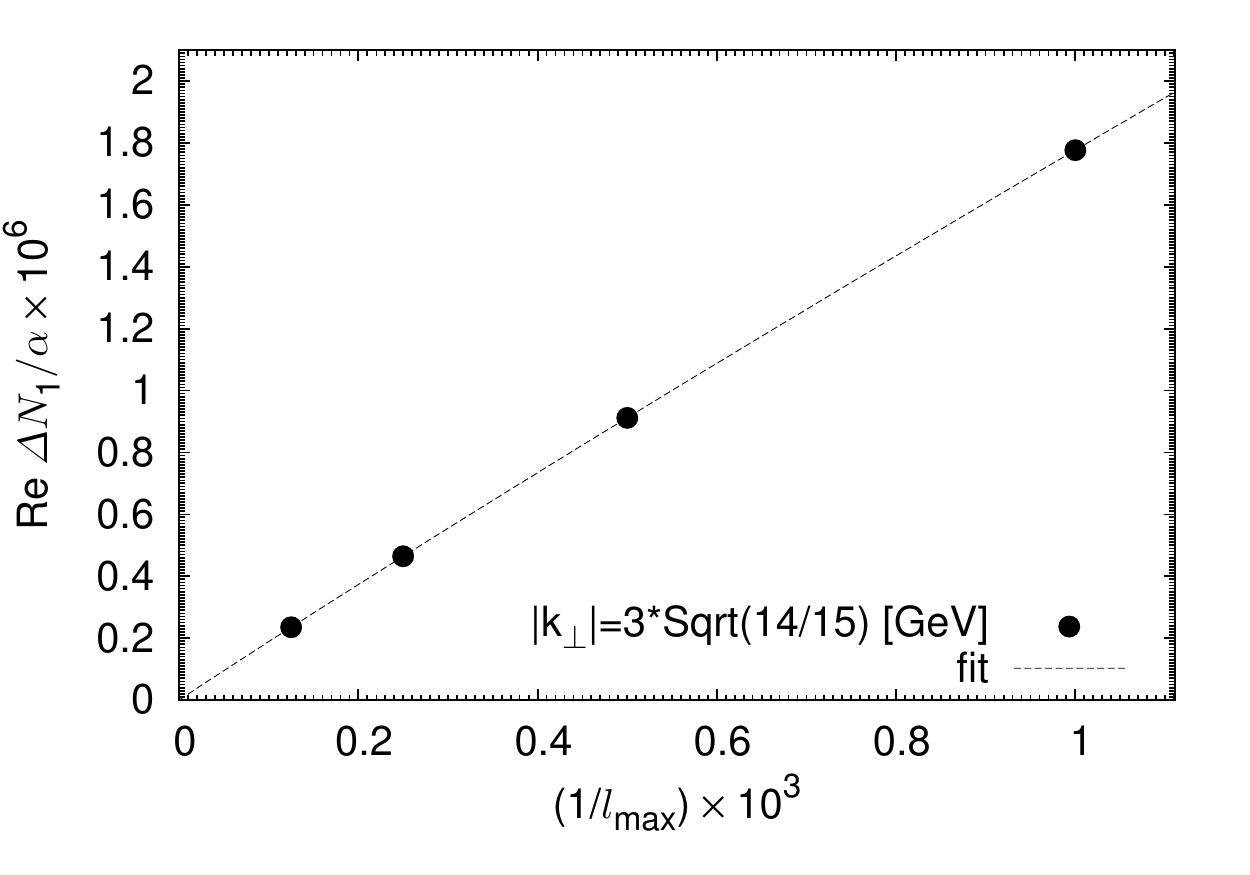}
}
\centerline{
\includegraphics[width=\figscale]{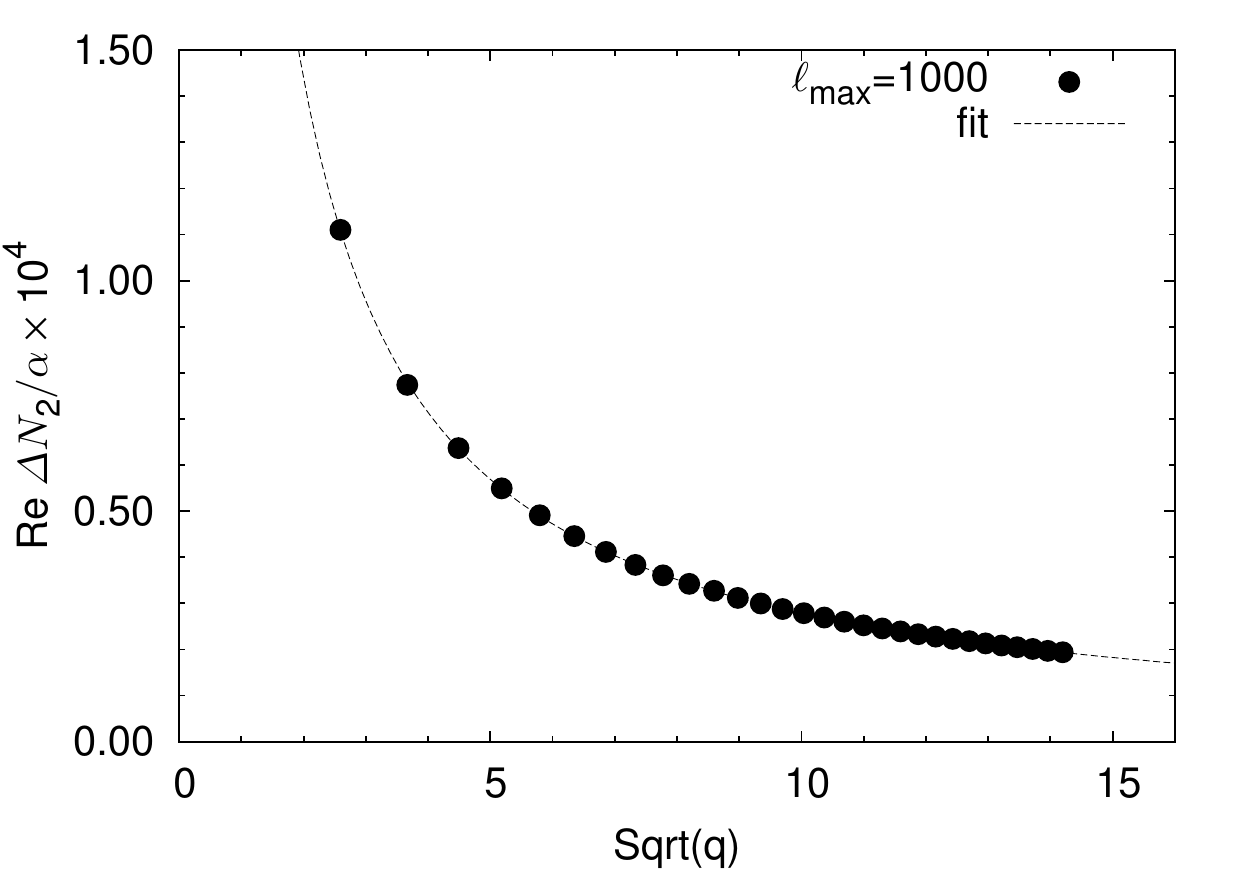}
\includegraphics[width=\figscale]{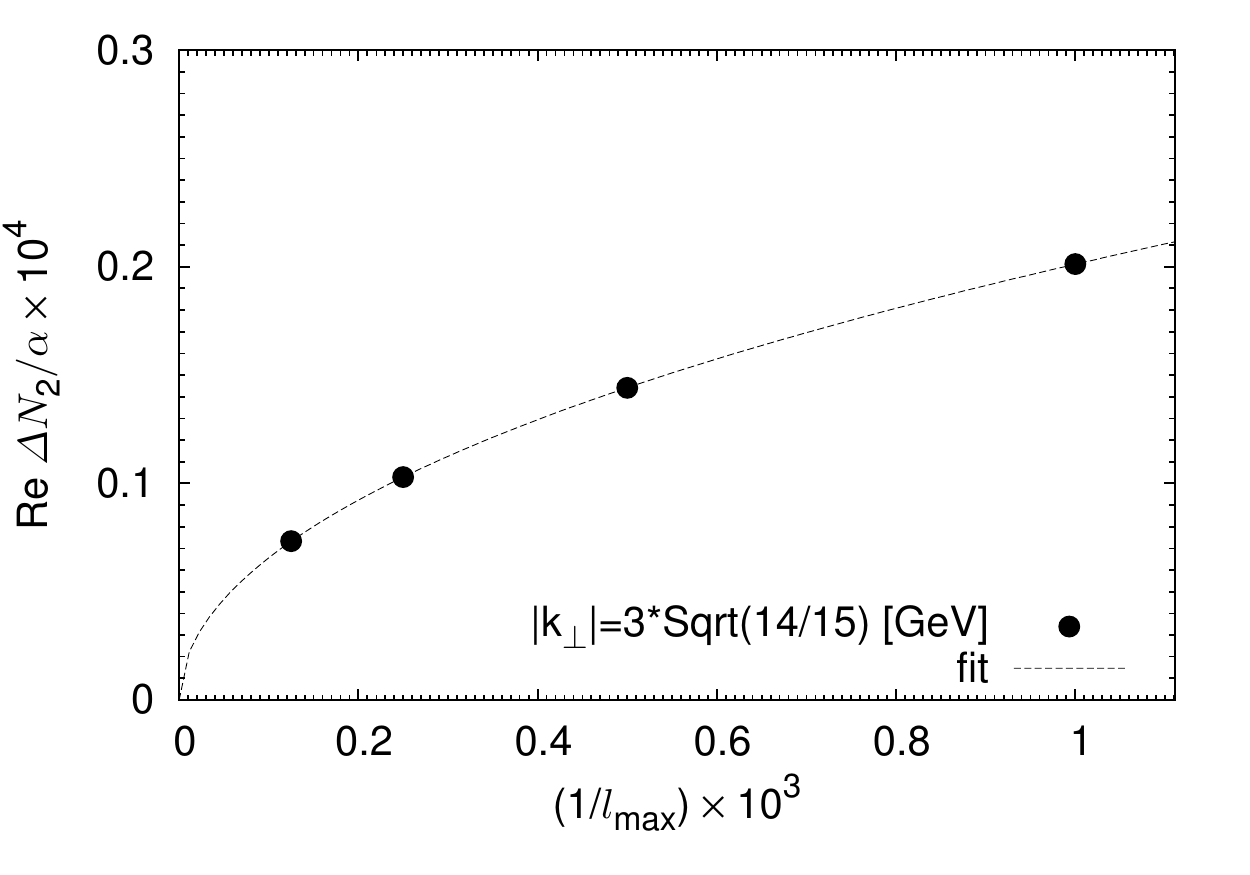}
}
\vspace*{-4pt}
\caption{$q$ dependence (left panels) and 
         $\ell_{\mathrm{max}}$ (right panels) dependence 
         of truncation errors at $r=0.8$ (case [a-1]).}
\label{fig:qandelldeperror}
\end{figure}

The $q$ dependence of the truncation error at $r=0.8$ with 
$\ell_{\mathrm{max}}=1000$ is shown in the left panels of
Fig.~\ref{fig:qandelldeperror} for the case [a-1].
$\Delta N_0$ and $\Delta N_1$ behave as a linear function of $\sqrt{q}$ while
$\Delta N_2$ behaves as a linear function of $1/\sqrt{q}$
as shown by the fit lines in the figures.
The $\ell_{\mathrm{max}}$ dependence of the truncation errors at 
$r=0.8$ and $|\bm{k}_{\perp}|=3\sqrt{14/15}$ [GeV] is shown in 
the right panels of Fig.~\ref{fig:qandelldeperror} for the case [a-1].
The truncation error for $N_0$ and $N_1$ can be fitted with
$(c+d/\sqrt{\ell_{\mathrm{max}}})/\ell_{\mathrm{max}}$.
$\Delta N_2$ can be fitted with $c/\sqrt{\ell_{\mathrm{max}}}+d/\ell_{\mathrm{max}}$.
The same behavior is observed for other $q$ in $r < 1$.
This behavior cannot be understood from the asymptotic behavior on $\ell$ at a fixed $n$
because it involves the truncation effect on the $n$ summation.

With the global analysis for all cases shown in Table~\ref{tab:params},
we find that the truncation error can be well expressed by
\begin{eqnarray}
    \Delta N_{j}/\alpha &=& \left(c_{j}+d_{j}\sqrt{\frac{q}{\ell_{\mathrm{max}}}}\right)
                     \frac{r}{\ell_{\mathrm{max}}}
    \quad\quad \quad\quad (\mbox{$j=0$ and $1$}),
\label{eq:TruncFitN0N1} \\
    \Delta N_{2}/\alpha &=& \left(c_2 + d_2 \sqrt{\frac{\ell_{\mathrm{max}}}{q}} \right)
                     \frac{r}{\ell_{\mathrm{max}}},
\label{eq:TruncFitN2}
\end{eqnarray}
with $r<1$ derived from check (A), and by
\begin{equation}
    \Delta N_{j}/\alpha = e_{j}\frac{r}{\ell_{\mathrm{max}}}
    \quad\quad \quad\quad (\mbox{$j=0$ and $1$}),
\label{eq:TruncAboveFitN0N1}
\end{equation}
with $1<r$ derived from check (B).

Table~\ref{tab:TruncErrorFit} shows the coefficients for
Eqs.~(\ref{eq:TruncFitN0N1})-(\ref{eq:TruncAboveFitN0N1})
obtained by fitting all data from the parameter sets shown in Table~\ref{tab:params}.
For electrons in strong magnetic fields of $eB=10m_{\pi}^2$ and $m_{\pi}^2$,
we cannot determine $c_1$ and $d_1$ properly by fitting
because $|\Delta N_1/N_1|$ in the region $r<1$ reaches on the double precision limit at $10^{-14}$--$10^{-15}$.
We note that $\Delta N_0$ and $\Delta N_1$ with $q=0$ give an upper bound for 
the truncation errors since the coefficients $d_0$ and $d_1$ are negative.
As seen from Table~\ref{tab:TruncErrorFit}, $c_0$ and $e_0$ ($c_1$ and $e_1$) are consistent
except for the cases at $eB=m_{\pi}^2/10$ (cases [c-1] and [c-2]).
This is because the truncation error deviates from the function form 
Eq.~(\ref{eq:TruncAboveFitN0N1}).
Thus the truncation errors for $N_0$ and $N_1$ can be directly estimated from 
the comparison (B) in the region of $1<r$ for sufficiently strong fields.
We extend Eq.~(\ref{eq:TruncFitN2}) determined by fitting in the region of $r<1$ 
to estimate the truncation error $\Delta N_2$ in the region of $1<r$.
This gives an upper bound for $\Delta N_2$ because it monotonically decreases
with increasing $q$.

A practical algorithm to compute the form factors in a strong magnetic field 
comparable to or stronger than the critical field is summarized as follows;
\begin{enumerate}
\item Use double integral forms for all $N_j$ in the region of $r<1$.
\item Use Landau level summation forms for $N_0$ and $N_1$ in the region of $1<r$
      with the truncation control by Eq.~(\ref{eq:TruncAboveFitN0N1}).
\item Use Landau level summation forms for $N_2$ in the region of $1<r$ 
      with the truncation control by Eq.~(\ref{eq:TruncFitN2}).
\end{enumerate}
For muons in strong magnetic fields of $eB= m_{\pi}^2$--$10 m_{\pi}^2$, 
summation up to $\ell_{\mathrm{max}}\simeq$ 10000--20000 yields a $\sim 10^{-4}$ accuracy 
for $N_0$ and $N_1$, and a $\sim 10^{-2}$ accuracy for $N_2$
in the kinematic region with
$1$ [GeV] $< |\bm{k}_{\perp}|$ and $0 < k_{\parallel}^2 < 4^2$ [GeV$^{2}$].

For the form factors, especially for $N_2$, in weaker magnetic fields 
or with more precise values, 
it becomes difficult to obtain the accurate form factors with our naive summation method. 
We might need to apply series acceleration techniques.

\begin{table}[t]
\tbl{Fit results for Eqs.~(\ref{eq:TruncFitN0N1})-(\ref{eq:TruncAboveFitN0N1}).}
{
\begin{tabular}{cccc} 
\toprule
 Case & [a-1] & [b-1] & [c-1] \\ \colrule
$c_0$ & 
  $1.820 \times 10^{-3}$ &  $1.794 \times 10^{-2}$ &  $1.636 \times 10^{-1}$ \\
 
$d_0$ &
 $-2.748 \times 10^{-4}$ & $-7.002 \times 10^{-3}$ & $-1.174 \times 10^{-1}$ \\
                            
$e_0$ &
  $1.828 \times 10^{-3}$ &  $1.877 \times 10^{-2}$ &  $3.132 \times 10^{-1}$ \\
   
$c_1$ &
  $2.425 \times 10^{-3}$ &  $2.380 \times 10^{-2}$ &  $2.116 \times 10^{-1}$ \\

$d_1$ &
 $-4.720 \times 10^{-4}$ & $-1.202 \times 10^{-2}$ & $-2.048 \times 10^{-1}$ \\
                            
$e_1$ &
  $2.439 \times 10^{-3}$ &  $2.513 \times 10^{-2}$ &  $2.652 \times 10^{-1}$ \\
                             
$c_2$ &
 $-1.324 \times 10^{-3}$ & $-1.231 \times 10^{-2}$ & $-1.203 \times 10^{-1}$ \\
                            
$d_2$ &
  $1.145 \times 10^{-2}$ &  $3.609 \times 10^{-2}$ &  $1.137 \times 10^{-1}$ \\\botrule

\toprule
 Case & [a-2] & [b-2] & [c-2] \\ \colrule
$c_0$ & 
  $4.256 \times 10^{-8}$ &  $4.196 \times 10^{-7}$ &  $3.830 \times 10^{-6}$ \\
                              
$d_0$ &
 $-3.105 \times 10^{-11}$& $-7.921 \times 10^{-10}$& $-1.329 \times 10^{-8}$ \\
                            
$e_0$ &
  $4.275 \times 10^{-8}$ &  $4.391 \times 10^{-7}$ &  $7.360 \times 10^{-6}$ \\
                             
$c_1$ &
      -                  &      -                  &  $4.959 \times 10^{-6}$ \\
$d_1$ &
      -                  &      -                  & $-2.324 \times 10^{-8}$ \\
$e_1$ &
  $5.705 \times 10^{-8}$ &  $5.880 \times 10^{-7}$ &  $1.076 \times 10^{-5}$ \\

$c_2$ &
 $-3.096 \times 10^{-8}$ & $-2.879 \times 10^{-7}$ & $-2.820 \times 10^{-6}$ \\

$d_2$ &
  $5.539 \times 10^{-5}$ &  $1.746 \times 10^{-4}$ &  $5.505 \times 10^{-4}$ \\\botrule

\end{tabular}
\label{tab:TruncErrorFit}
}
\end{table}

\section{Summary}
\label{sec:Summary}
We investigated the vacuum polarization tensor in constant background magnetic fields 
based on the Hattori-Itakura's Landau-level summation formula with an appropriate UV subtraction 
method we constructed. 
We could reproduce the numerical values computed with the Landau-level summation form 
consistent with those with known formulae.
The Landau-level summation was truncated and we estimated the truncation error in a range of 
the parameter sets for muons and electrons.
In very strong magnetic fields of $eB = m_{\pi}^2$--$10 m_{\pi}^2$, we could evaluate the form factors
with a practically acceptable accuracy in the limited kinematic region with 
$1$ [GeV] $< |\bm{k}_{\perp}|$ and $0 < k_{\parallel}^2 < 4^2$ [GeV$^2$] for muons and electrons.
This kinematic region is accessible provided by a small invariant mass
in the heavy ion collision experiments at RHIC\cite{Adare:2009qk} and LHC\cite{Alessandro:2006yt}
where such a strong magnetic field exists in the early stage of the heavy ion collisions.
The propagation of a real or a virtual photon emitted in the early stage of the collisions 
could receive a large asymmetry due to the direction dependent polarization tensor originating from 
the pair creation phase space suppression due to the Landau-level bound states.
Hadronic contributions to the vacuum polarization tensor must be incorporated
before phenomenologically applying the propagator to investigate the effect
of strong magnetic fields.
However we expect that the polarization tensor estimated in this paper is partly applicable 
to prove the existence of strong magnetic fields via the photon propagation in the heavy ion 
collisions at LHC experiments. 

\section*{Acknowledgments}
We thank Koich Hattori and Kazunori Itakura for valuable discussions.
The numerical computations have been done with the PC cluster
at INSAM (Institute for Nonlinear Science and Applied Mathematics) Hiroshima University.
This work is supported in part by JSPS KAKENHI Grant Number 23654091 (Grant-in-Aid for Challenging Exploratory Research).

\appendix
\section{Integrals for Eqs.~(\ref{eq:FuncF})-(\ref{eq:FuncH})}
\label{sec:apdxA}
We follow the notations given by Ref.~\citen{Hattori:2012je} except for
the dimensionless parameters $r$ and $\mu$ 
(correspondence to Ref.~\citen{Hattori:2012je} is 
$r \leftrightarrow r_{\parallel}^2$, $\mu \leftrightarrow B_r$).
The analytic expression for Eq.~(\ref{eq:FuncF}) is
\begin{eqnarray}
    F^n_{\ell}(r,\mu)&=&
\left\{
    \begin{array}{lcr}
\displaystyle     
\frac{1}{\sqrt{\cal D}} 
\ln\left|\frac{a-c-\sqrt{{\cal D}}}{a-c+\sqrt{{\cal D}}}\right|
&\quad& \mbox{($r < s_{-}^{\ell n}$)}\\
\displaystyle     
\frac{2}{\sqrt{|{\cal D}|}}
 \left[
   \arctan\left(\frac{b+2a}{\sqrt{|{\cal D}|}}\right)
  -\arctan\left(\frac{b-2a}{\sqrt{|{\cal D}|}}\right)
 \right]
&& \mbox{($s_{-}^{\ell n}< r < s_{+}^{\ell n}$)}\\
\displaystyle     
\frac{1}{\sqrt{{\cal D}}}
\left[
\ln\left|\frac{a-c-\sqrt{{\cal D}}}{a-c+\sqrt{{\cal D}}}\right| + 2\pi i
\right]
&& \mbox{($s_{+}^{\ell n}< r$)}\\
    \end{array}
\right., 
\nonumber\\
&&
\label{eq:AnlFuncF}
\end{eqnarray}
\begin{equation}
s^{\ell n}_{\pm} \equiv
\frac{1}{4} \left( \sqrt{1+2\ell \mu} \pm \sqrt{1+2(\ell + n)\mu} \right)^2,
\end{equation}
where $a \equiv r$, $b \equiv -n\mu$, 
$c \equiv 1-r+(2\ell+n)\mu$, and ${\cal D} \equiv b^2-4ac$.

Eqs.~(\ref{eq:FuncG}) and (\ref{eq:FuncH}) are given by
\begin{eqnarray}
  G^n_{\ell}(r,\mu)&=& \frac{1}{2r} \left[ \Xi^n_{\ell}(\mu) + n\mu F^n_{\ell}(r,\mu) \right],\\
  H^n_{\ell}(r,\mu)&=& \frac{1}{r} 
\left[ 2 +\frac{n\mu}{2r}\Xi^n_{\ell}(\mu) + \frac{b^2-2ac}{2a}F^n_{\ell}(r,\mu) \right],\\
\Xi^n_{\ell}(\mu)&\equiv & \ln\left|\frac{1+2\ell \mu}{1+2(\ell+n)\mu}\right|.
\end{eqnarray}

When evaluating these functions numerically, the naive implementation causes a loss of significant figures near $r=0$.
We use 8th order Taylor expansion forms when $|r/(n\mu)| < 10^{-3}$ for $n>0$ 
and $|r/(1+2\ell \mu)| < 10^{-3}$ for $n=0$.

\section{Form factors with $q=0$}
\label{sec:apdxB}

When $q=0$ case we can integrate $z$ analytically for Eq.~(\ref{eq:Nmaster})
using the residue theorem and the reflection formula of the DiGamma function.
The expression for $N_1$ has been obtained in Ref.~\citen{Karbstein:2011ja}.
We give similar expressions for $N_0$ and $N_2$ in order to compare 
the numerical values with the Landau-level summation formula with
$q=0$ as a consistency check.

After integrating for $z$ in $N_0$ we obtain
\begin{eqnarray}
    N_0 &=& -\frac{\alpha}{4\pi}
\left\{
\int_{-1}^{1}dv 
\frac{1}{2}\Biggl[ -2 v^2 - 2(1-v^2)\ln\left(2\mu\right)
\right.
\nonumber\\
&&
\hspace*{6em}
-\left(1+v \left(\frac{S^1_{0}(v;r,\mu)}{\mu}-1\right)\right)\Psi\left(\frac{S^1_{K+1}(v;r,\mu)}{2\mu}\right)
\nonumber\\
&&
\hspace*{6em}
-\left(1-v \left(\frac{S^1_{0}(-v;r,\mu)}{\mu}-1\right)\right)\Psi\left(\frac{S^1_{K+1}(-v;r,\mu)}{2\mu}\right)\Biggr]
\nonumber\\
&&
 + 2 \sum_{k\ge 0}^{K}\biggl[
  \frac{1}{a}\left\{ -2b+(c-a)b F^1_k(r,\mu) \right.
\nonumber\\
&&\hspace{6em}
\left. +(b^2-a^2-ac+a)G^1_{k}(r,\mu)\right\} + b H^1_{k}(r,\mu)\biggr]\Biggr\},
\label{eq:N0q0}
\end{eqnarray}
where $a\equiv r$, $b\equiv -\mu$, and $c \equiv 1-r+(2k+1)\mu$. The shift integer $K$ is given by
\begin{equation}
K =
\left\{
    \begin{array}{lcr}
        -\mathrm{Ceiling}[A^1_{0}] &\quad\quad\quad& \mbox{($|\mu/(2r)|<1$ and $A^1_{0} \le 0$)}\\
        -1                   && \mbox{(otherwise)}
    \end{array}\right.,
\end{equation}
where $S^n_\ell $ and $A^n_\ell$ are given by Eq.~(\ref{eq:SPropLN}) and Eq.~(\ref{eq:Amin}) 
respectively.

Similarly we have
\begin{eqnarray}
    N_1 &=& -\frac{\alpha}{4\pi}
\left\{
  \int_{-1}^{1}dv (1-v^2)
  \left[ -\ln\left(2\mu\right)
    -\Psi\left(\frac{S^0_{K+1}(v;r,\mu)}{2\mu}\right)
  \right] 
\right.
\nonumber\\
&&\hspace{2em}\left. 
-\mu\left(F^0_{0}(r,\mu)-H^0_{0}(r,\mu)\right)
+\sum_{k\ge 0}^{K} 2 \mu 
               \left(F^0_{k}(r,\mu)-H^0_{k}(r,\mu)\right)\right\},
\label{eq:N1q0}
\\
K&=&
\left\{
    \begin{array}{lcr}
        -\mathrm{Ceiling}[A^0_{0}] &\quad\quad\quad& \mbox{($A^0_{0} \le 0$)}\\
        -1                   && \mbox{(otherwise)}
    \end{array}\right.,
\end{eqnarray}
\begin{eqnarray}
    N_2 &=& -\frac{\alpha}{4\pi}
\left\{
  \int_{-1}^{1}dv 
  \frac{1}{2}\left[ -1-3v^2-2(1-v^2)\ln\left(2\mu\right)+2\frac{S^0_0(r,\mu)}{\mu}
\right.\right.
\nonumber\\
&&\hspace{7em}
   -2\left(\frac{S^0_{0}(v;r,\mu)}{\mu}\right)^2 \Psi\left(\frac{S^0_{1+J+1}(v;r,\mu)}{2\mu}\right)
\nonumber\\
&&\hspace{7em}
   +\frac{S^1_{0}( v;r,\mu)}{\mu} \left(\frac{S^1_{0}( v;r,\mu)}{\mu}-2\right) 
    \Psi\left(\frac{S^1_{K+1}(v;r,\mu)}{2\mu}\right)
\nonumber\\
&&\hspace{7em}\left.
   +\frac{S^1_{0}(-v;r,\mu)}{\mu} \left(\frac{S^1_{0}(-v;r,\mu)}{\mu}-2\right) 
\Psi\left(\frac{S^1_{K+1}(-v;r,\mu)}{2\mu}\right)\right]
\nonumber\\
&&\hspace{2em}      +\sum_{j\ge 0}^{J} 
                      2\left[ \left(2-\frac{4r}{3}\right)\frac{1}{\mu}
                              -4(j+1)+4(j+1)^2 \mu F^0_{j+1}(r,\mu)\right]
\nonumber\\
&&\hspace{2em}\left.+\sum_{k\ge 0}^{K}
                      2\left[-\left(2-\frac{4r}{3}\right)\frac{1}{\mu}
                              +2(2k+1)-4k(k+1) \mu F^1_{k}(r,\mu)\right]\right\},
\label{eq:N2q0}
\end{eqnarray}
\begin{eqnarray}
J&=&
\left\{
    \begin{array}{lcr}
        -\mathrm{Ceiling}[A^0_{1}] &\quad\quad\quad& \mbox{($A^0_{1} \le 0$)}\\
        -1                   && \mbox{(otherwise)}
    \end{array}\right. ,
\\
K&=&
\left\{
    \begin{array}{lcr}
        -\mathrm{Ceiling}[A^1_{0}] &\quad\quad\quad& \mbox{($|\mu/(2r)|<1$ and $A^1_{0} \le 0$)}\\
        -1                   && \mbox{(otherwise)}
    \end{array}\right. .
\end{eqnarray}

\section{Three term recurrence for $C^n_{\ell}$ and $dC^n_{\ell}/d\eta$}
\label{sec:apdxC}

When we evaluate $C^n_{\ell}$ and $dC^n_{\ell}/d\eta$ for a large Landau level ($n$,$m$)
with a naive implementation using the three term recurrence formula for Laguerre polynomials,
we encounter arithmetic overflow or underflow in double precision arithmetic.
In order to tame the numerical overflow and underflow we employ
a modified recurrence formula with rescaling and quadruple precision arithmetic.

We define $f^n_{\ell}$ and $df^n_{\ell}$ satisfying the following recurrence formula;
\begin{equation}
  f^n_0= 1, \quad f^n_{-1} = 0,
\quad
 df^n_0= 0,\quad  df^n_{-1} = 0,
\label{eq:initialcond}
\end{equation}
\begin{eqnarray}
  f^n_{\ell} &=& (\alpha^n_{\ell} f^n_{\ell-1} + \beta^n_{\ell} f^n_{\ell-2} ) \gamma^n_{\ell},
\\
 df^n_{\ell} &=& (\alpha^n_{\ell}df^n_{\ell-1} + \beta^n_{\ell}df^n_{\ell-2} - f^n_{\ell-1} ) \gamma^n_{\ell},
\\
\alpha^n_{\ell} &=& (2 \ell + n - 1 - \eta),
\\
 \beta^n_{\ell} &=&  (1 - \ell - n) \sqrt{(\ell - 1)/(\ell - 1 + n)},
\\
\gamma^n_{\ell} &=& \sqrt{\ell/(\ell + n)}/\ell,
\end{eqnarray}
for $1 \le \ell$. $\eta$ is the argument of $C^n_\ell$ and $dC^n_{\ell}/d\eta$.
$f^n_{\ell}$ and $df^n_{\ell}$ are proportional to $\sqrt{\ell!/((\ell+n)!)}L^n_{\ell}(\eta)$ 
and its derivative respectively. When either of $|f^n_{\ell}|$ or $|df^{n}_{\ell}|$ takes a value larger than 
$10^{100}$ or smaller than $10^{-100}$ 
during the recurrence, intermediate states, $(f^n_{\ell},f^n_{\ell-1},f^n_{\ell-2},df^{n}_{\ell},df^{n}_{\ell-1},df^{n}_{\ell-1})$,
are rescaled by multiplying the inverse of $\max(|f^n_{\ell}|,|df^{n}_{\ell}|)$ or $\min(|f^n_{\ell}|,|df^{n}_{\ell}|)$ and 
the scaling factor is stored for later use below.

The coefficients $C^n_\ell$ and $dC^n_{\ell}/d\eta$ are derived by
\begin{eqnarray}
    C^n_{\ell}(\eta)&=& \left(h_n f^n_{\ell}\right)^2, \\
\frac{d C^n_{\ell}}{d\eta}(\eta)&=&
  \left[2 \left(h_n f^n_{\ell}\right) 
          \left(h_n df^n_{\ell}\right)
         -\left(h_n f^n_{\ell}\right)^2 \right]
      +n  \left( g_{n} f^n_{\ell}\right)^2,
\end{eqnarray}
\begin{eqnarray}
h_n &=& \left\{
    \begin{array}{lcc}
\displaystyle
          e^{-\left(\eta + \sum_{k\ge 1}^{n}\log(k) - 2\sum_{j\ge 1}^{N_{\mathrm{scale}}}\log(S_j) - n \log(\eta) \right)/2} && \mbox{(for $\eta >1$)}\\
          e^{-\left(\eta + \sum_{k\ge 1}^{n}\log(k) - 2\sum_{j\ge 1}^{N_{\mathrm{scale}}}\log(S_j) \right)/2} \eta^n         && \mbox{(for $\eta \le1 $)}\\
    \end{array}
\right.,\\
g_n &=& \left\{
    \begin{array}{lcc}
\displaystyle
          e^{-\left(\eta + \sum_{k\ge 1}^{n}\log(k) - 2\sum_{j\ge 1}^{N_{\mathrm{scale}}}\log(S_j) - (n-1) \log(\eta) \right)/2} && \mbox{(for $\eta >1$)}\\
          e^{-\left(\eta + \sum_{k\ge 1}^{n}\log(k) - 2\sum_{j\ge 1}^{N_{\mathrm{scale}}}\log(S_j) \right)/2} \eta^{n-1}         && \mbox{(for $\eta \le1 $)}\\
    \end{array}
\right.,
\end{eqnarray}
where $S_j$'s are the rescaling factors stored during the recurrence.
Finally the coefficients $C^n_\ell$ and $dC^n_{\ell}/d\eta$ are converted to double 
precision numbers.


\section{Zero field limit of the imaginary parts of  Eqs.~(\ref{eq:N0q0}) and (\ref{eq:N1q0})  with $q=0$}
\label{sec:apdxD}
The vacuum polarization tensor in vacuum is written by
\begin{equation}
\Pi(k^2)=
\frac{\alpha}{3\pi}\left\{
       \frac{1}{3}+\left(2 + \frac{1}{y}\right)
       \left[\sqrt{1/y-1} \cot^{-1}\left(\sqrt{1/y-1}\right) -1 \right]\right\}
\label{eq:PiBelowThZeroB}
\end{equation}
for $y< 1$ and
\begin{equation}
\Pi(k^2)=
\frac{\alpha}{3\pi}\left\{
       \frac{1}{3}+\left(2 + \frac{1}{y}\right)
       \left[\sqrt{1-1/y} \tanh^{-1}\left(\sqrt{1-1/y}\right)-1 -i \frac{\pi}{2}\sqrt{1-1/y}\right]\right\},
\label{eq:PiAboveThZeroB}
\end{equation}
for $1<y$ with $y\equiv k^2/(2m)^2$. The imaginary part is thus
\begin{equation}
    \mathrm{Im}\Pi(k^2) = -\frac{\alpha}{6}\left(2+1/y\right)\sqrt{1-1/y}.
\end{equation}
The zero field limit for the imaginary parts of Eqs.~(\ref{eq:N0q0}) and (\ref{eq:N1q0}) 
can be taken as follows.

\begin{eqnarray}
\lim_{eB\rightarrow 0} \mathrm{Im}N_0 &=&
\lim_{eB\rightarrow 0} 
\left[
 -\frac{\alpha}{4\pi} \mathrm{Im}
\left\{ 
  2 \sum_{k\ge 0}^{K}
\Biggl[
 \frac{1}{a}
    \left\{ -2b+(c-a)b F^1_k(r,\mu) 
\right.
\right.
\right.
\nonumber\\
&&\hspace*{11em}
\left.\left.\left.
    +(b^2-a^2-ac+a)G^1_{k}(r,\mu)\right\} + b H^1_{k}(r,\mu)
\Biggr]
\right\}
\right]
\nonumber\\
&=&
\lim_{\Delta \beta \rightarrow 0} 
\left[
 -\frac{\alpha}{4\pi} \frac{\pi}{2}
   \sum_{k\ge 0}^{\frac{1-1/r}{\Delta \beta}}
\left(1+\frac{1}{r}+\beta_k\right)\frac{\Delta \beta}{\sqrt{\beta_k}}
\right],
\end{eqnarray}
where $\Delta \beta \equiv 2\mu/r$ and $\beta_k\equiv 1-1/r-k\Delta\beta$. 
This is the rectangular approximation of integration and the limit leads
\begin{eqnarray}
\lim_{eB\rightarrow 0} \mathrm{Im}N_1 &=&
 -\frac{\alpha}{4\pi}\frac{\pi}{2}
 \int_{0}^{1-1/r}
\left(1+\frac{1}{r}+\beta \right)\frac{d \beta}{\sqrt{\beta}}
\nonumber\\
&=&
 -\frac{\alpha}{6}
\left(2+1/r\right) \sqrt{1-1/r} = \mathrm{Im}\Pi(k_{\parallel}^2).
\end{eqnarray}

Similarly we have
\begin{eqnarray}
\lim_{eB\rightarrow 0} \mathrm{Im}N_1 &=&
\lim_{eB\rightarrow 0} \left[
 -\frac{\alpha}{4\pi}
\mathrm{Im}
\left[
 \mu \sum_{k \ge 0}^{K} (2-\delta_{0k}) \left(F^0_k(r,\mu)-H^0_k(r,\mu)\right)
\right]
\right]
 \nonumber\\
&= &
\lim_{\Delta \beta \rightarrow 0} \left[
 -\frac{\alpha}{4\pi}
 \frac{\pi}{2} \sum_{k \ge 0}^{ \frac{1-1/r}{\Delta \beta}}
(2-\delta_{0k})
 \left(\frac{1- \beta_k}{\sqrt{\beta_k}}\right)\Delta \beta
\right]
\nonumber\\
&&=
 -\frac{\alpha}{4\pi}
 \pi \int_{0}^{1-1/r}
 \left(\frac{1- \beta}{\sqrt{\beta}}\right) d\beta \nonumber\\
&=&
 -\frac{\alpha}{6}
\left(2+1/r\right) \sqrt{1-1/r} = \mathrm{Im}\Pi(k_{\parallel}^2).
\end{eqnarray}

\section{Form factors with $eB=m_{\pi}^2$, $(1/10)m_{\pi}^2$ and 0}
\label{sec:apdxE}
\begin{figure}[ht]
\centerline{
\includegraphics[width=\figscale]{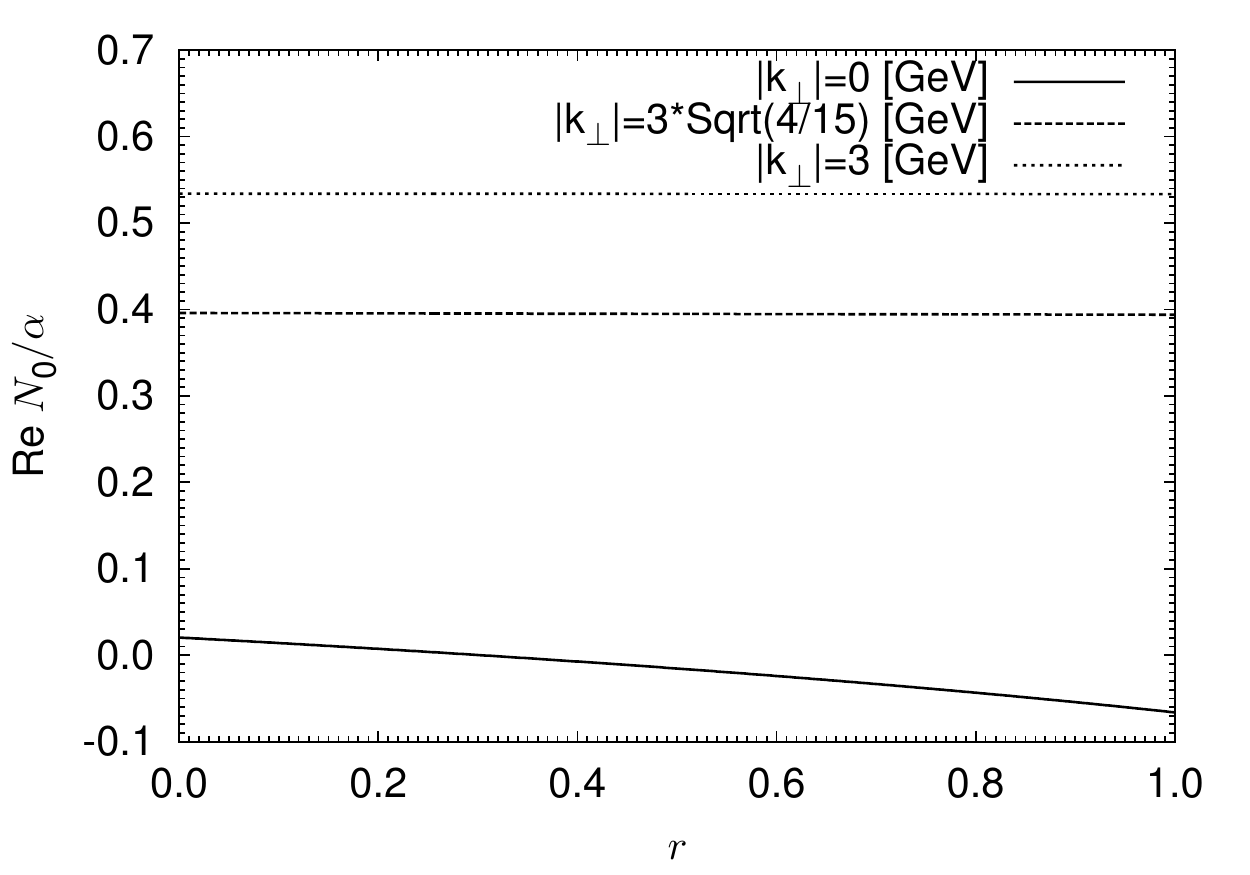}
\includegraphics[width=\figscale]{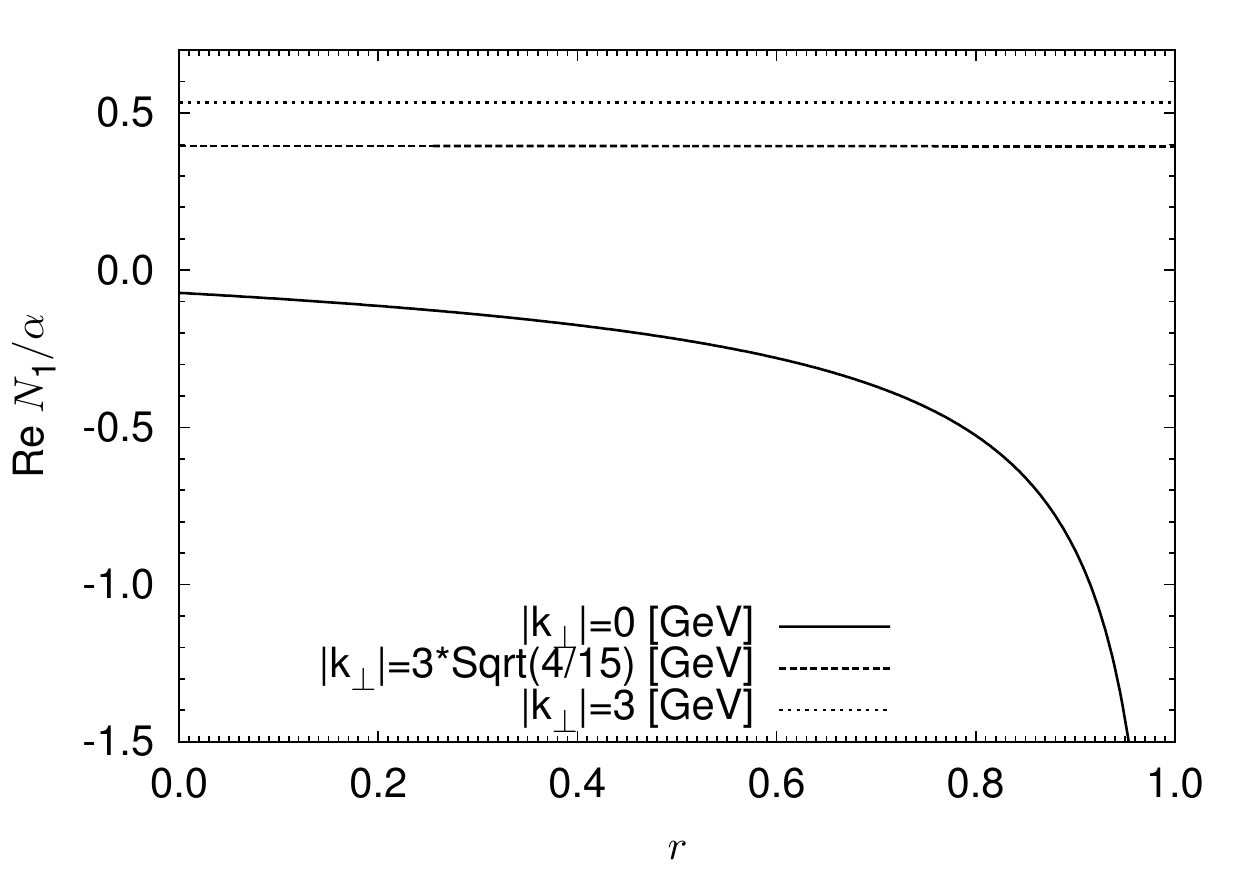}
}
\centerline{
\includegraphics[width=\figscale]{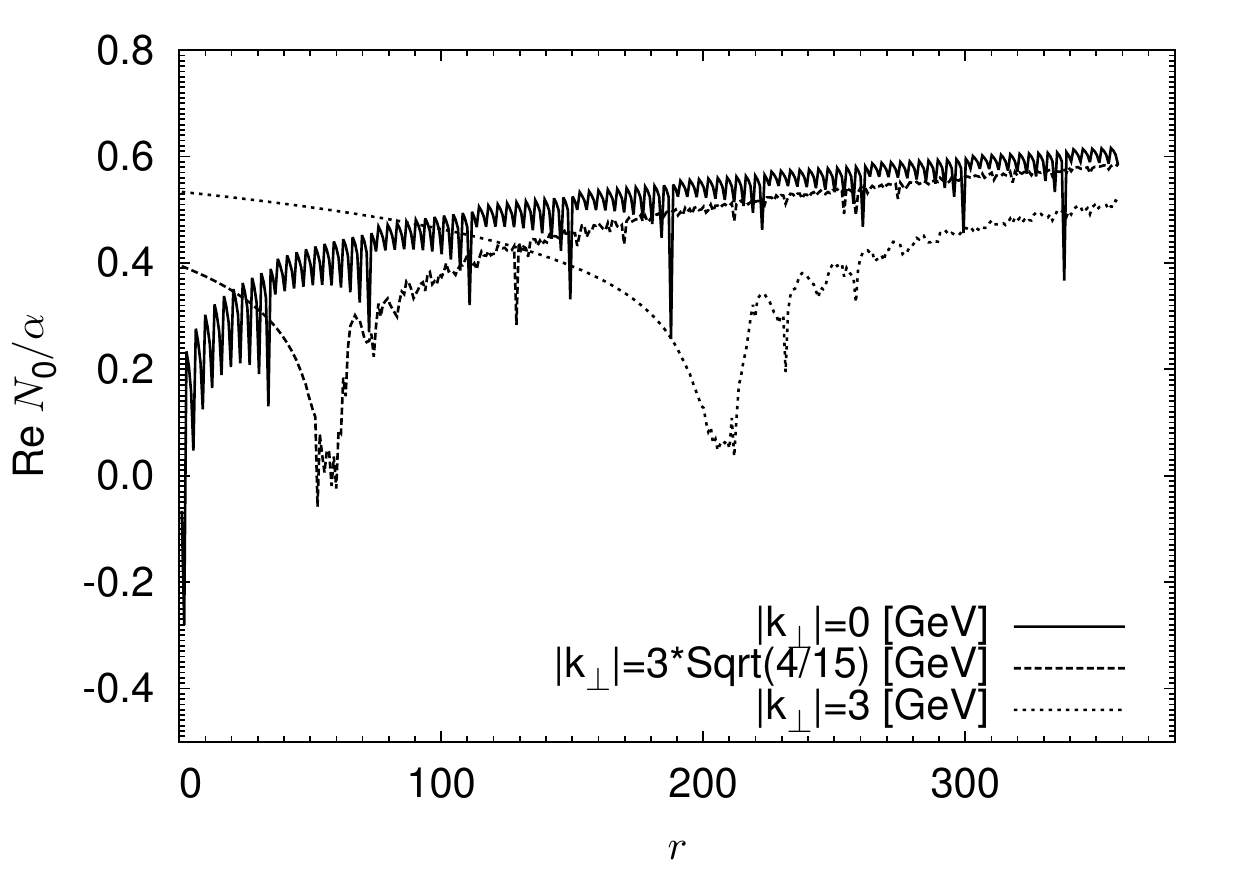}
\includegraphics[width=\figscale]{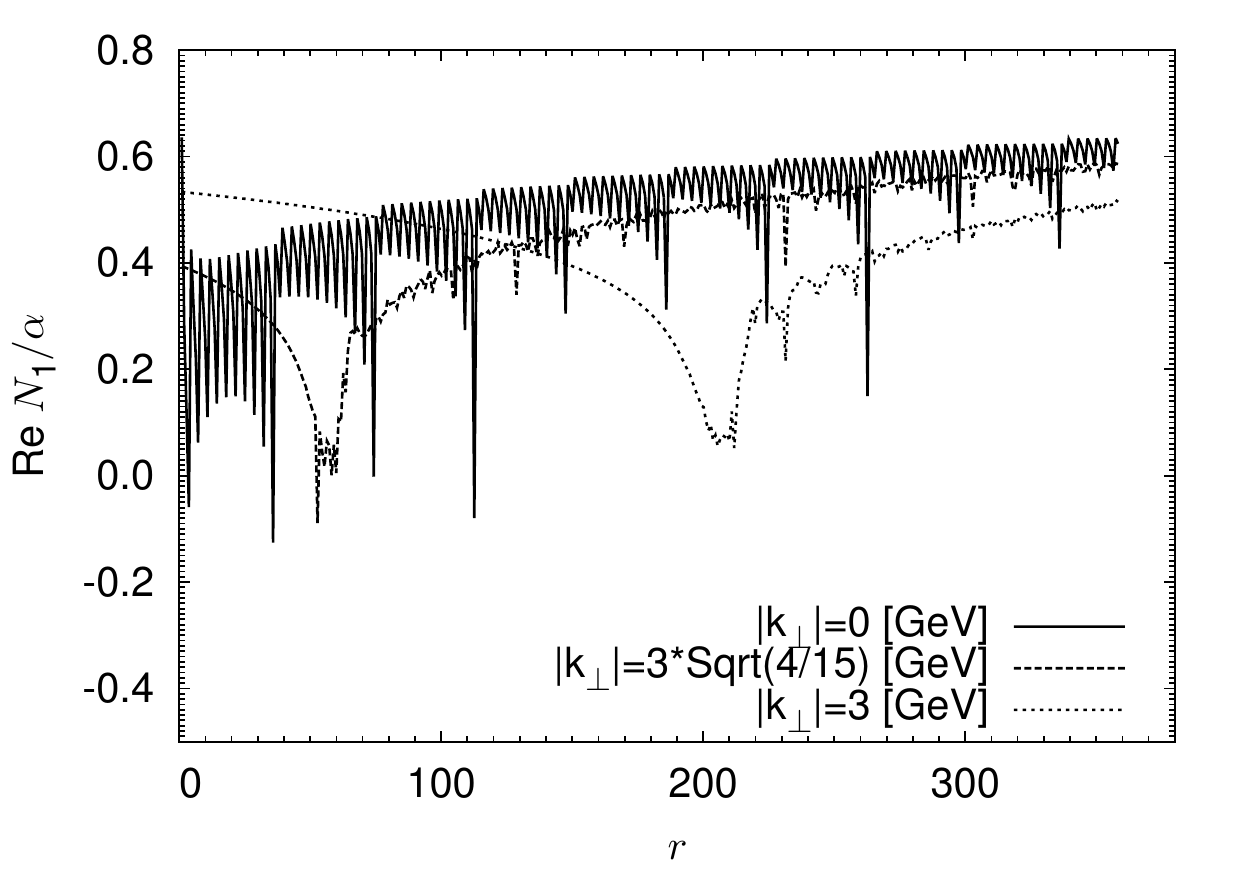}
}
\centerline{
\includegraphics[width=\figscale]{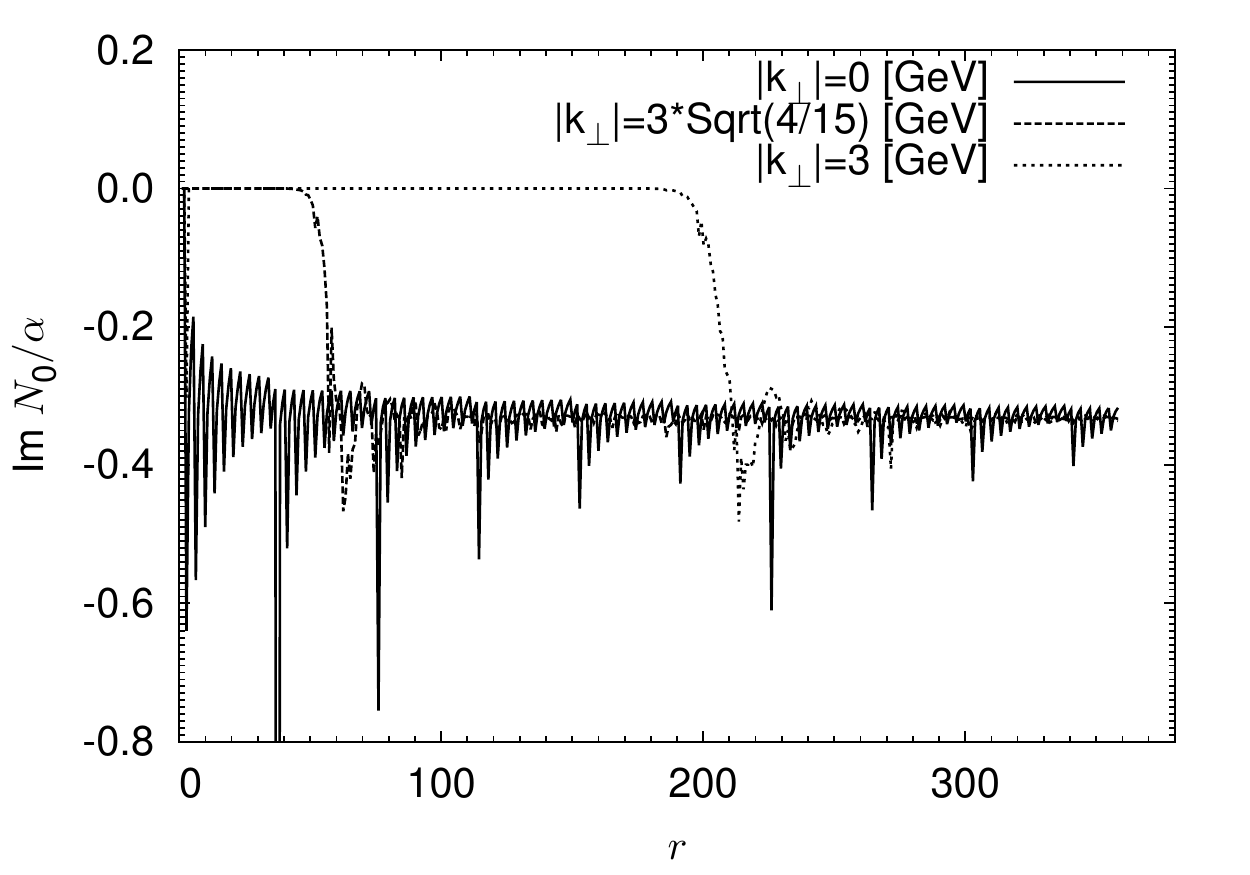}
\includegraphics[width=\figscale]{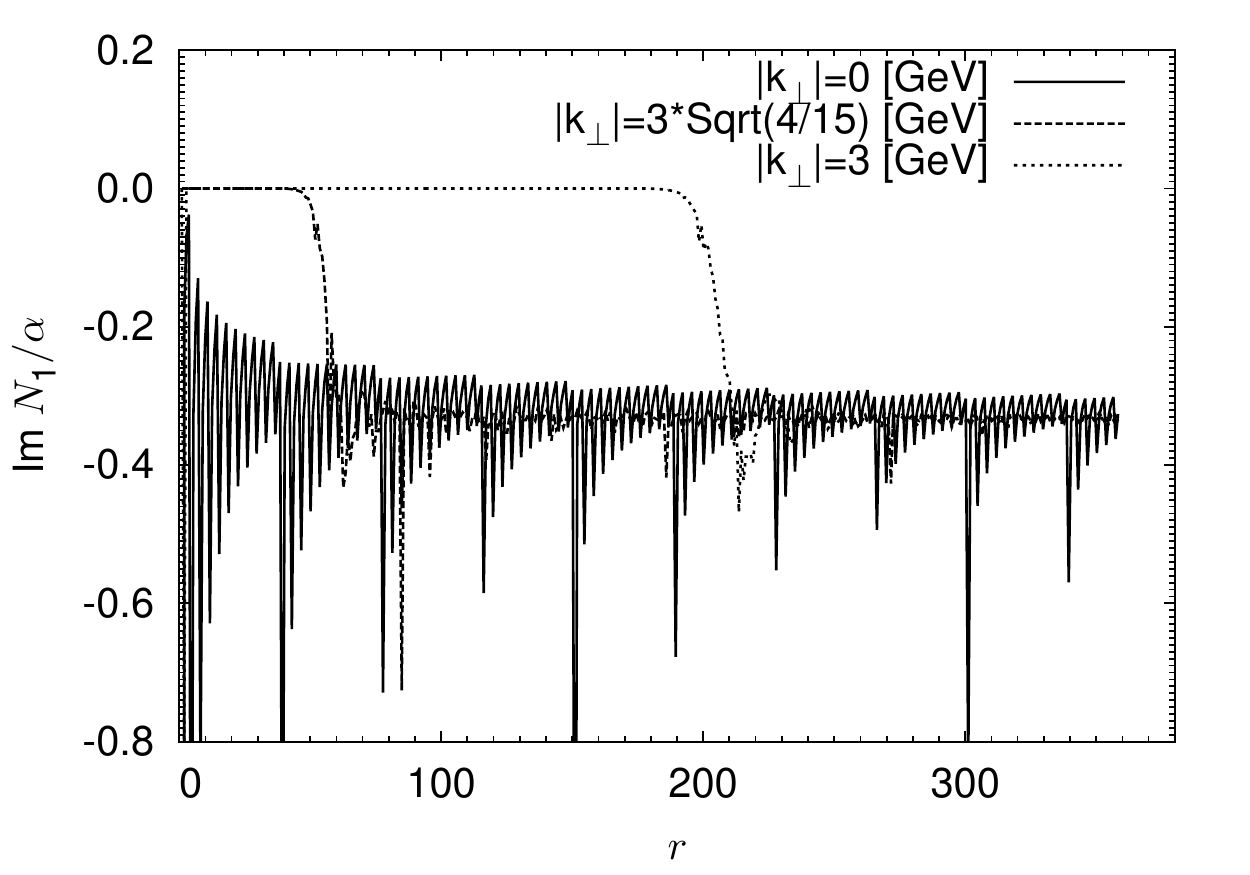}
}
\vspace*{-4pt}
\caption{Form factors $N_0$ (left) and $N_1$ (right) for muons (case [b-1])
with $\ell_{\mathrm{max}}=1000$
(top ($r<1$), middle (real part in $1<r$), and bottom (imaginary part in $1<r$).}
\label{fig:N0N1muonmpi}
\end{figure}

\begin{figure}[ht]
\centerline{
\includegraphics[width=\figscale]{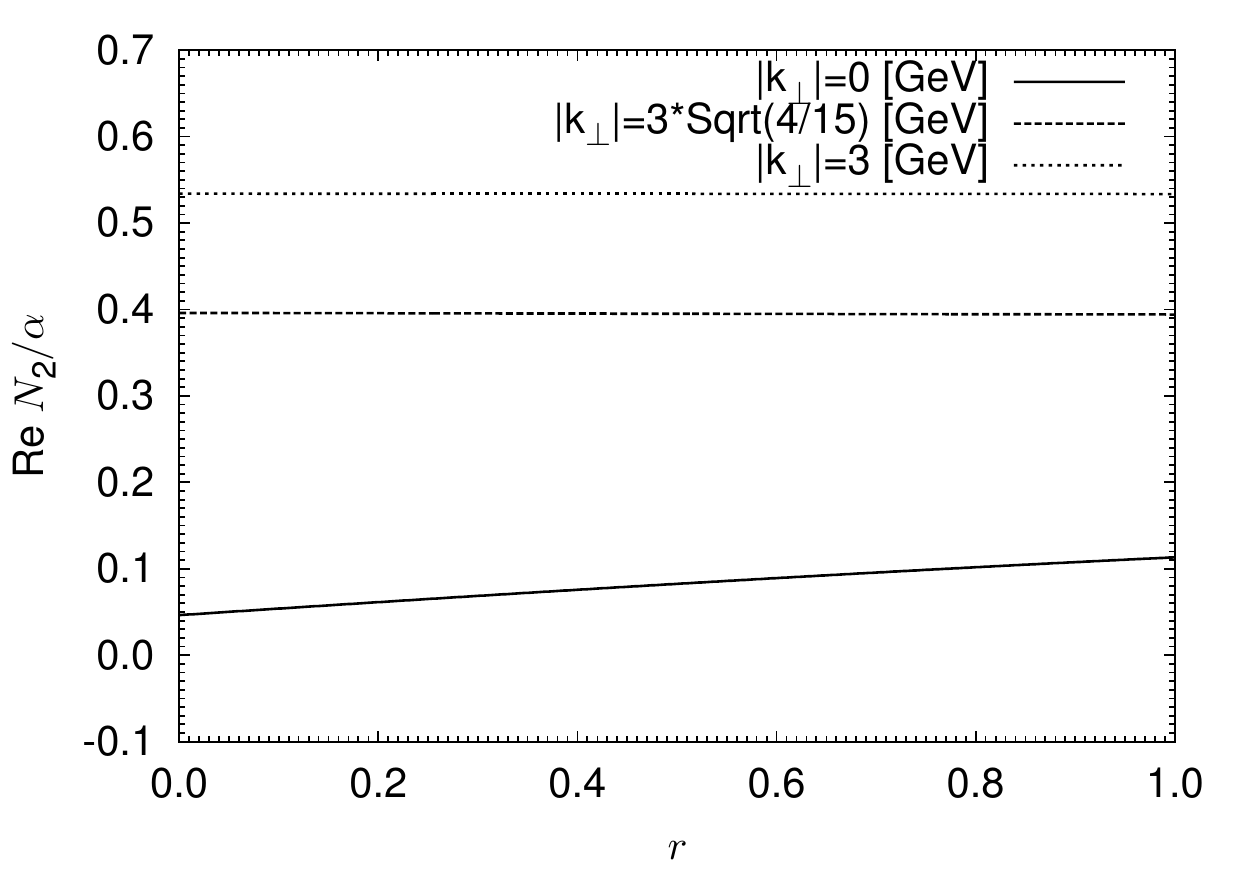}
}
\centerline{
\includegraphics[width=\figscale]{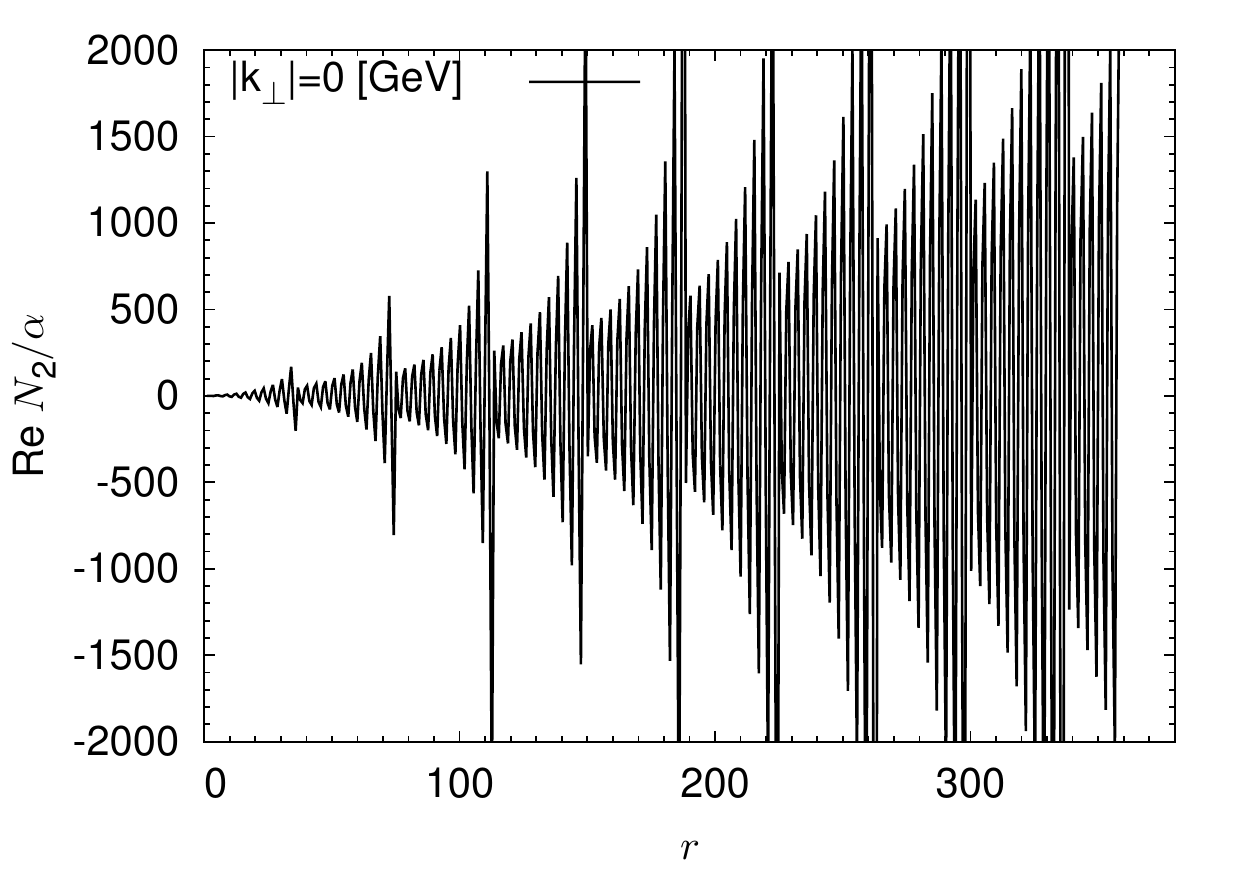}
\includegraphics[width=\figscale]{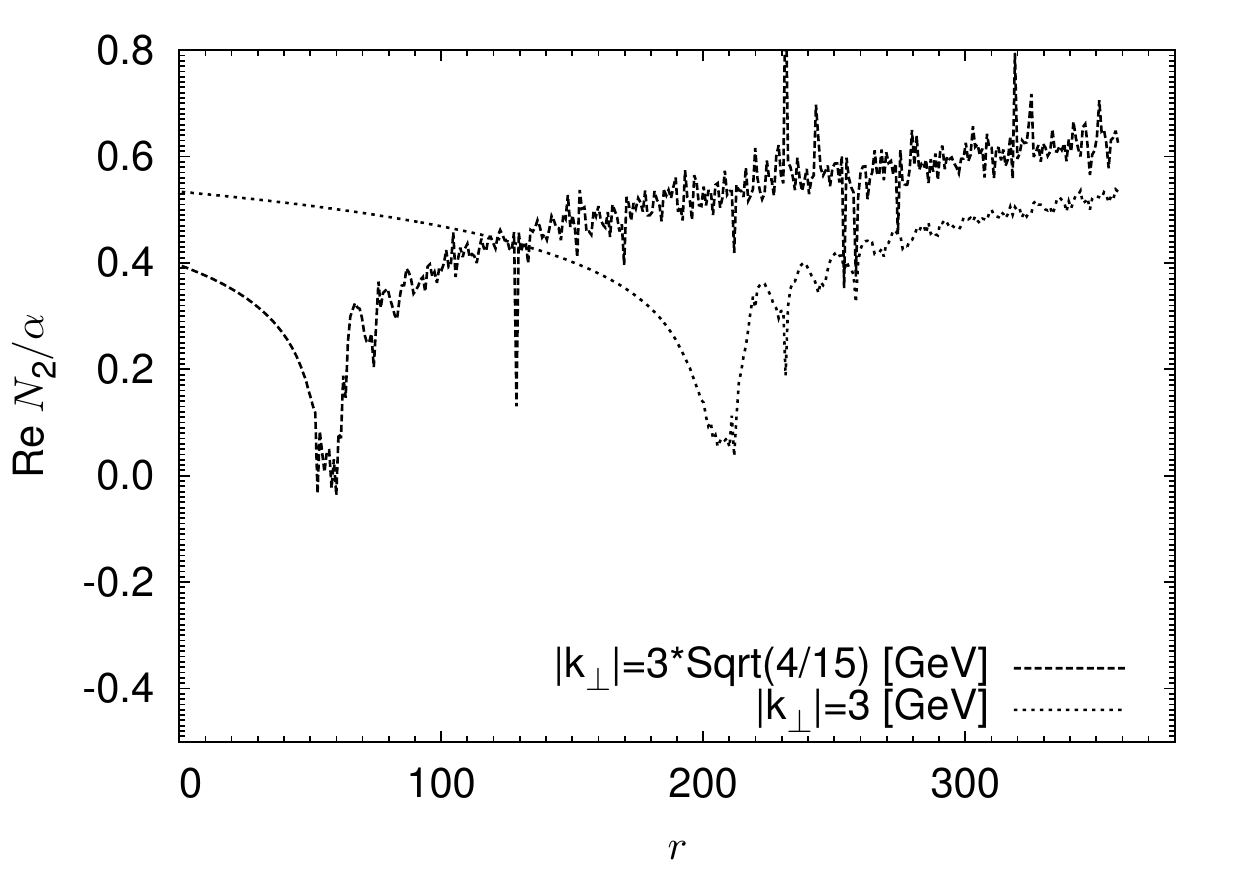}
}
\centerline{
\includegraphics[width=\figscale]{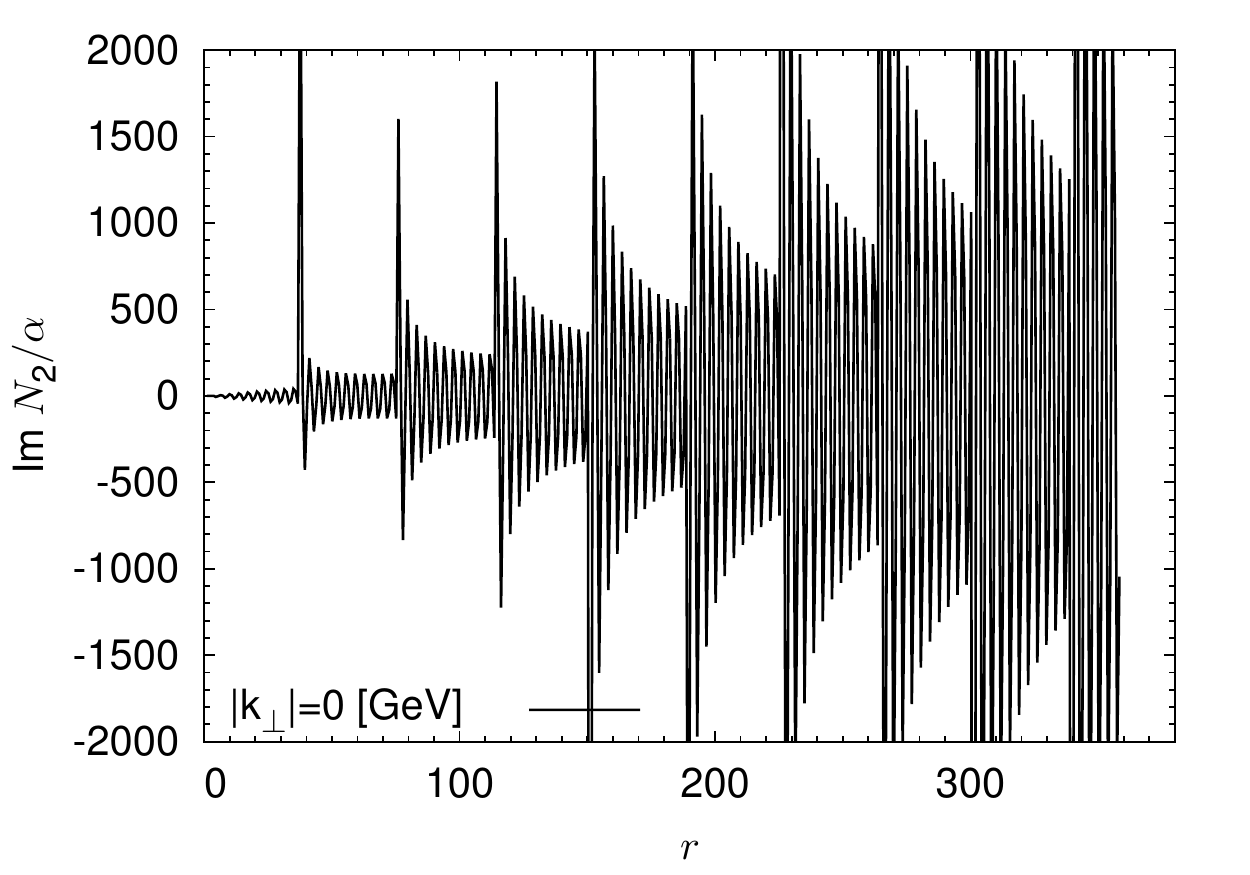}
\includegraphics[width=\figscale]{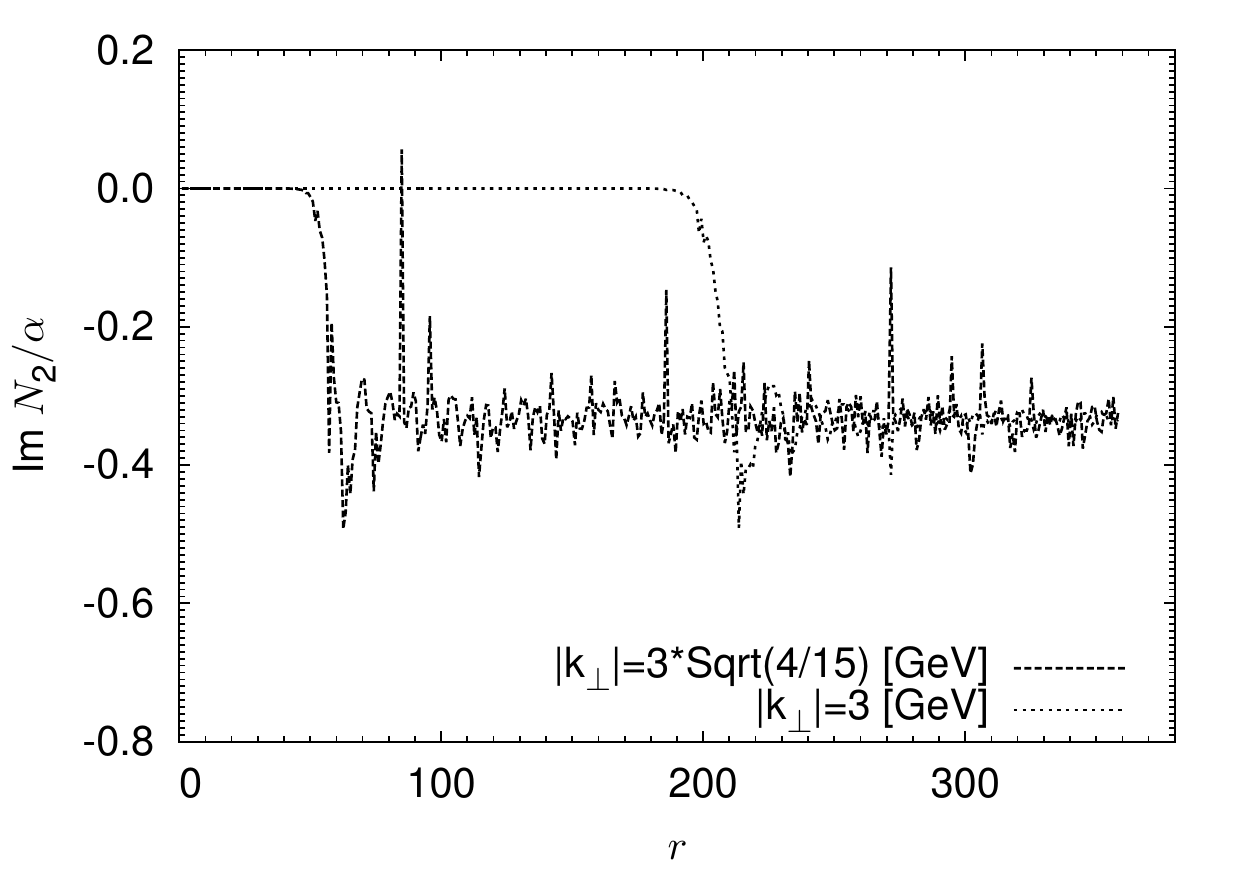}
}
\vspace*{-4pt}
\caption{Form factor $N_2$ for muons (case [b-1]) with $\ell_{\mathrm{max}}=1000$
(top ($r<1$), middle (real part in $1<r$) and bottom (imaginary part in $1<r$).}
\label{fig:N2muonmpi}
\end{figure}

\begin{figure}[ht]
\centerline{
\includegraphics[width=\figscale]{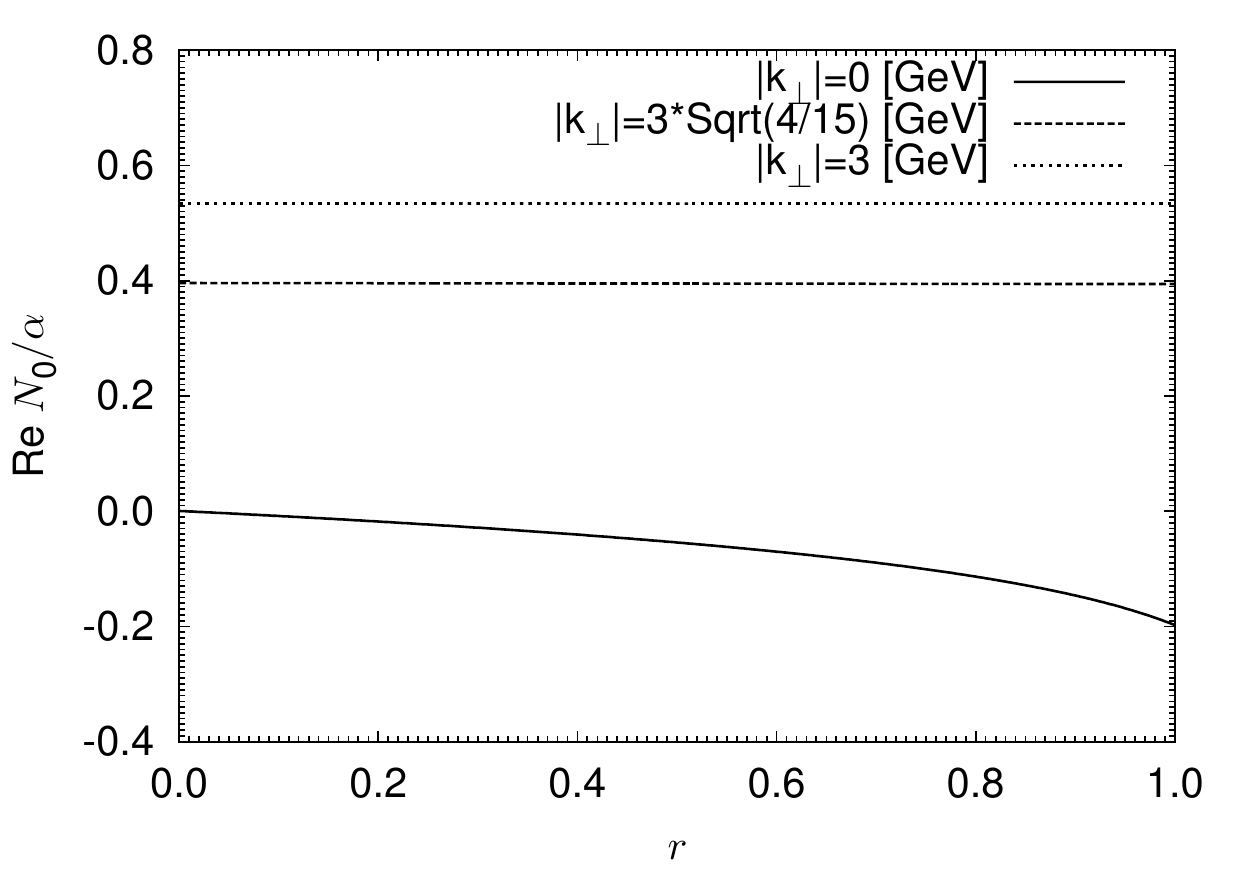}
\includegraphics[width=\figscale]{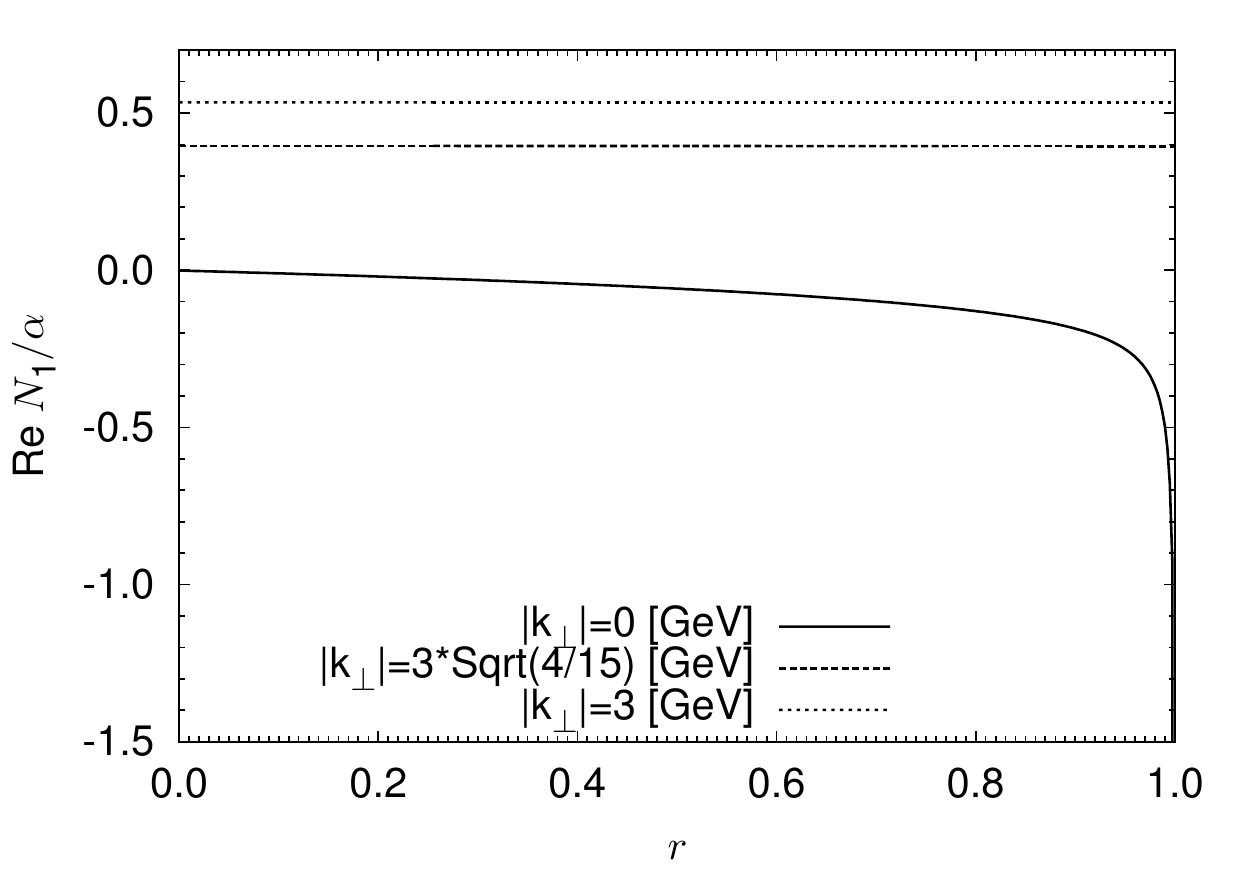}
}
\centerline{
\includegraphics[width=\figscale]{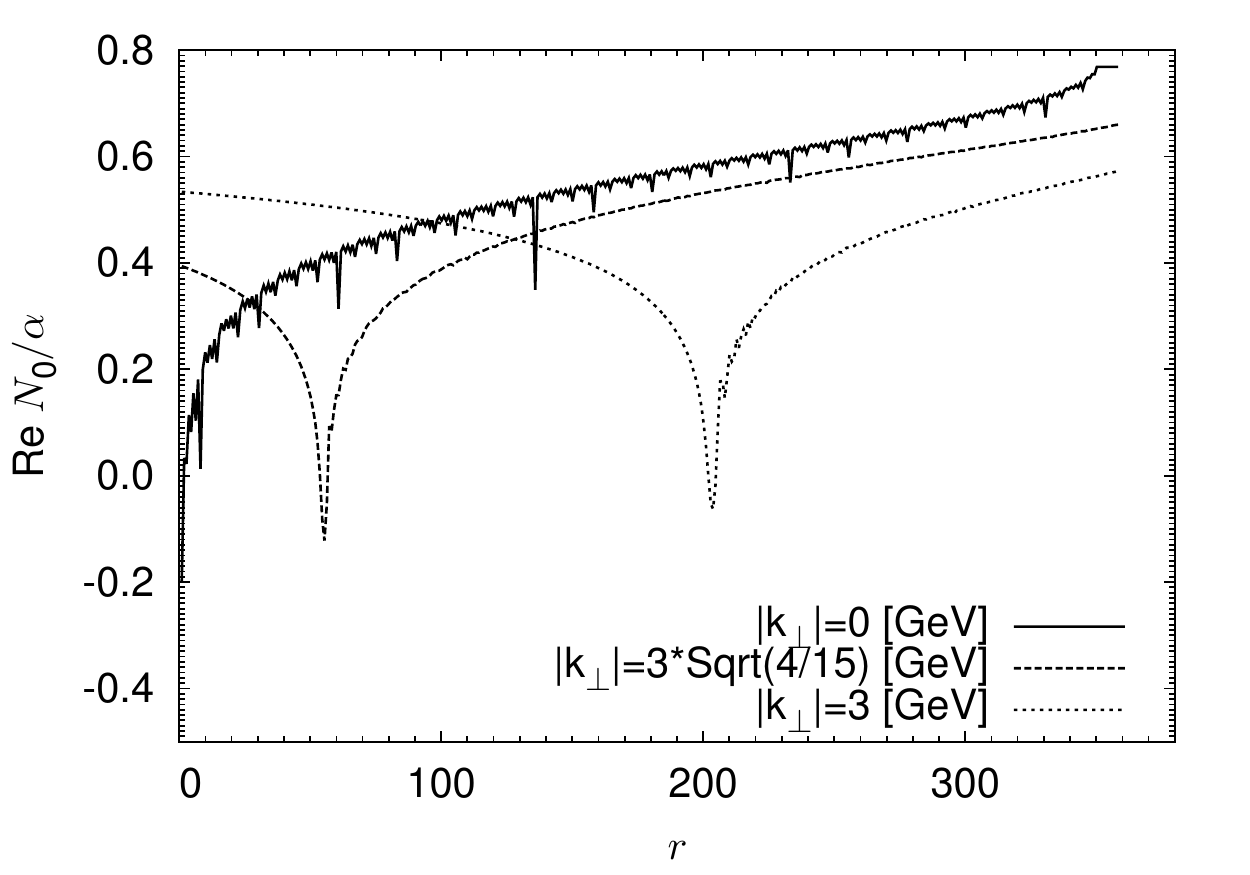}
\includegraphics[width=\figscale]{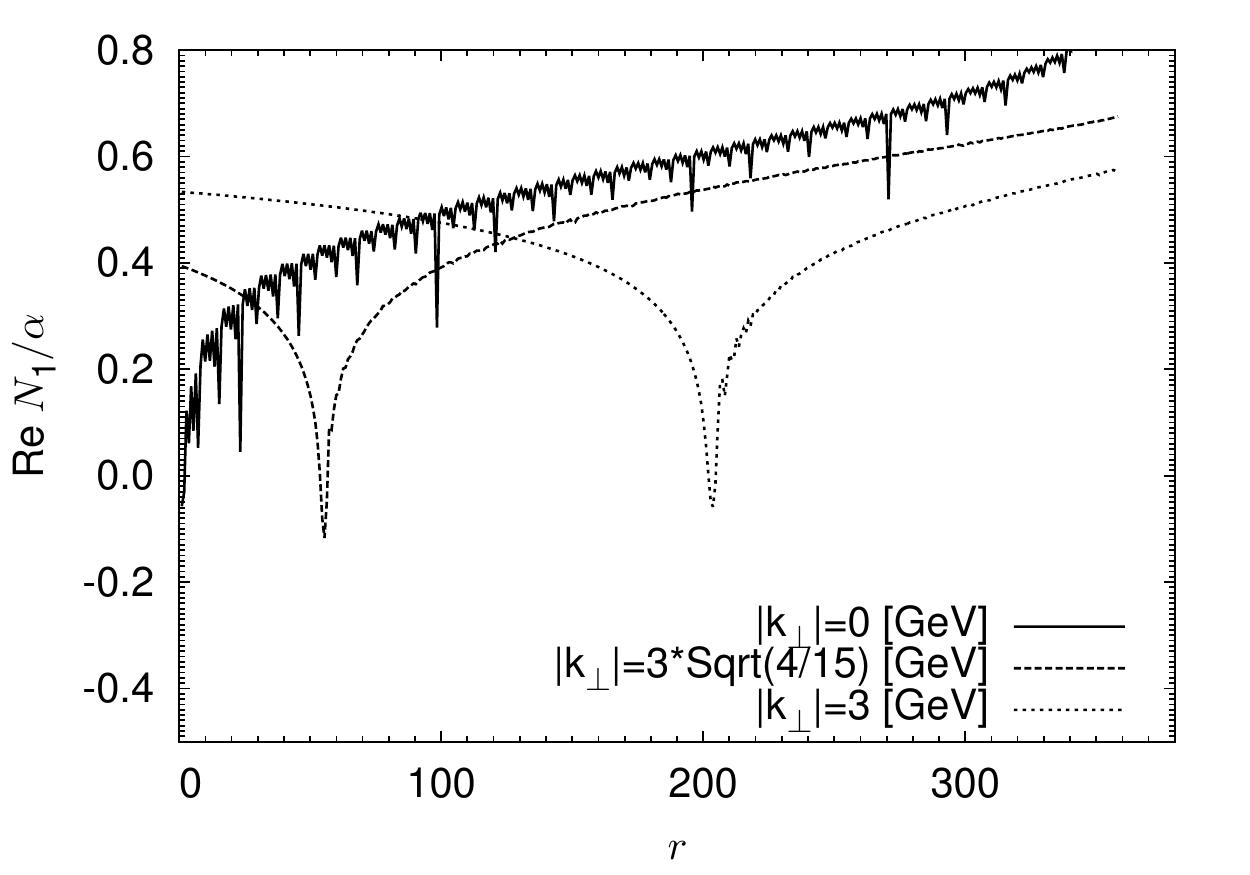}
}
\centerline{
\includegraphics[width=\figscale]{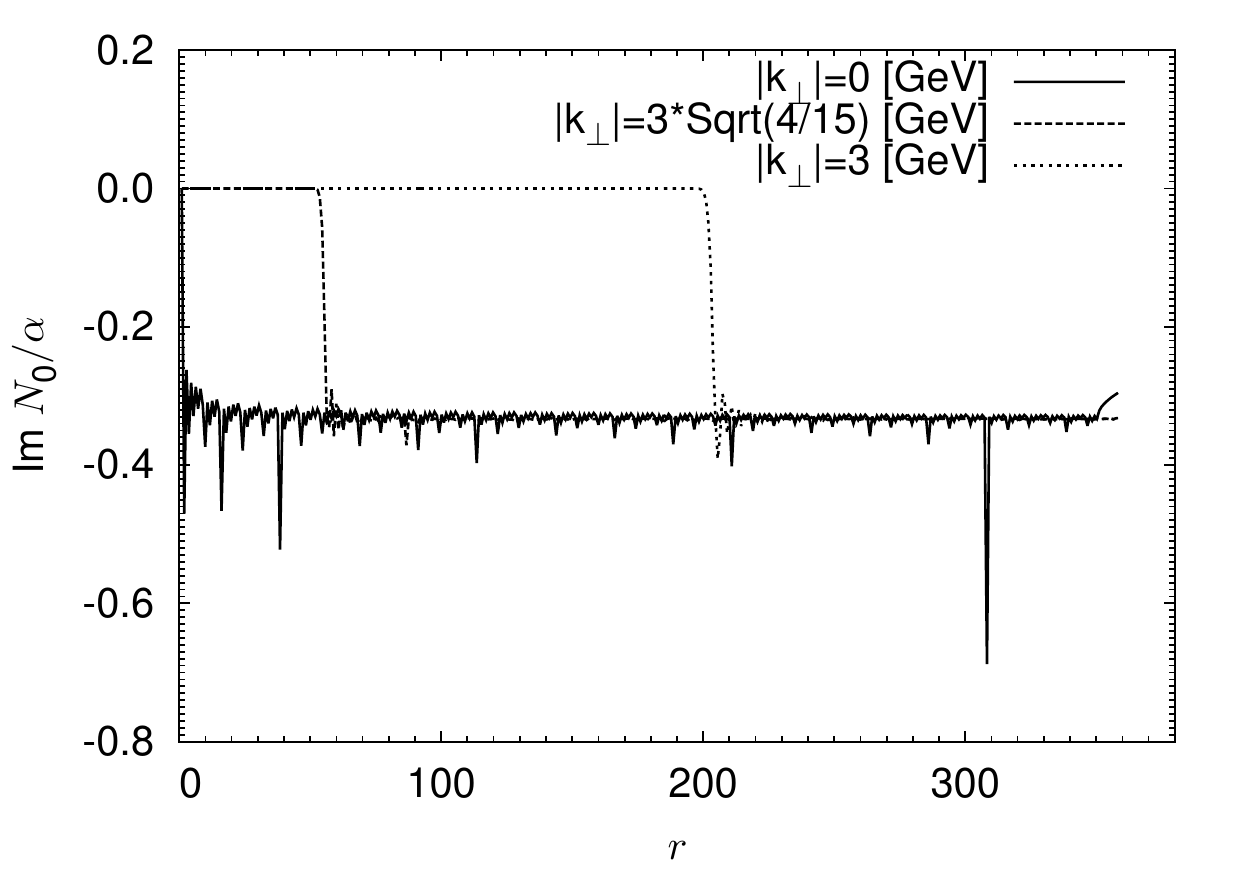}
\includegraphics[width=\figscale]{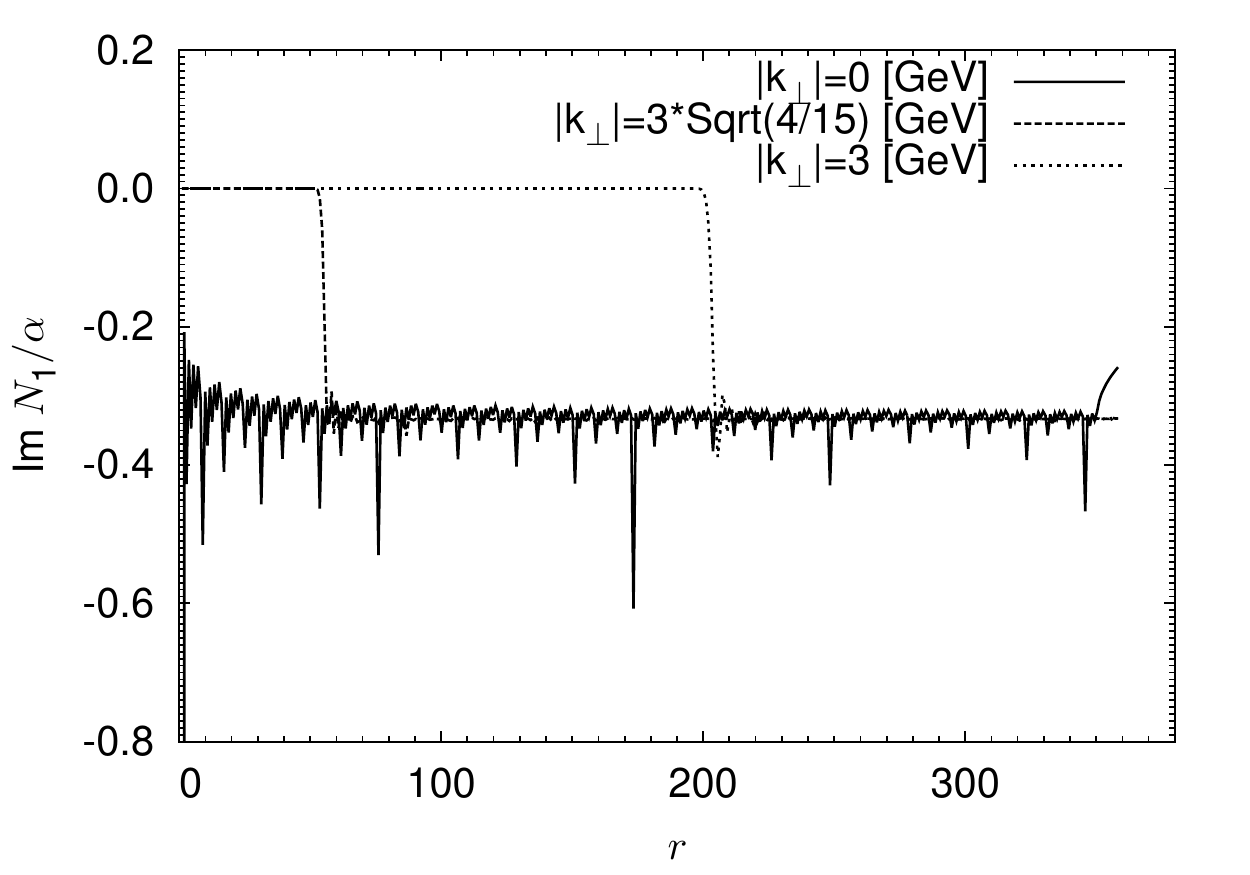}
}
\vspace*{-4pt}
\caption{Form factors $N_0$ (left) and $N_1$ (right) for muons (case [c-1])
with $\ell_{\mathrm{max}}=1000$
(top ($r<1$), middle (real part in $1<r$), and bottom (imaginary part in $1<r$).}
\label{fig:N0N1muon01mpi}
\end{figure}

\begin{figure}[ht]
\centerline{
\includegraphics[width=\figscale]{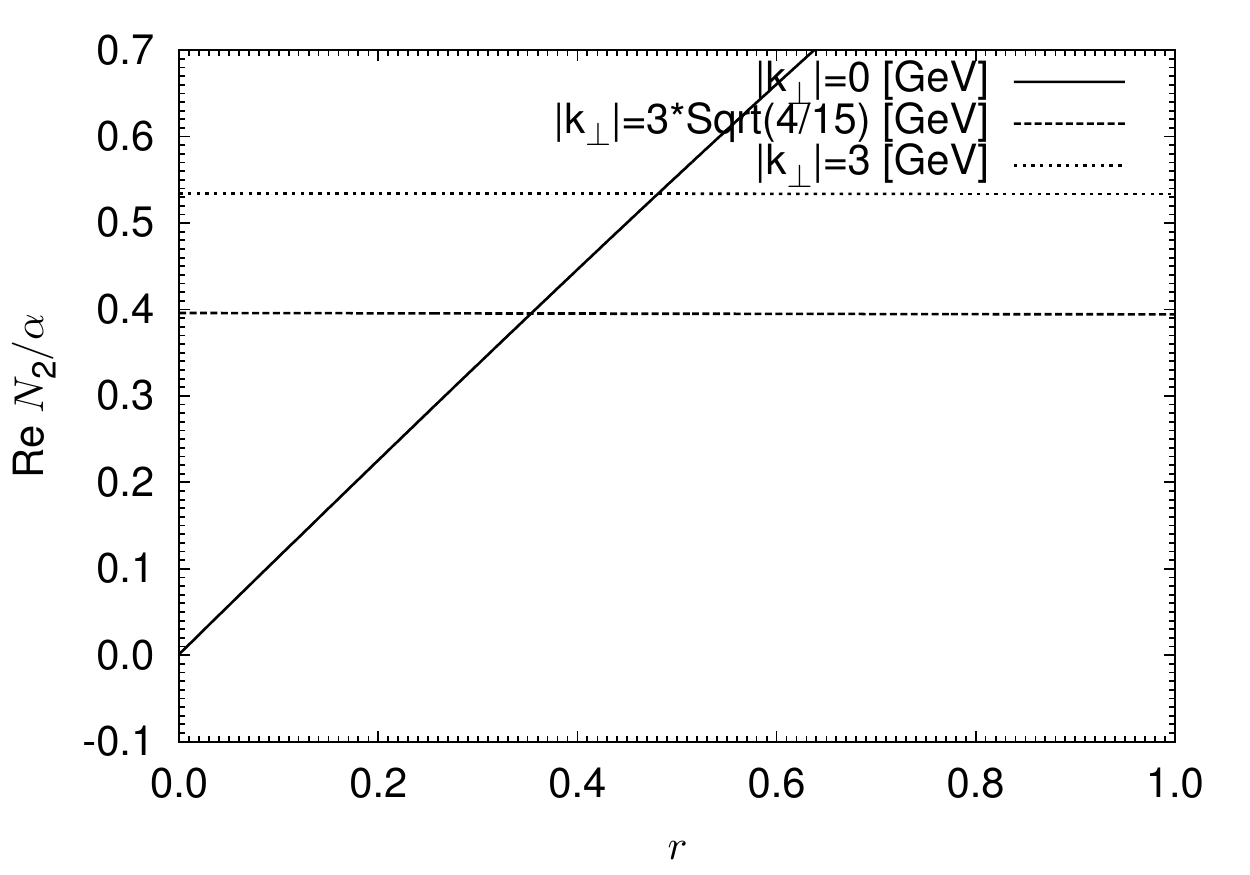}
}
\centerline{
\includegraphics[width=\figscale]{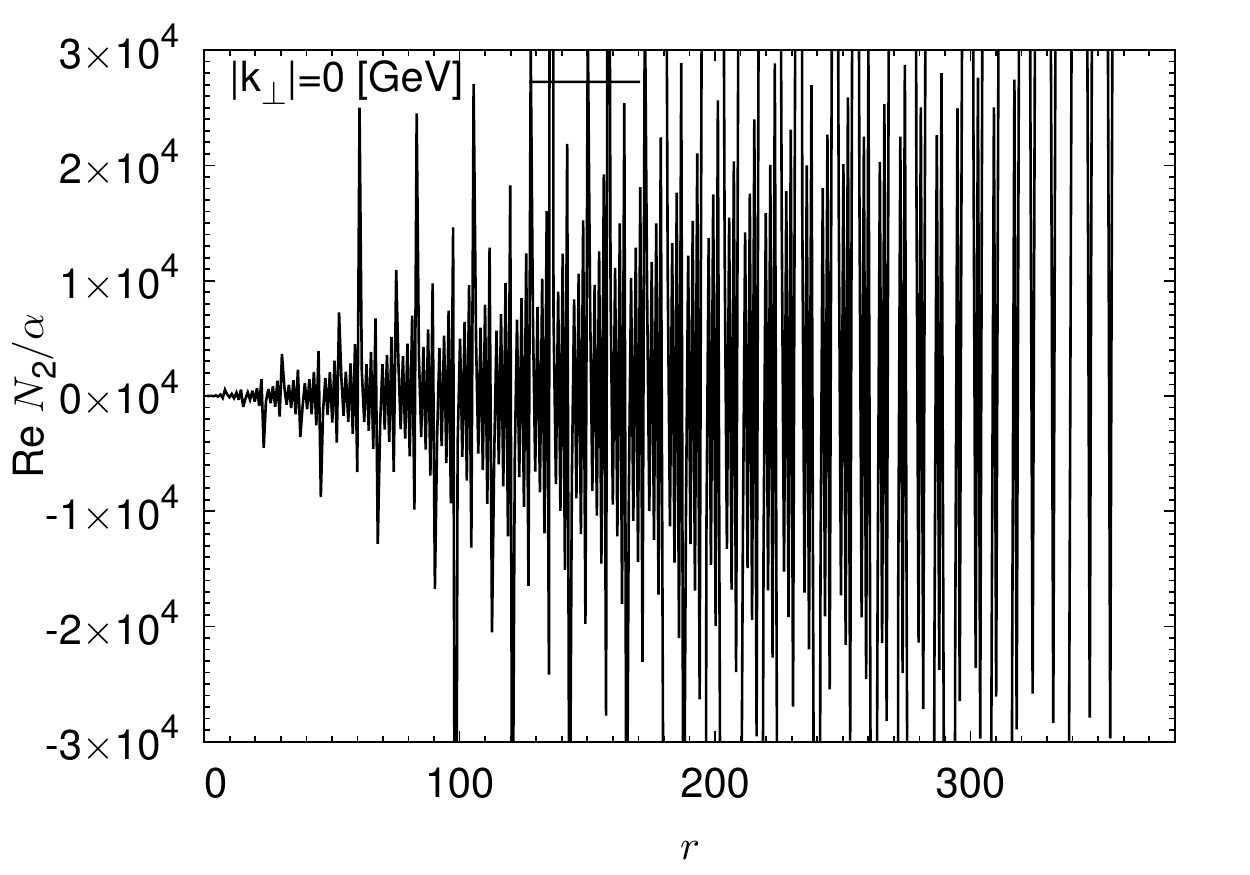}
\includegraphics[width=\figscale]{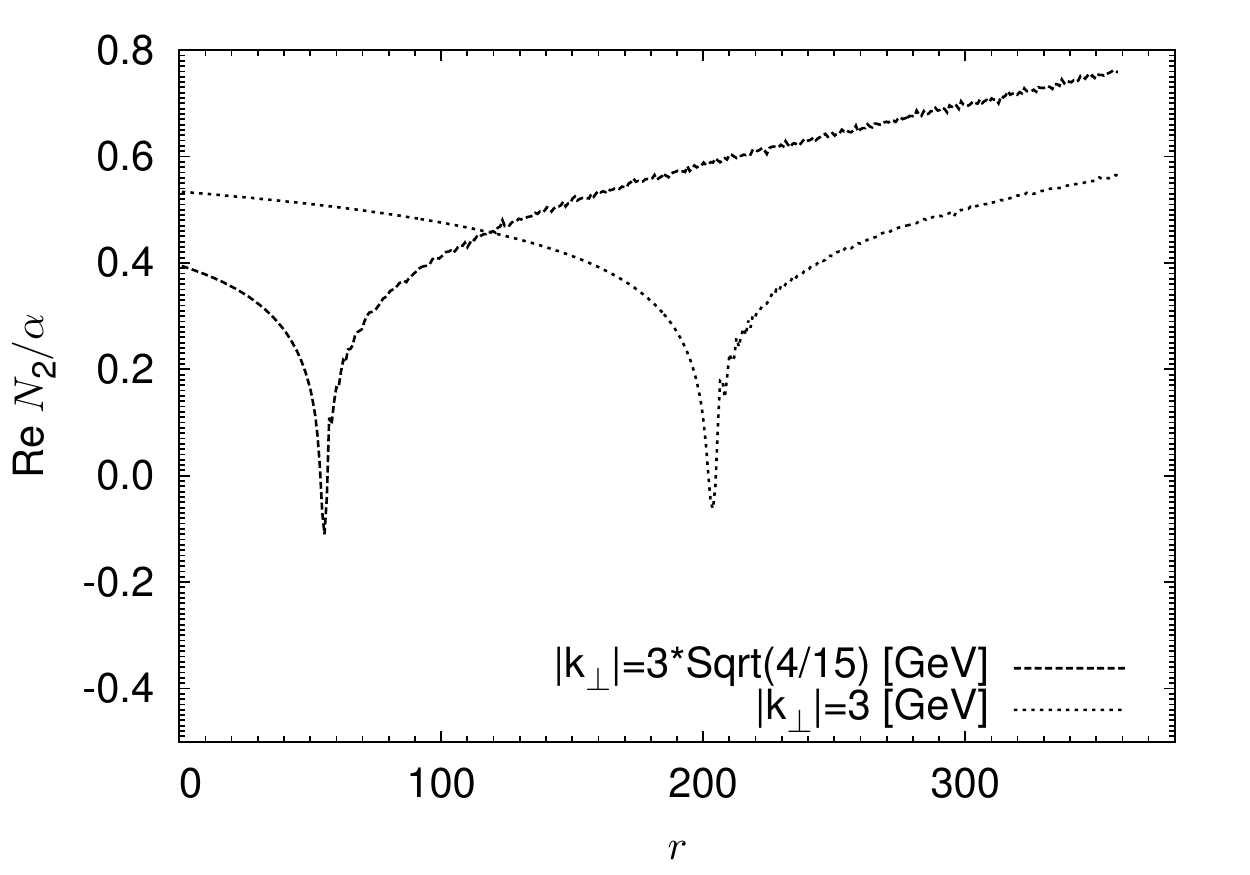}
}
\centerline{
\includegraphics[width=\figscale]{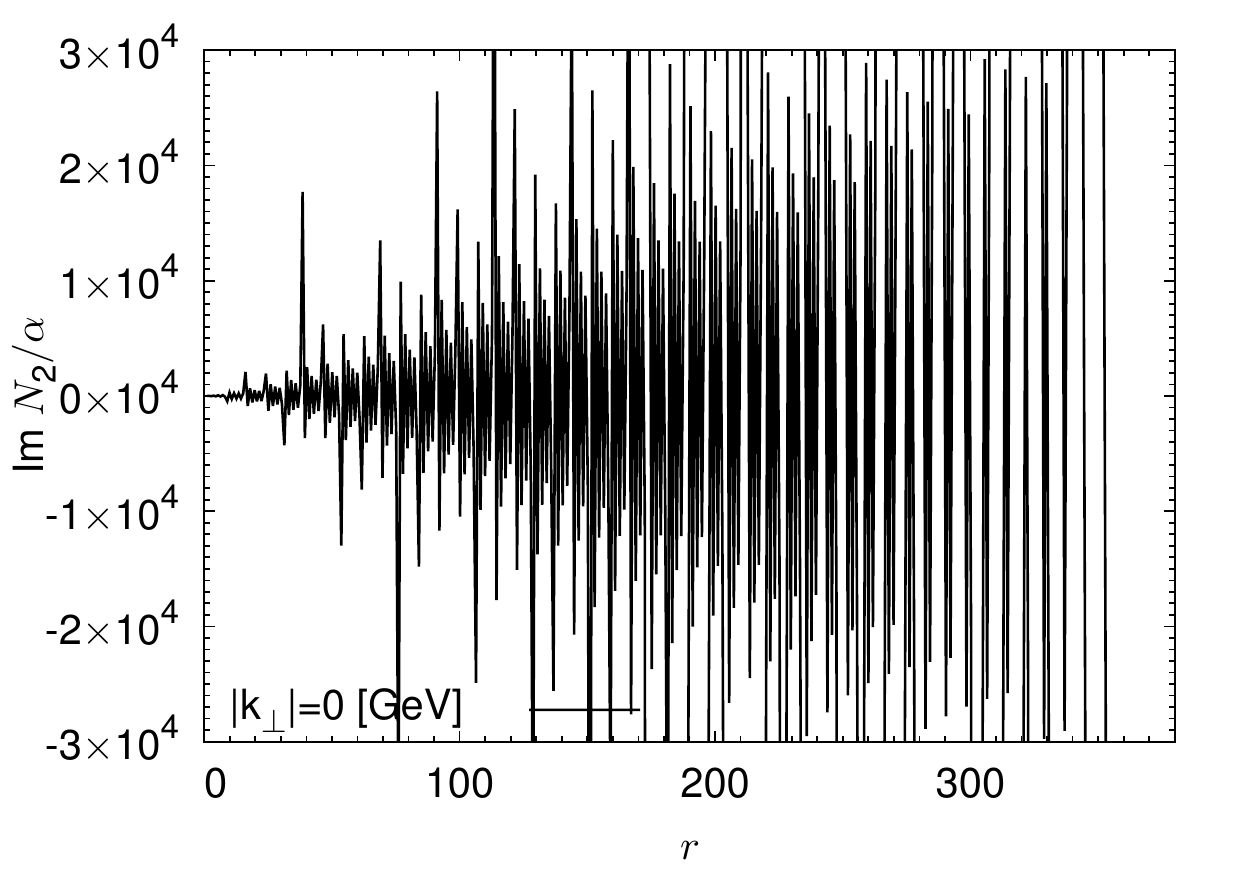}
\includegraphics[width=\figscale]{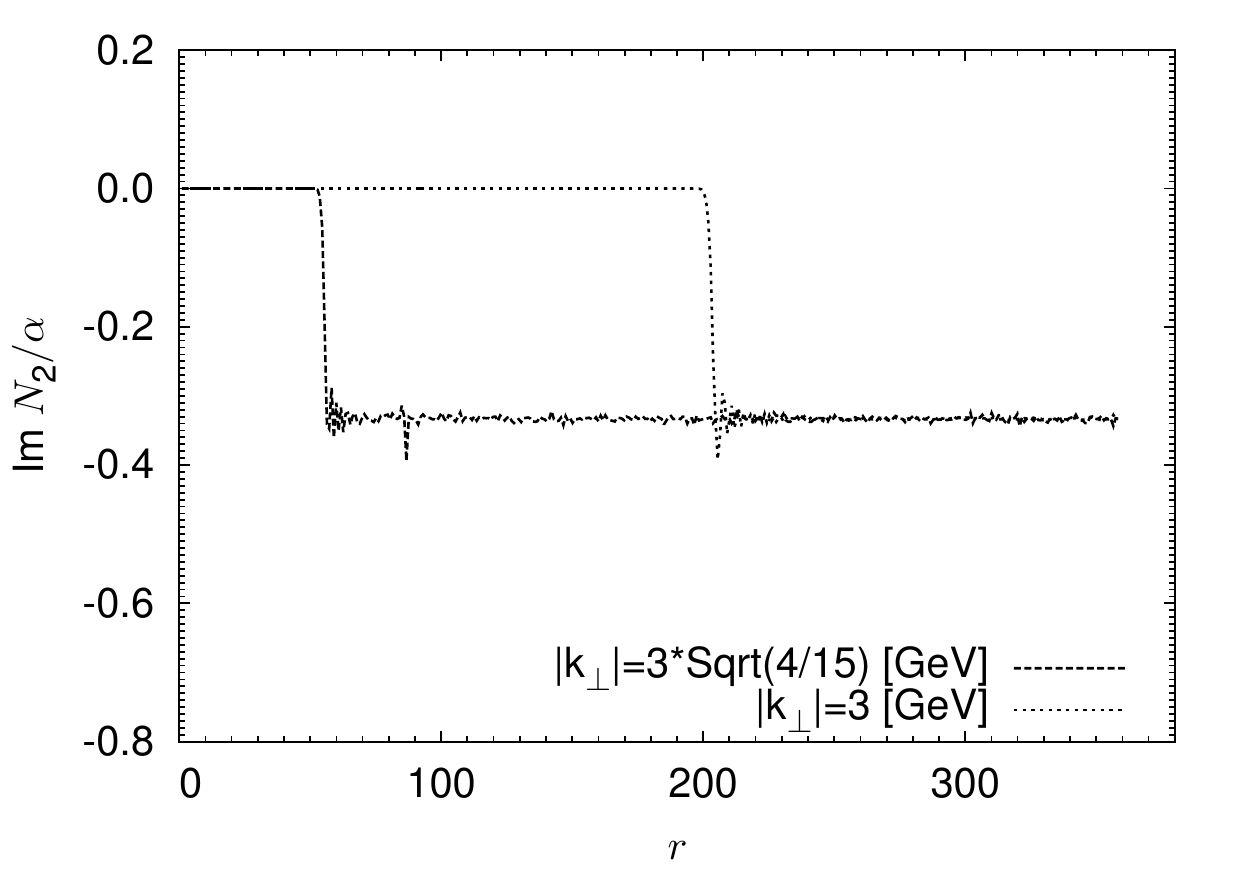}
}
\vspace*{-4pt}
\caption{Form factor $N_2$ for muons (case [c-1]) with $\ell_{\mathrm{max}}=1000$
(top ($r<1$), middle (real part in $1<r$) and bottom (imaginary part in $1<r$).}
\label{fig:N2muon01mpi}
\end{figure}

\begin{figure}[ht]
\centerline{
\includegraphics[width=\figscale]{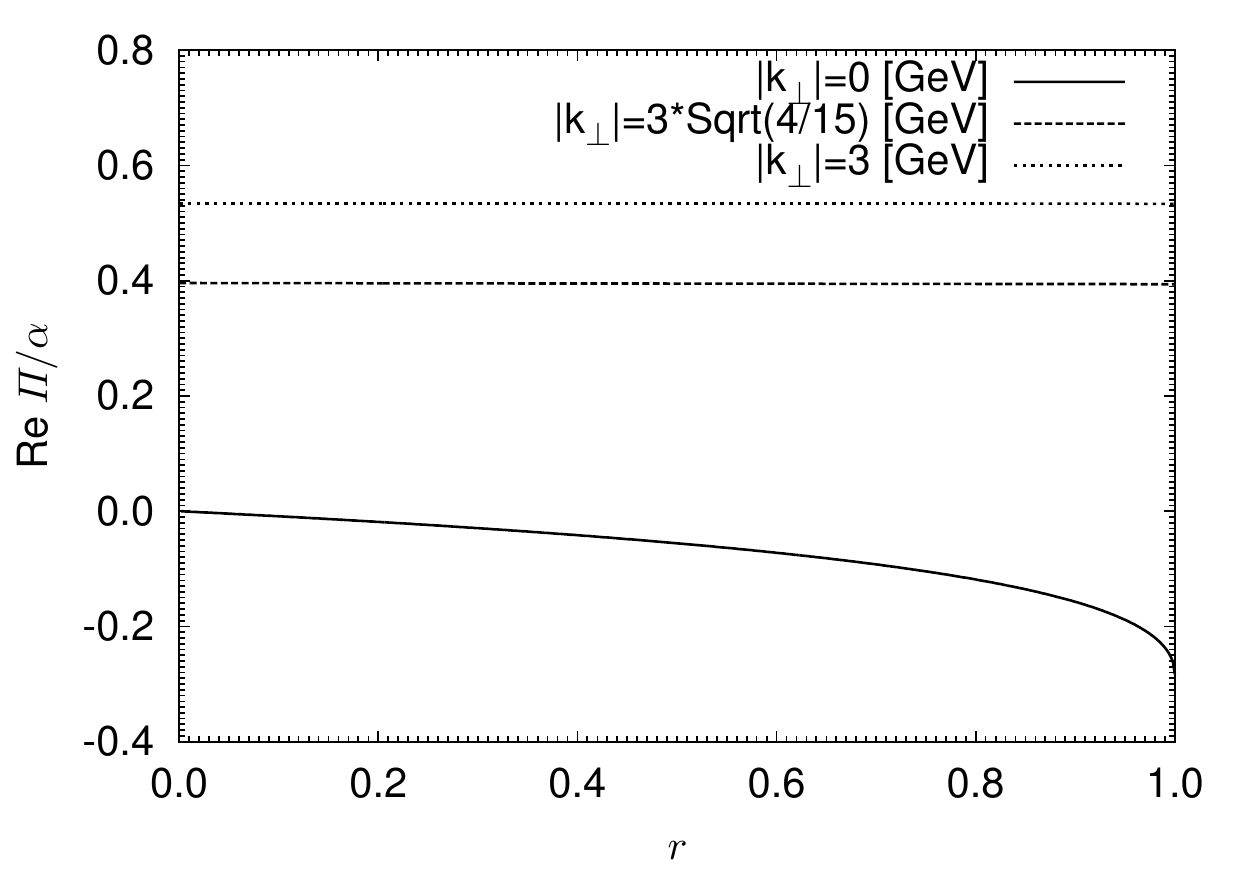}
}
\centerline{
\includegraphics[width=\figscale]{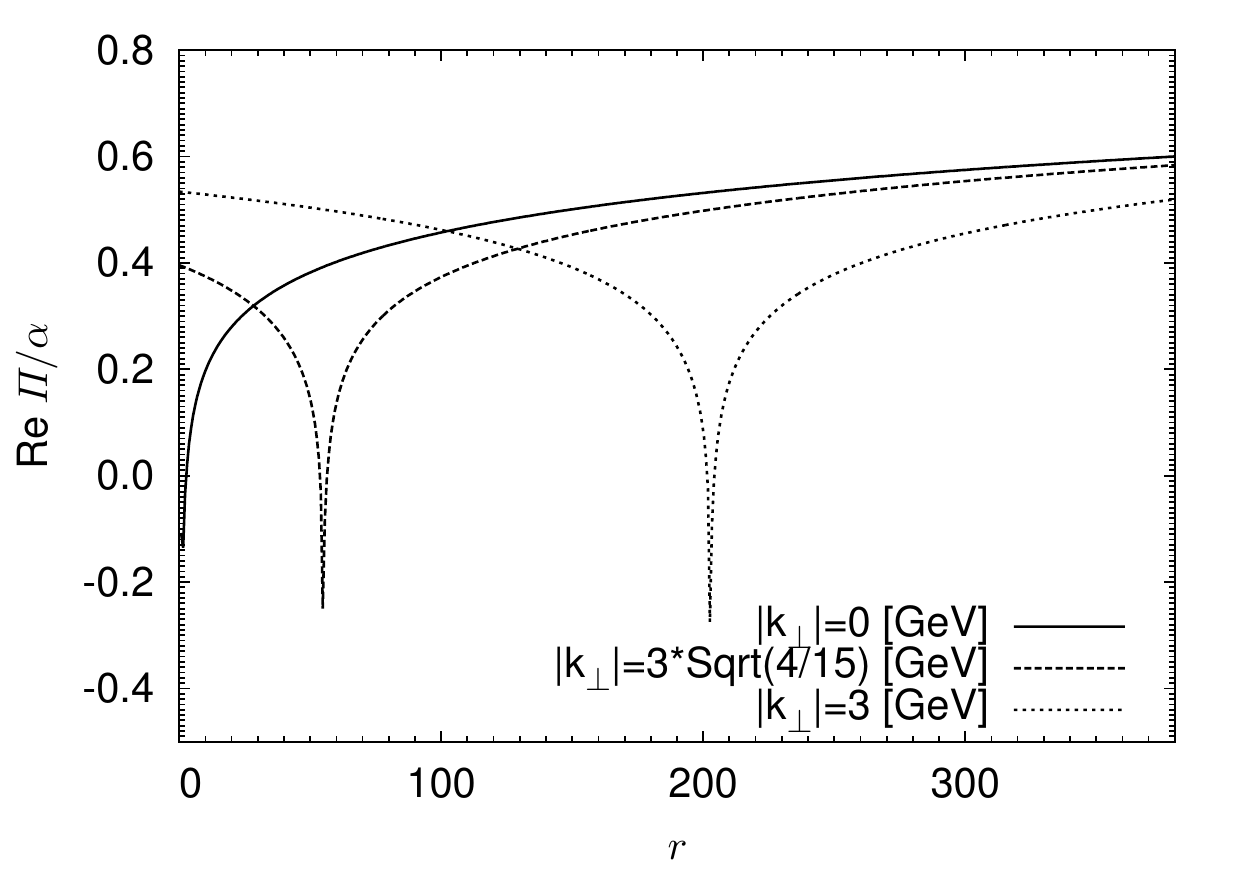}
\includegraphics[width=\figscale]{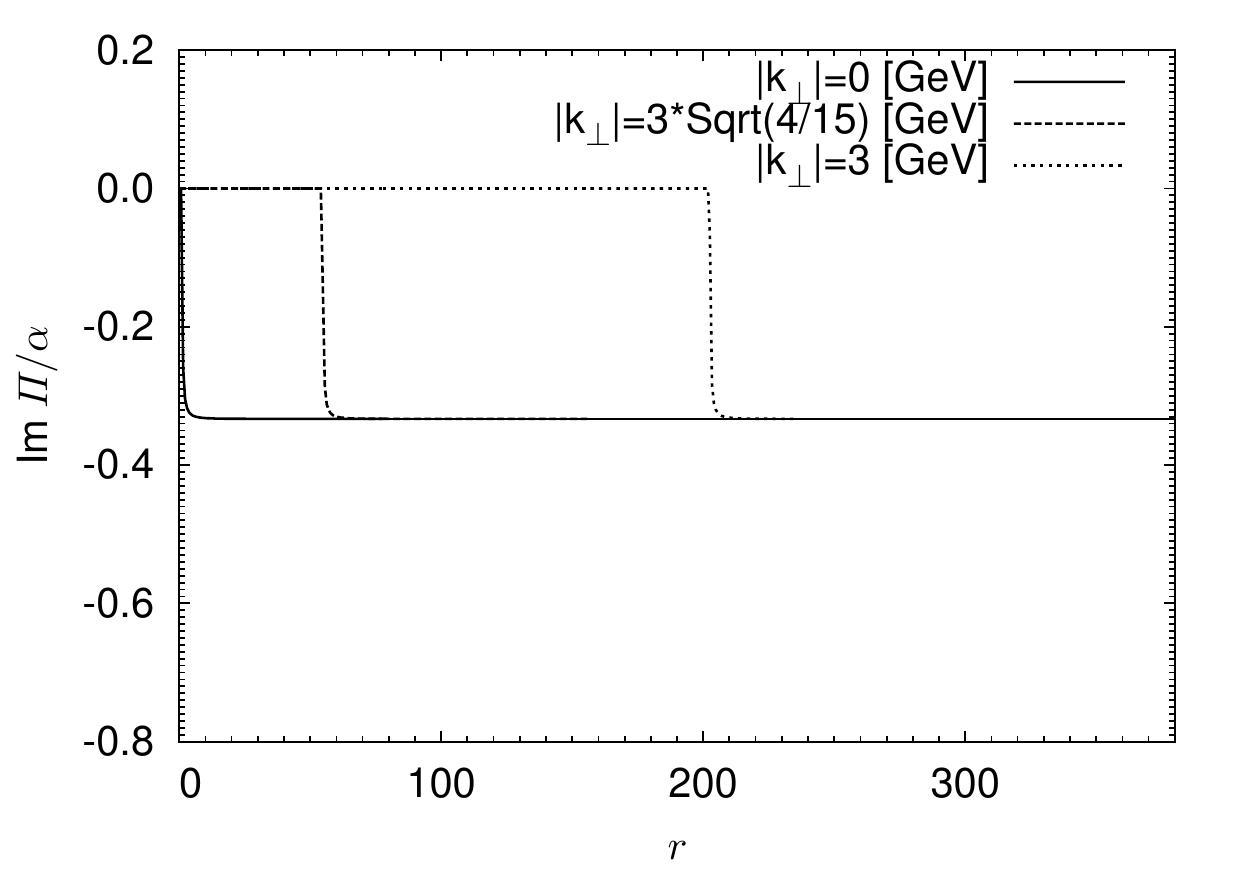}
}
\vspace*{-4pt}
\caption{Form factors $\Pi$, 
Eqs.(\ref{eq:PiBelowThZeroB}) and (\ref{eq:PiAboveThZeroB}), for muons
(top ($r<1$), bottom left (real part in $1<r$), and bottom right (imaginary part in $1<r$).}
\label{fig:PimuonZeroB}
\end{figure}

\begin{figure}[ht]
\centerline{
\includegraphics[width=\figscale]{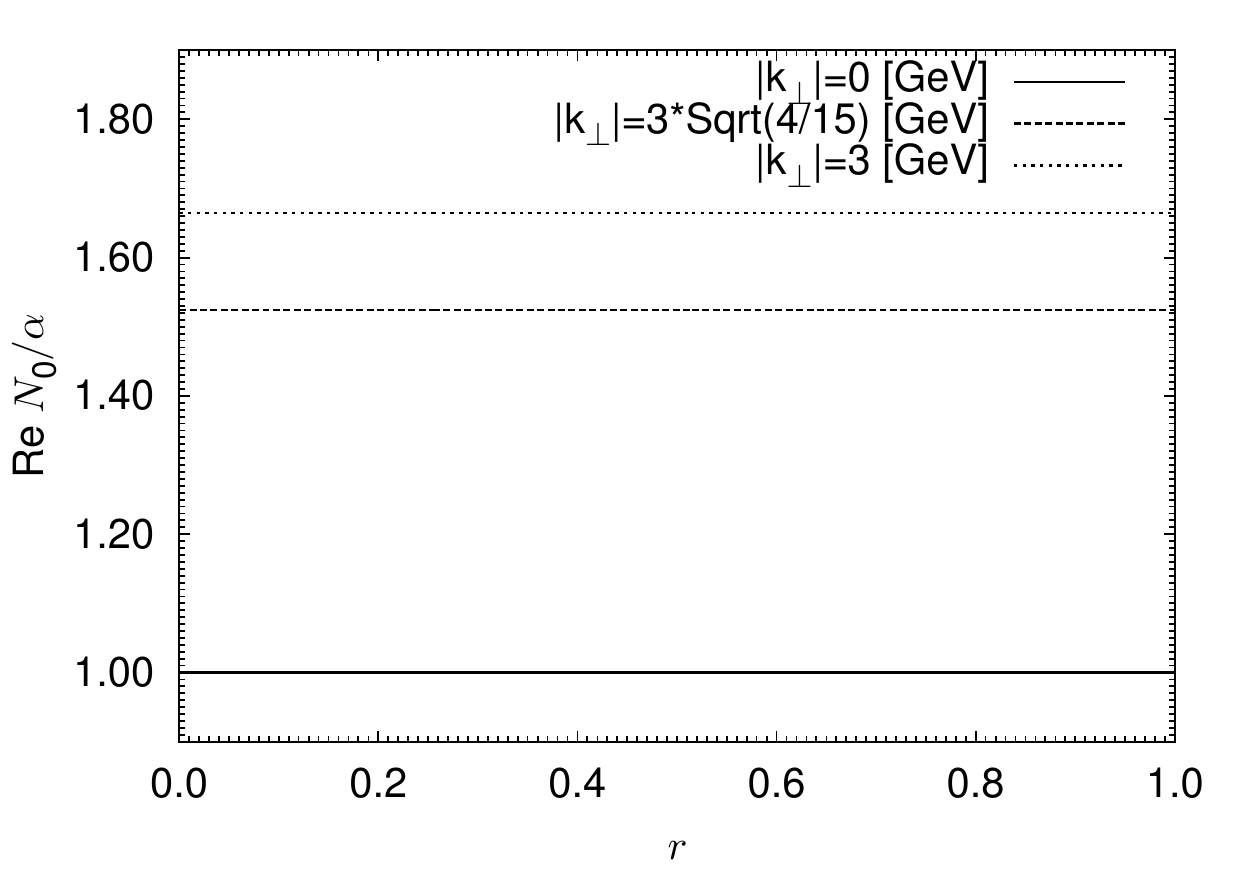}
\includegraphics[width=\figscale]{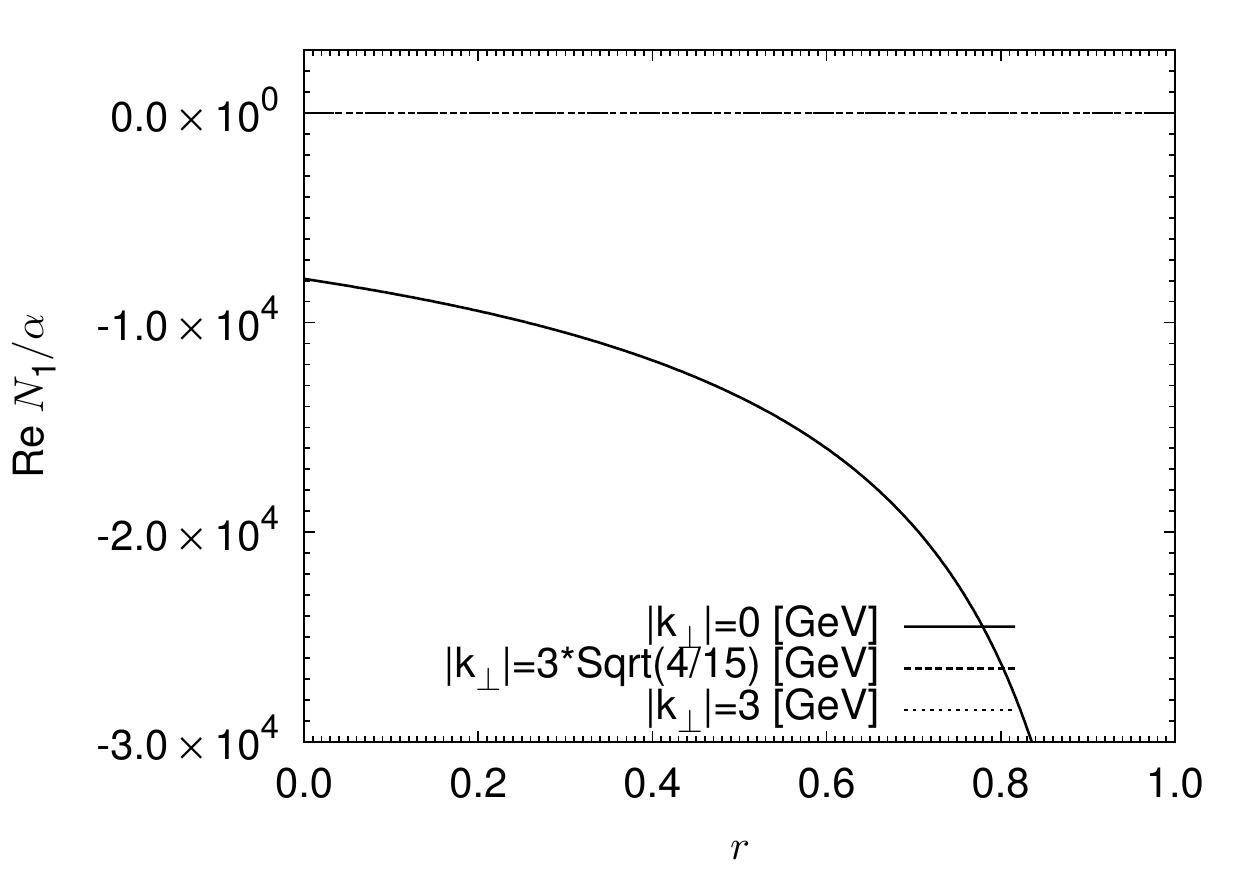}
}
\centerline{
\includegraphics[width=\figscale]{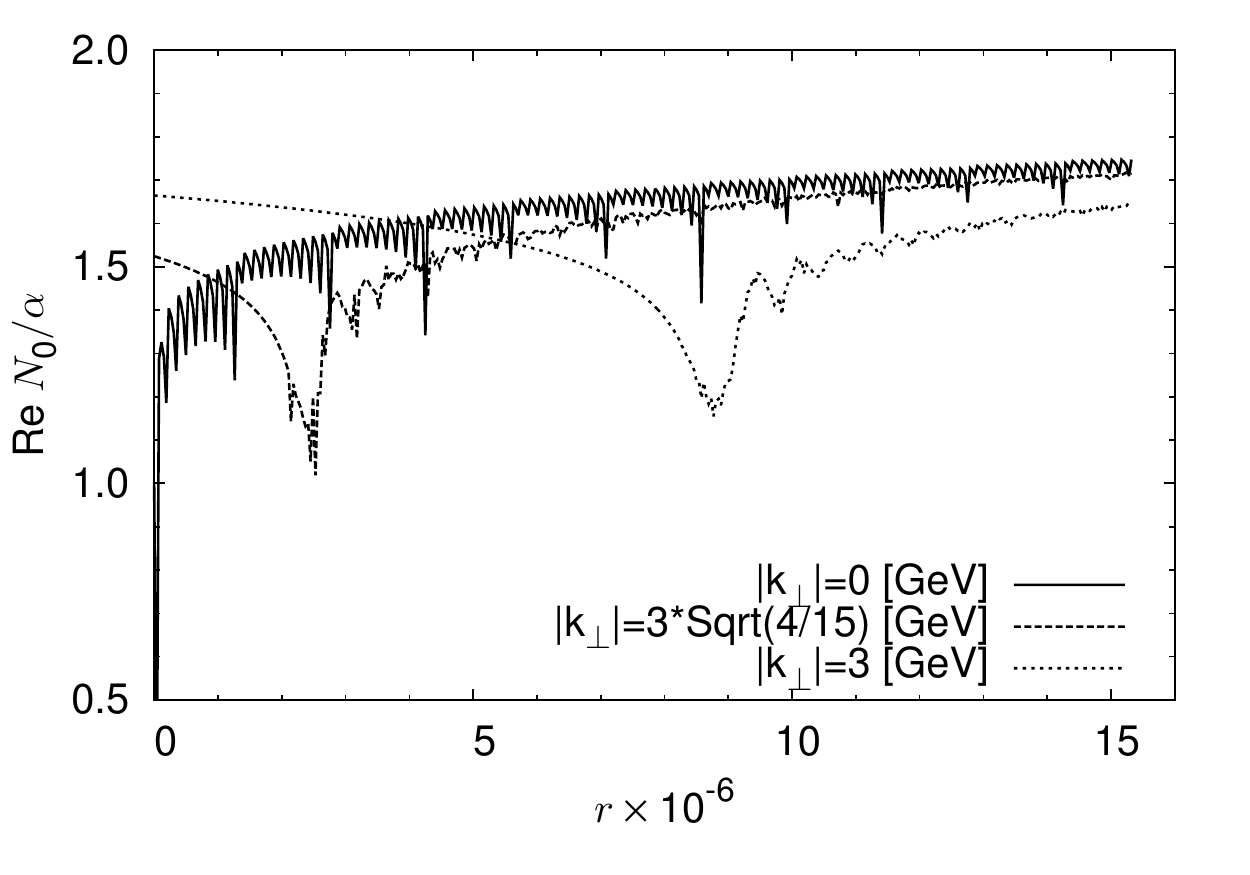}
\includegraphics[width=\figscale]{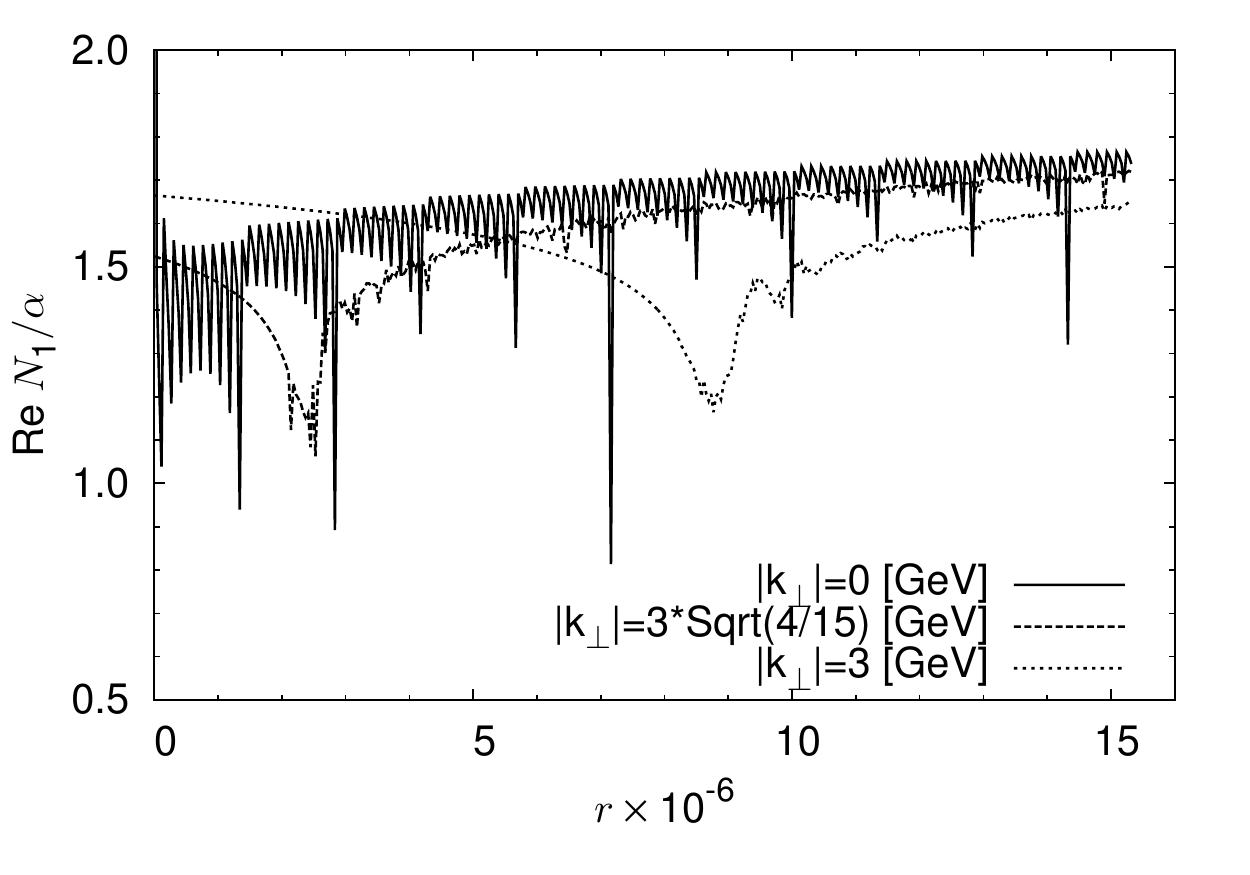}
}
\centerline{
\includegraphics[width=\figscale]{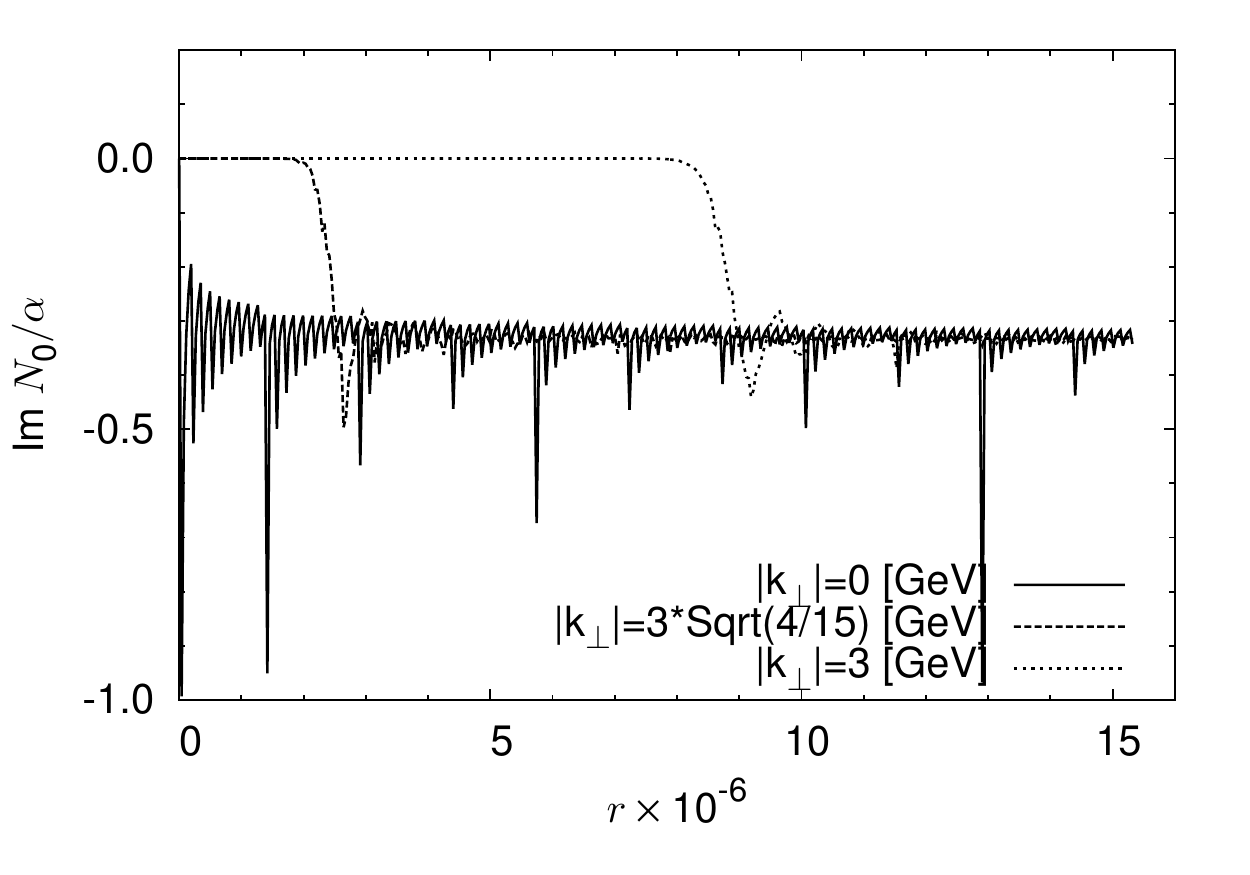}
\includegraphics[width=\figscale]{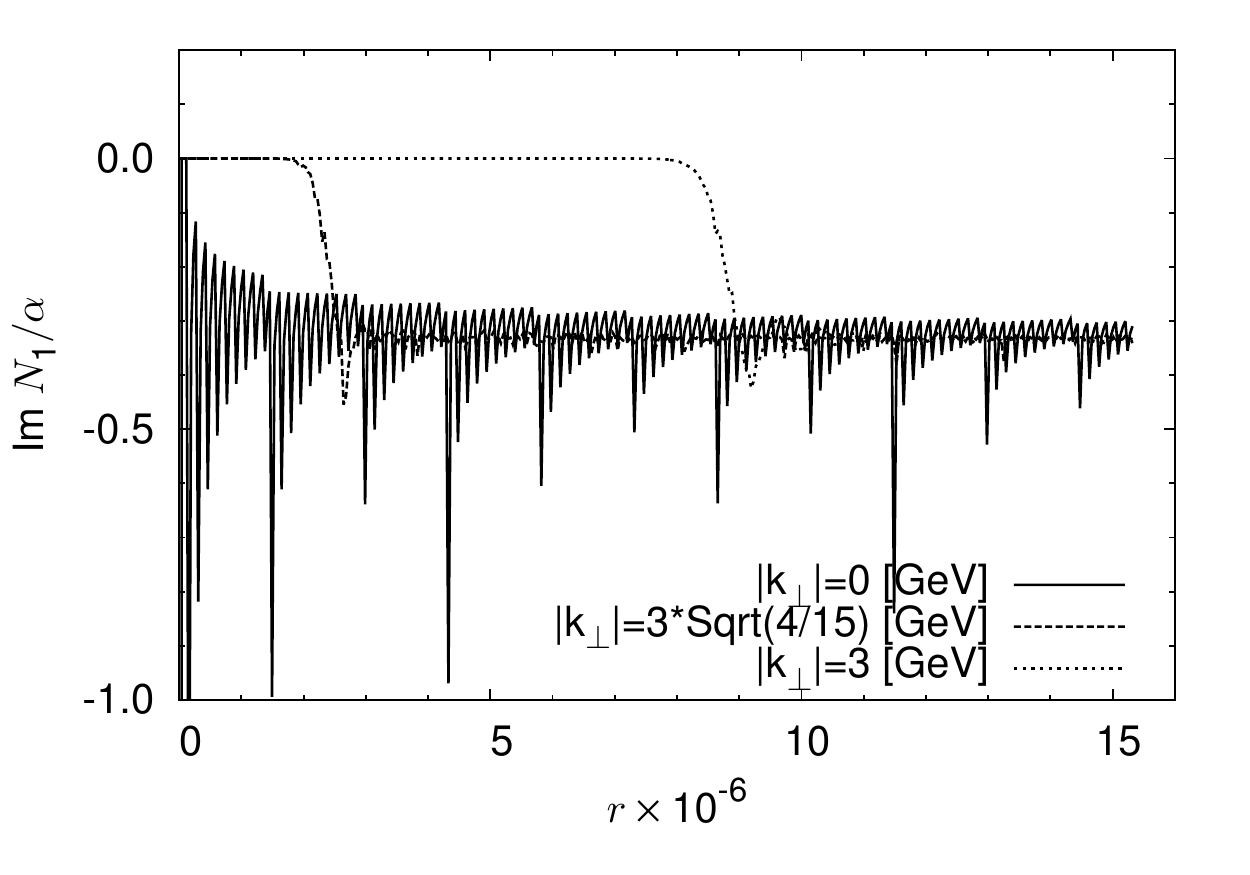}
}
\vspace*{-4pt}
\caption{Form factors $N_0$ (left) and $N_1$ (right) for electrons (case [b-2])
with $\ell_{\mathrm{max}}=1000$
(top ($r<1$), middle (real part in $1<r$), and bottom (imaginary part in $1<r$).}
\label{fig:N0N1electronmpi}
\end{figure}

\begin{figure}[ht]
\centerline{
\includegraphics[width=\figscale]{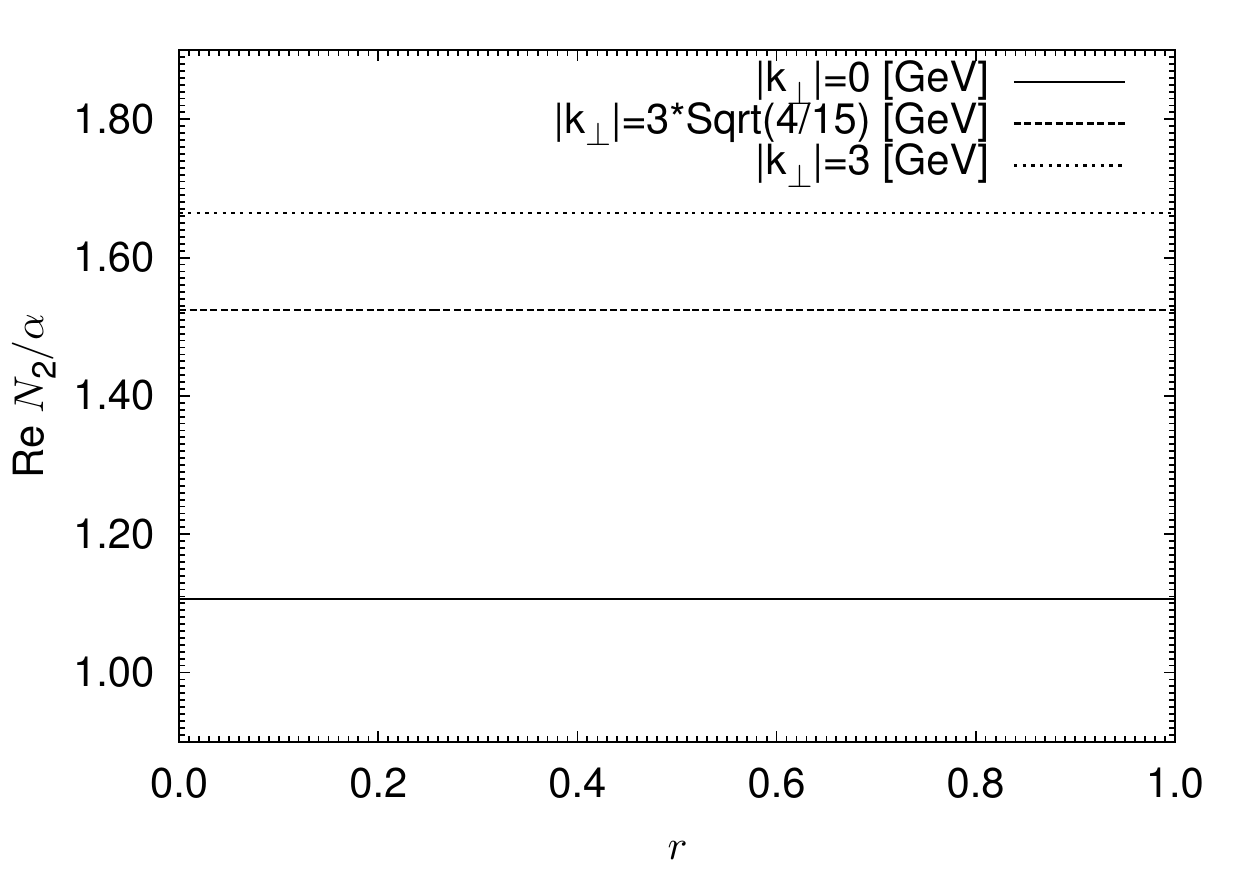}
}
\centerline{
\includegraphics[width=\figscale]{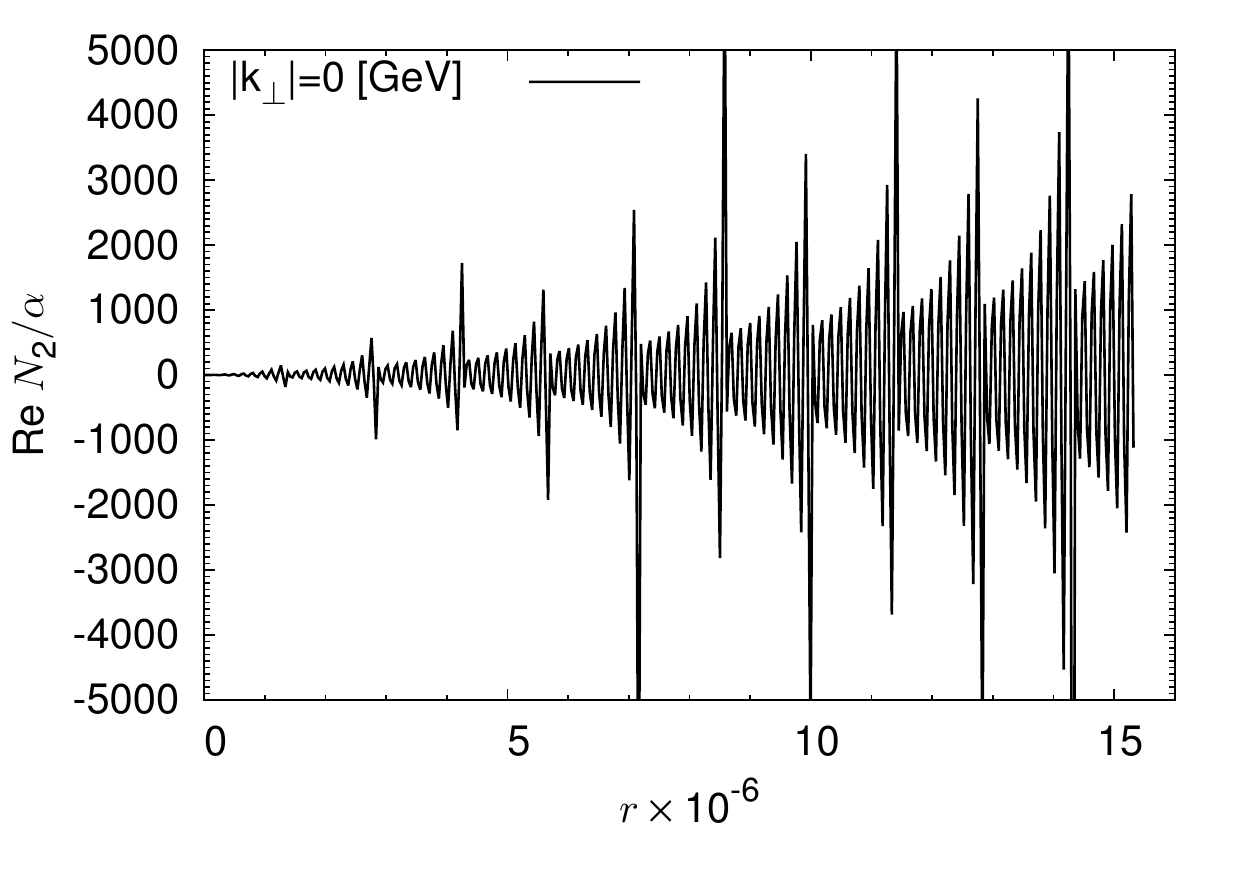}
\includegraphics[width=\figscale]{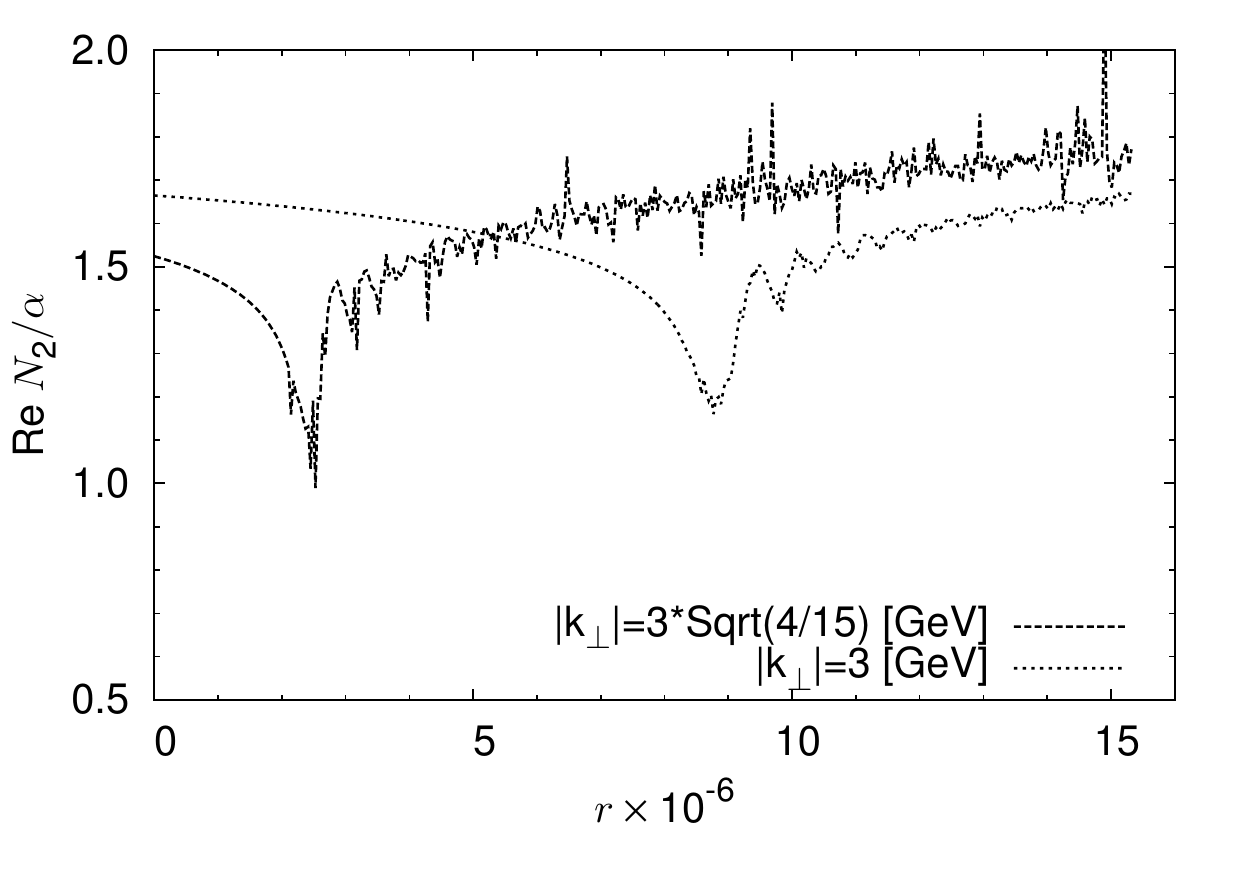}
}
\centerline{
\includegraphics[width=\figscale]{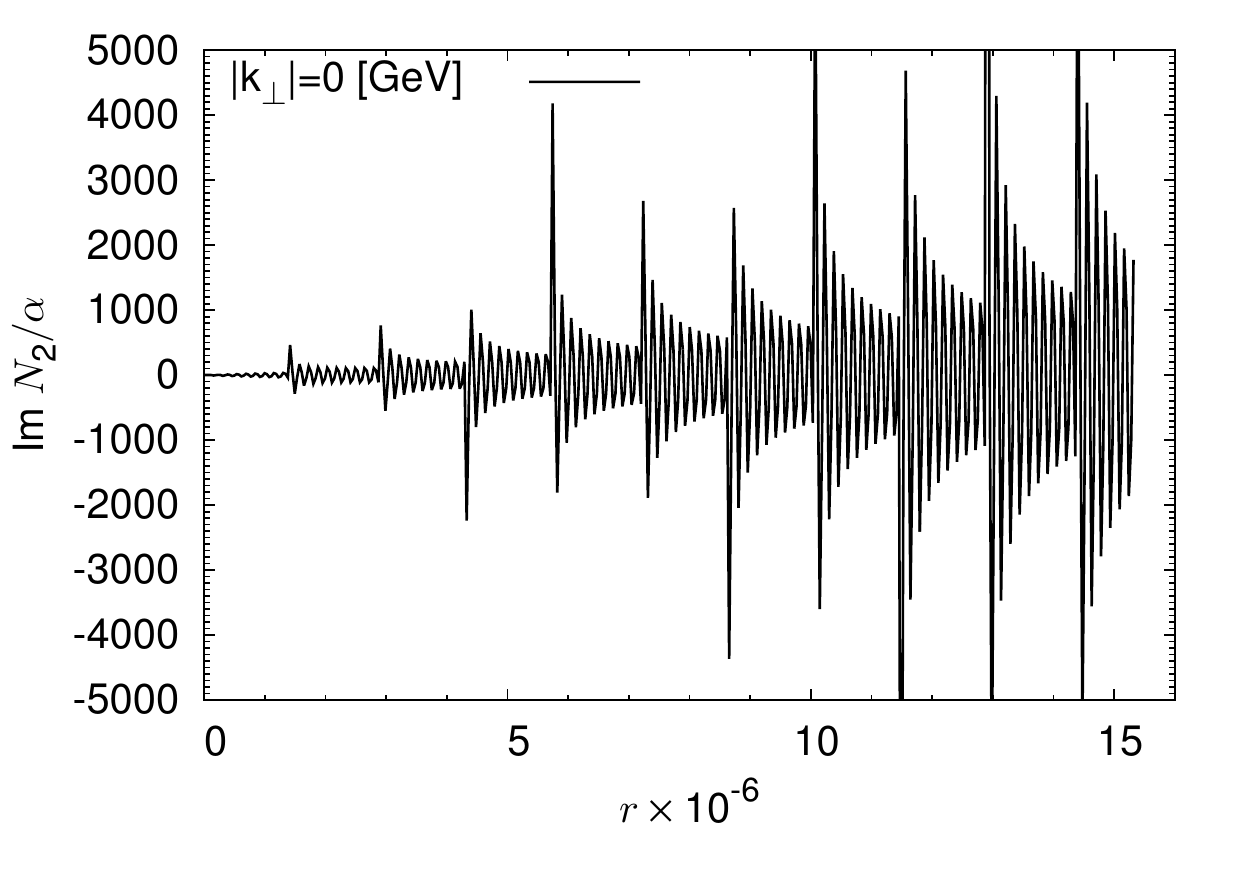}
\includegraphics[width=\figscale]{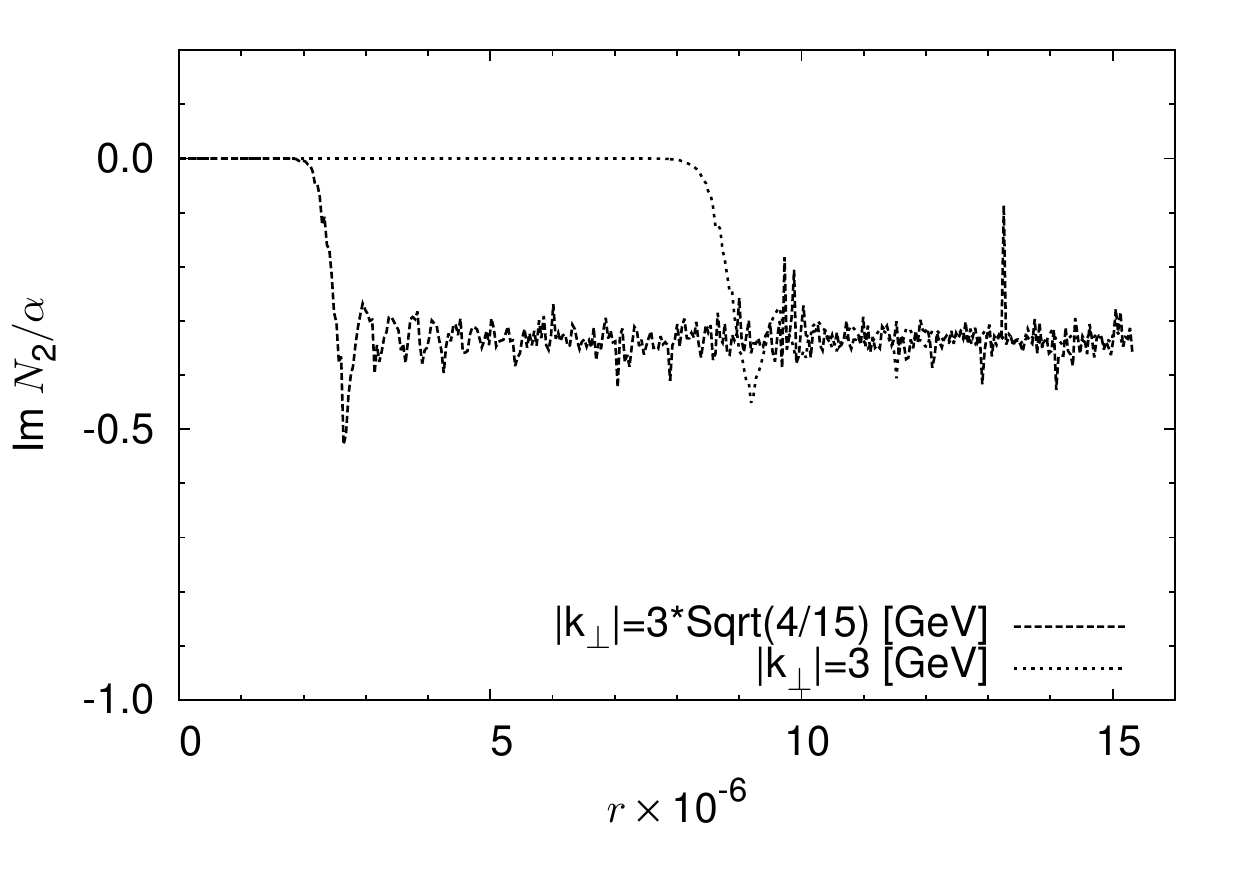}
}
\vspace*{-4pt}
\caption{Form factor $N_2$ for electrons (case [b-2]) with $\ell_{\mathrm{max}}=1000$
(top ($r<1$), middle (real part in $1<r$) and bottom (imaginary part in $1<r$).}
\label{fig:N2electronmpi}
\end{figure}

\begin{figure}[ht]
\centerline{
\includegraphics[width=\figscale]{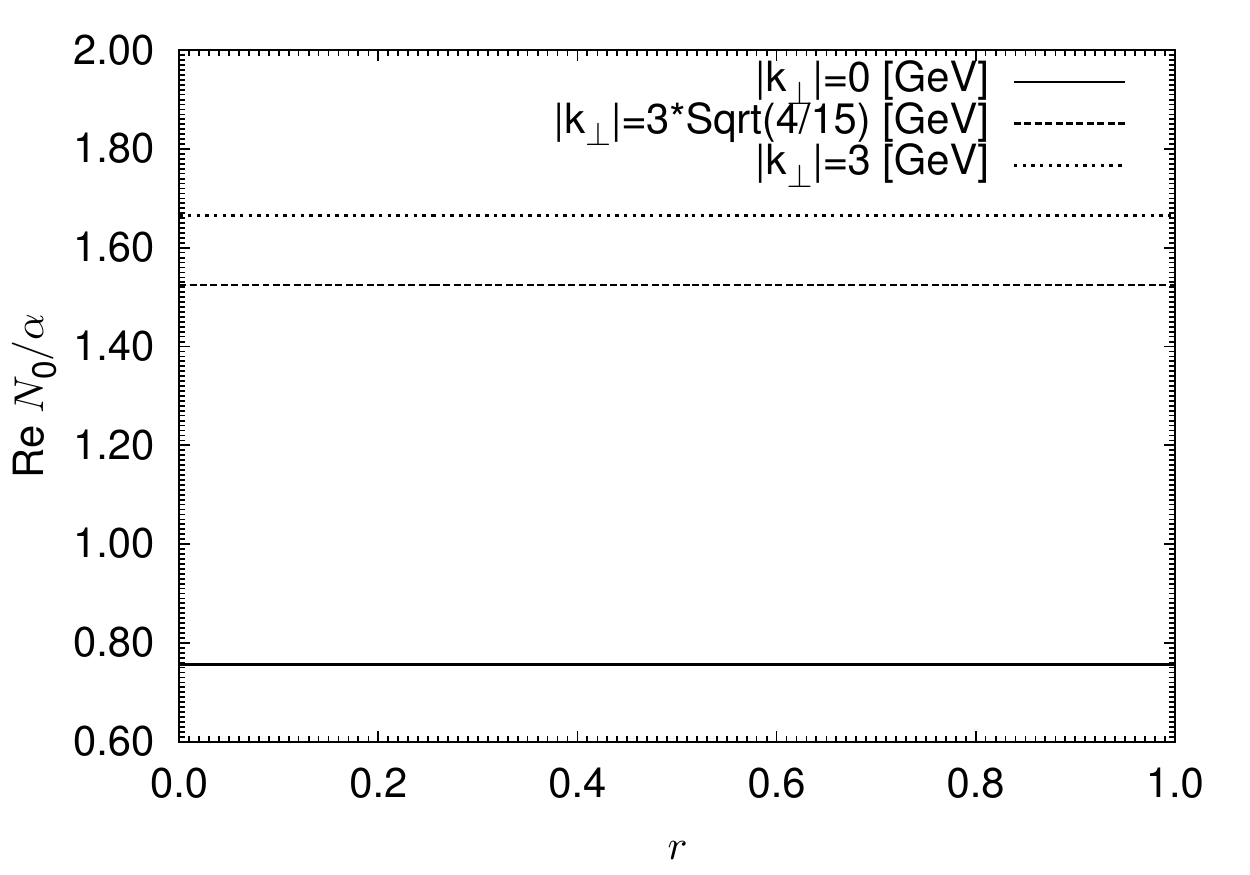}
\includegraphics[width=\figscale]{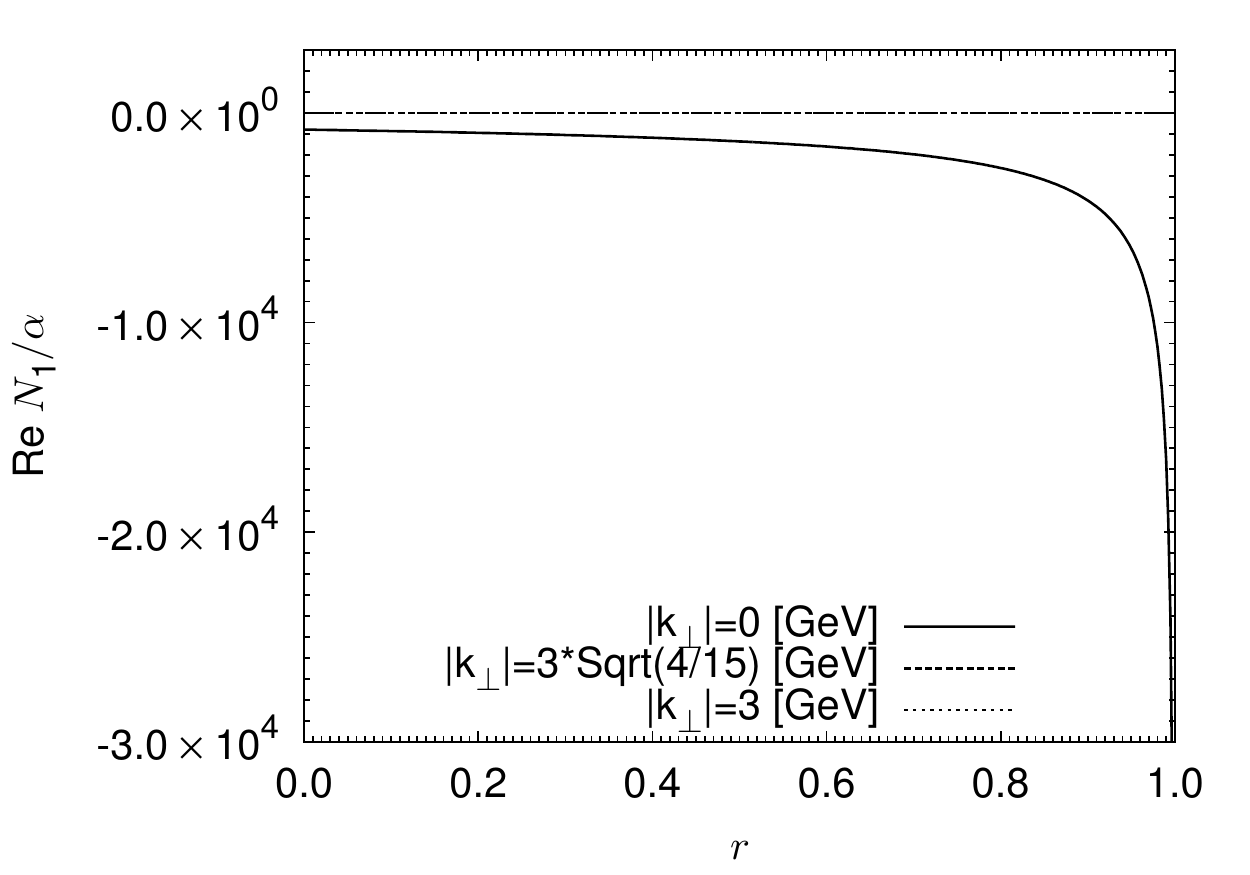}
}
\centerline{
\includegraphics[width=\figscale]{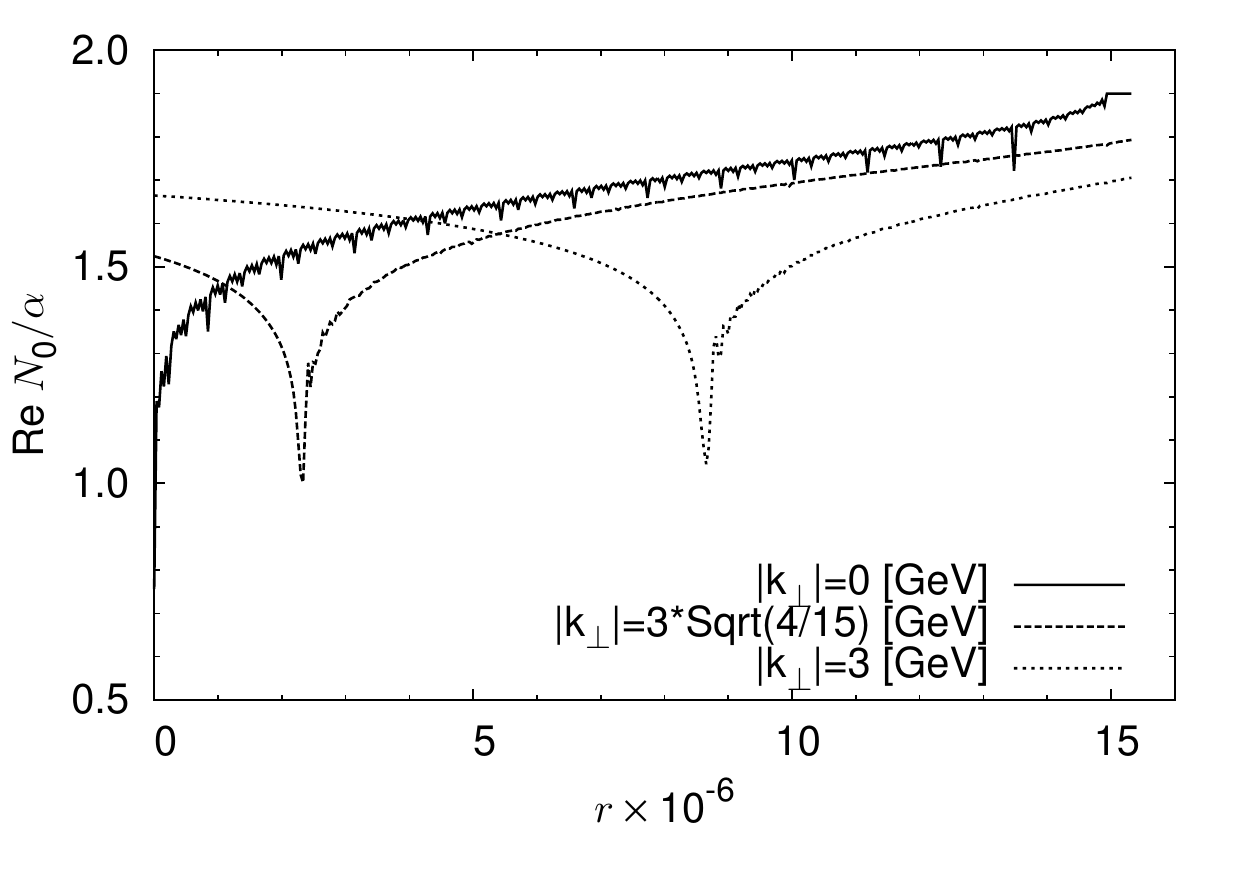}
\includegraphics[width=\figscale]{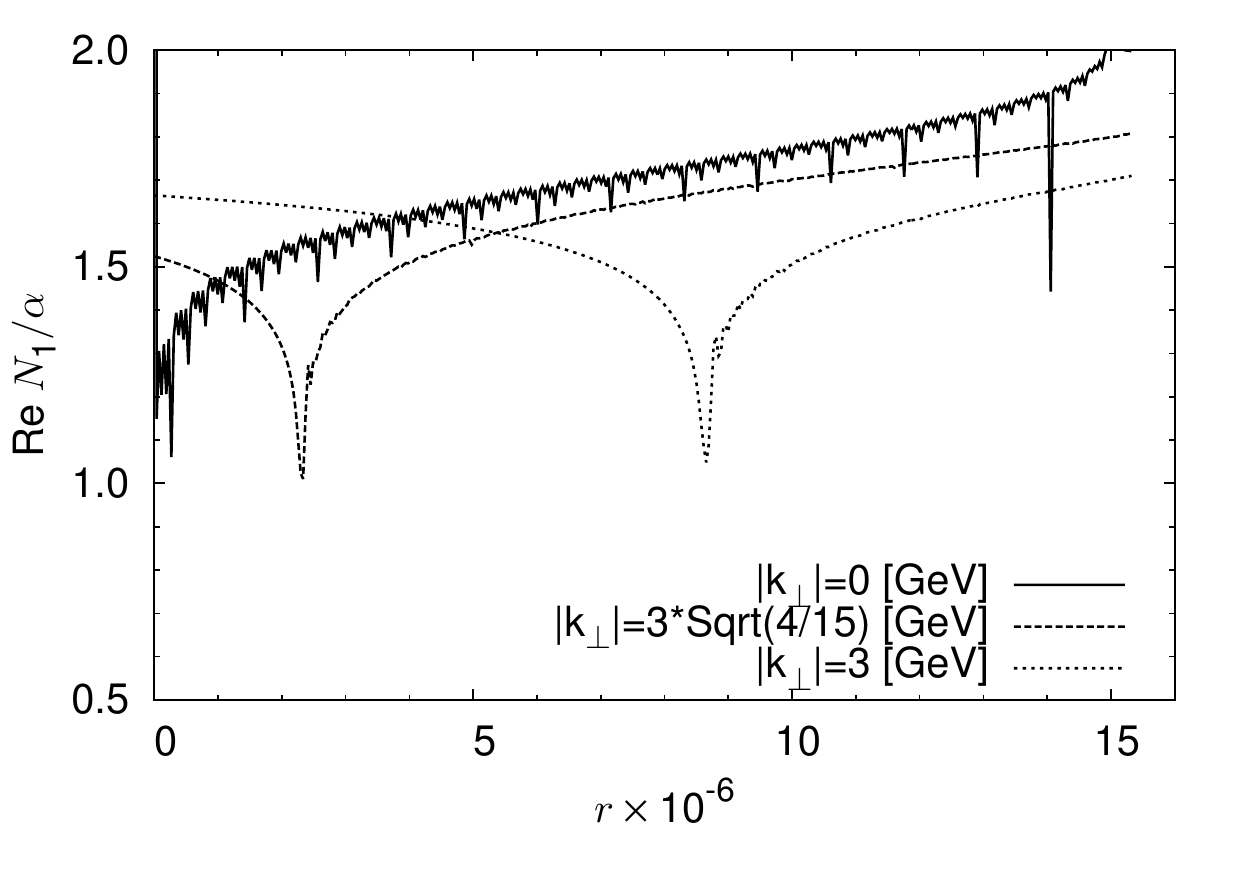}
}
\centerline{
\includegraphics[width=\figscale]{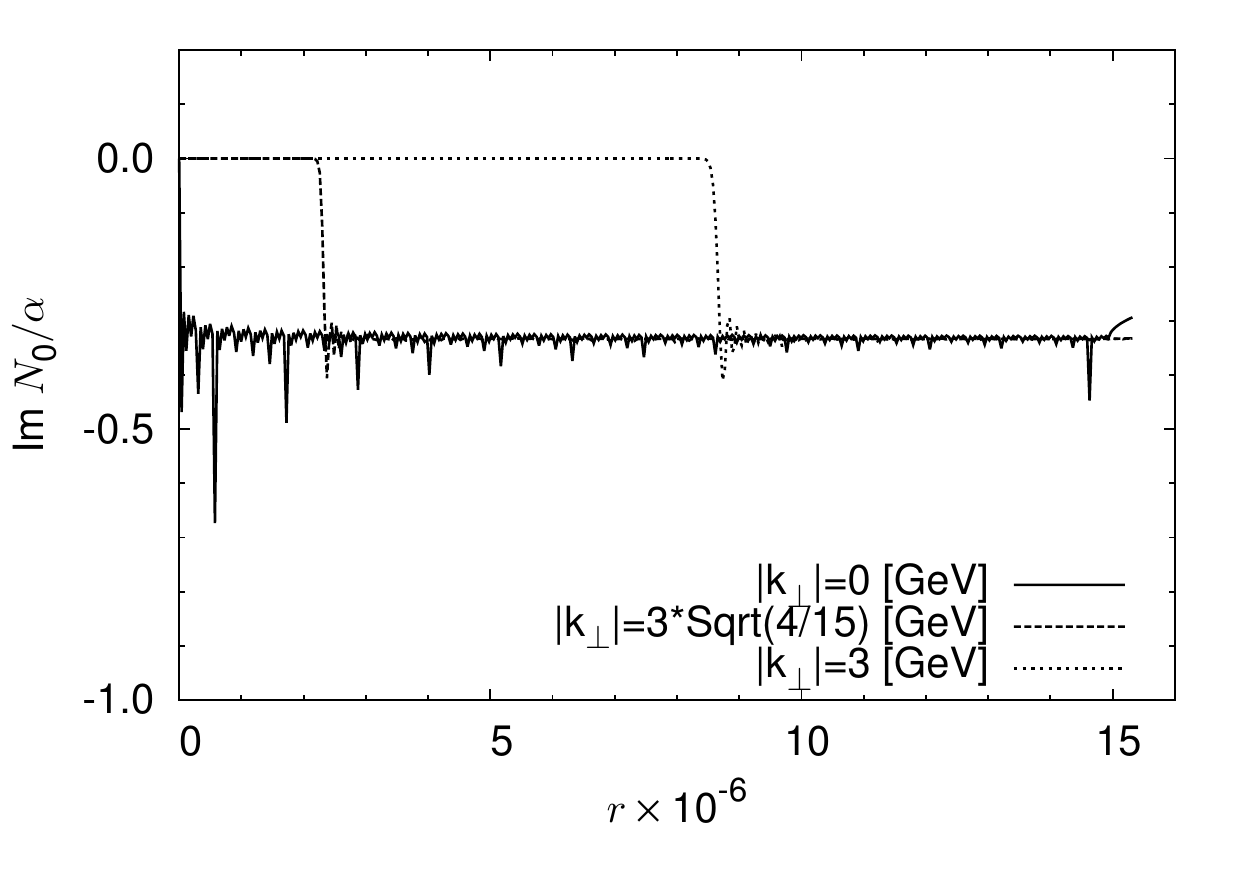}
\includegraphics[width=\figscale]{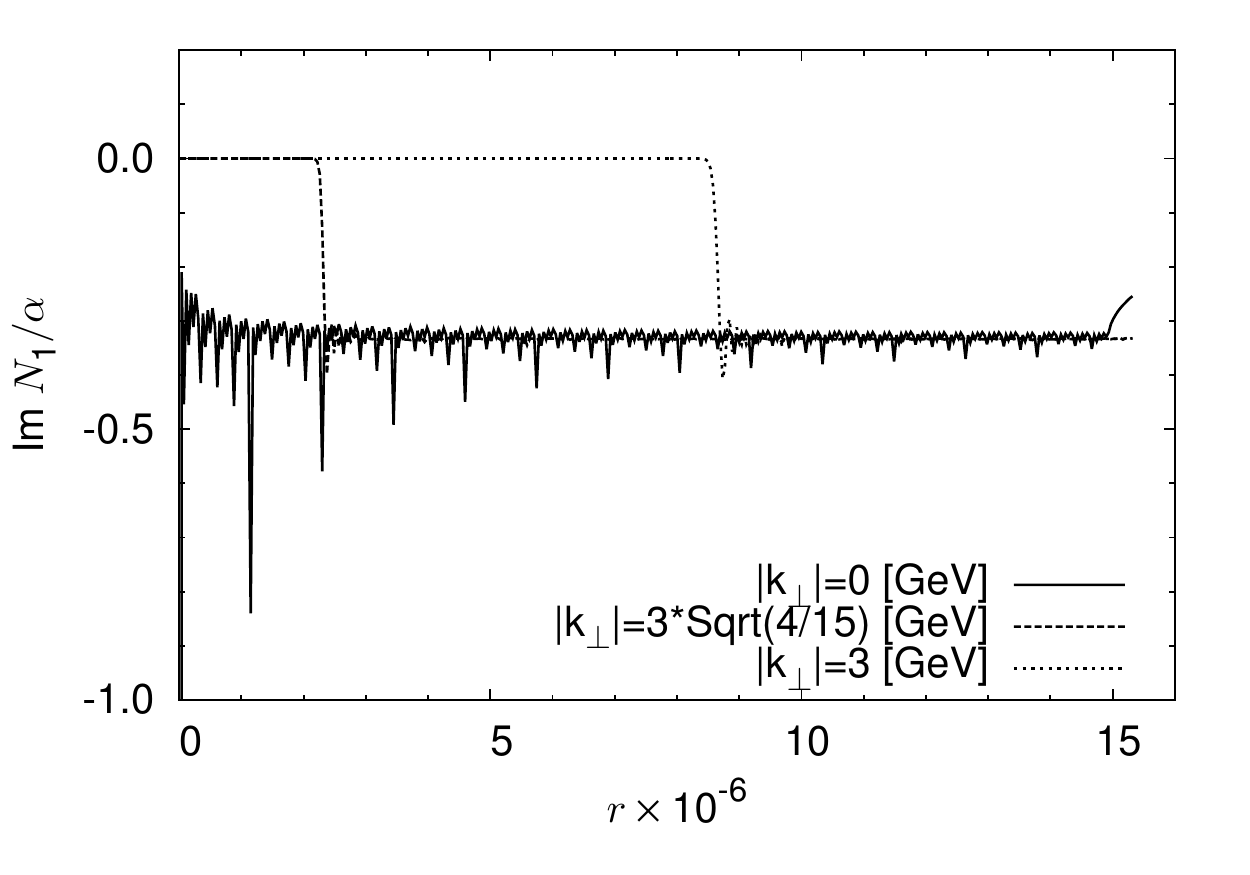}
}
\vspace*{-4pt}
\caption{Form factors $N_0$ (left) and $N_1$ (right) for electrons (case [c-2])
with $\ell_{\mathrm{max}}=1000$
(top ($r<1$), middle (real part in $1<r$), and bottom (imaginary part in $1<r$).}
\label{fig:N0N1electron01mpi}
\end{figure}

\begin{figure}[ht]
\centerline{
\includegraphics[width=\figscale]{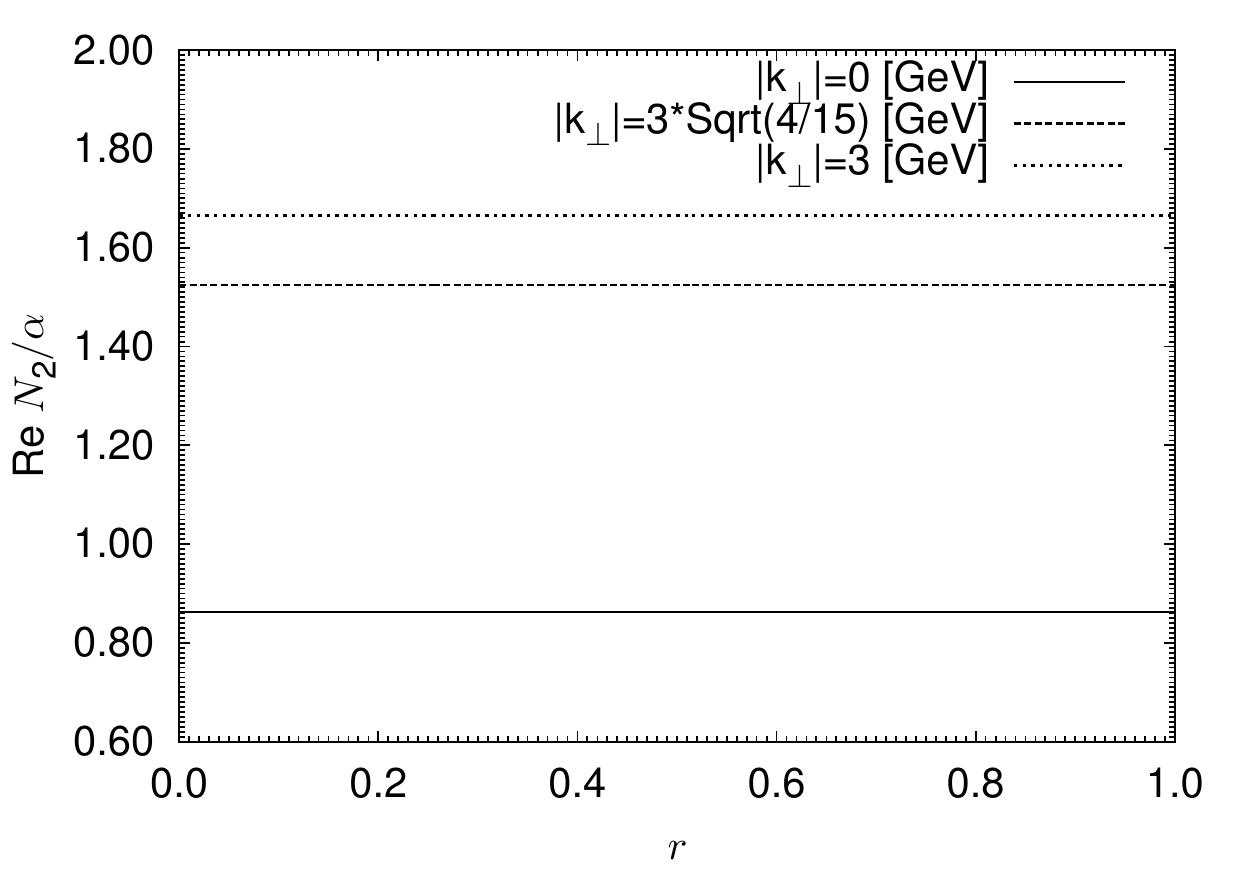}
}
\centerline{
\includegraphics[width=\figscale]{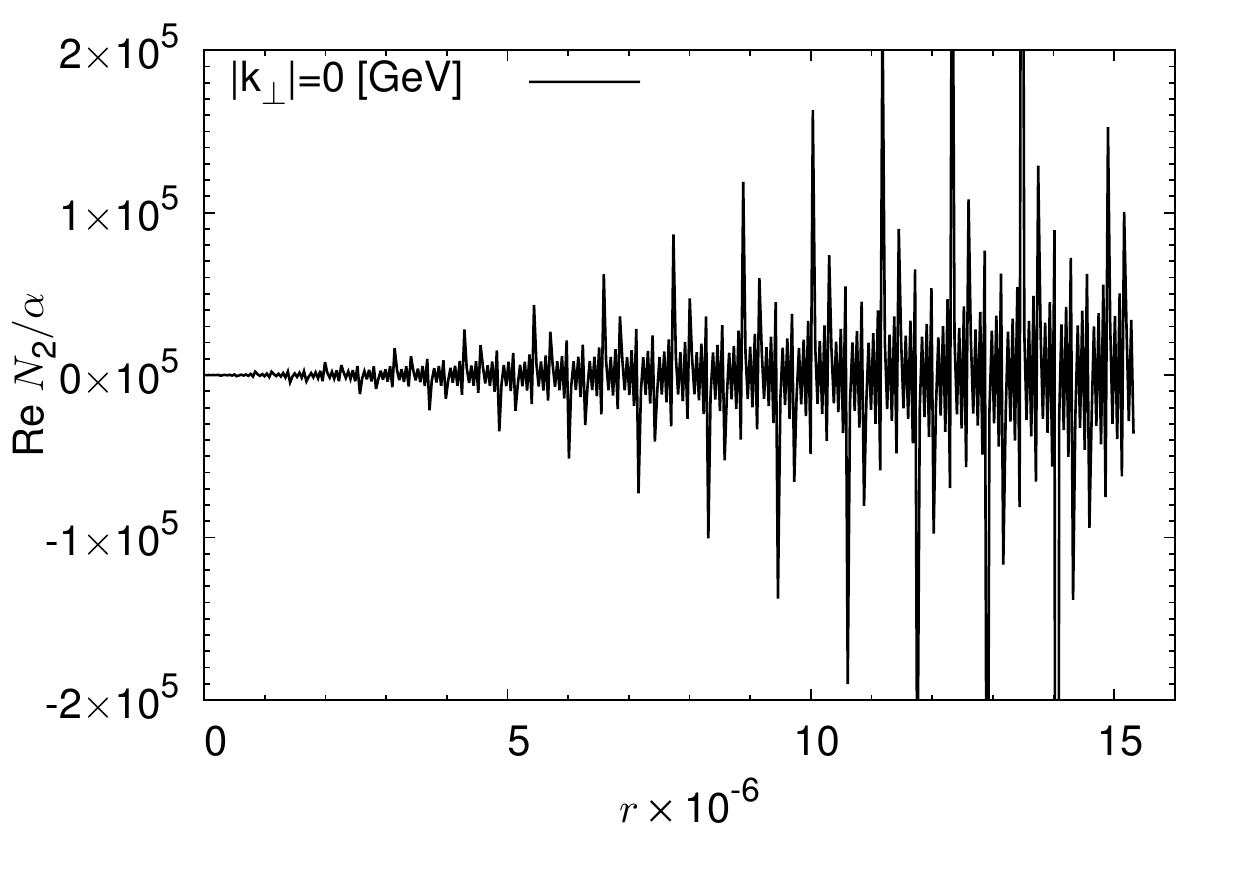}
\includegraphics[width=\figscale]{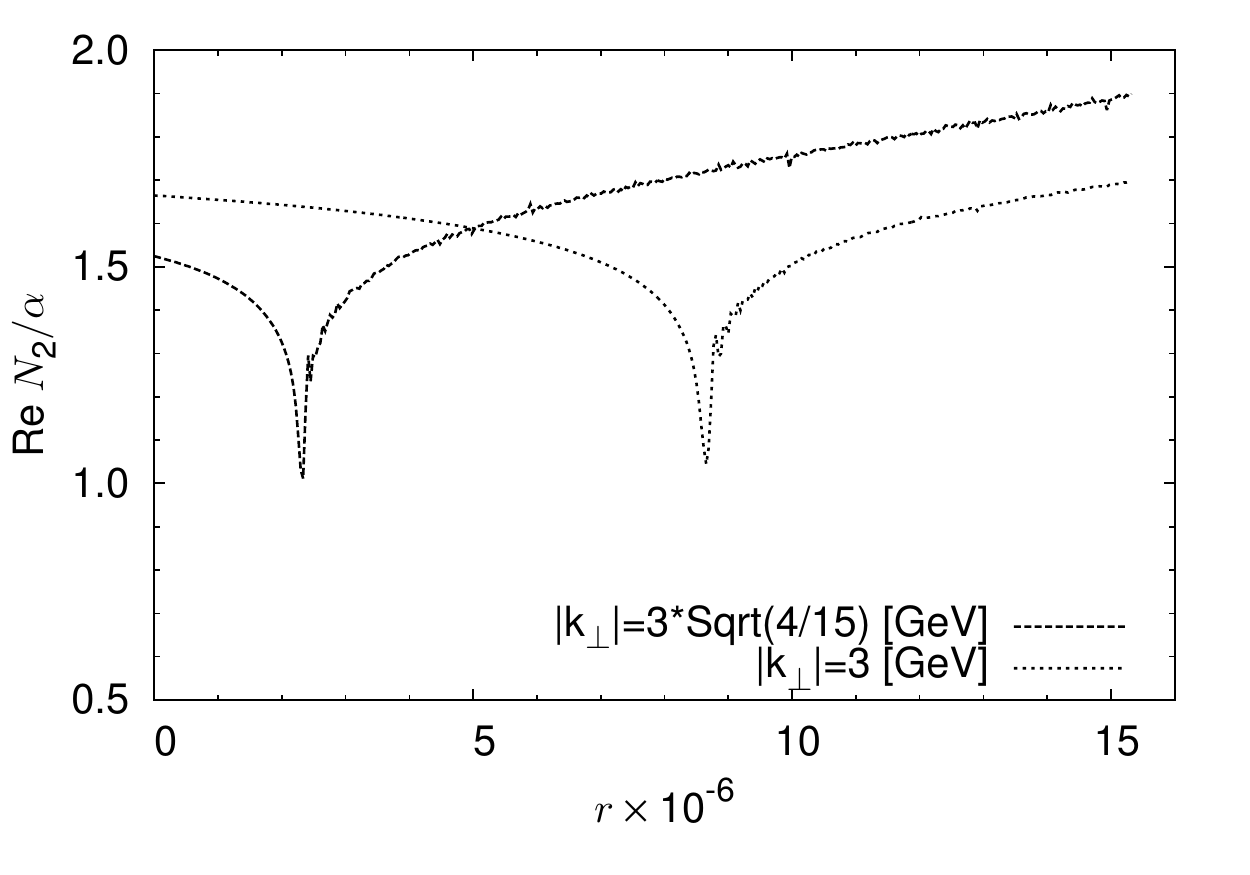}
}
\centerline{
\includegraphics[width=\figscale]{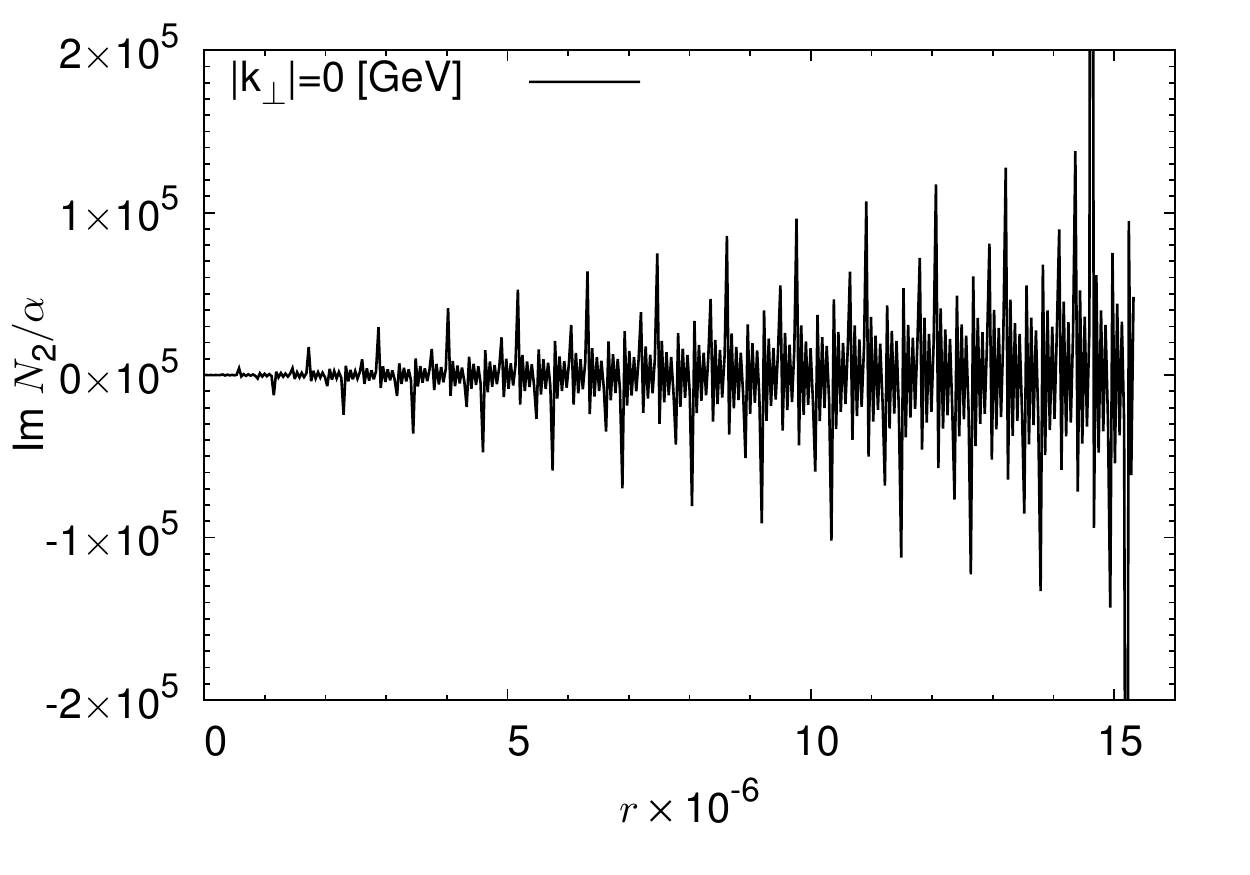}
\includegraphics[width=\figscale]{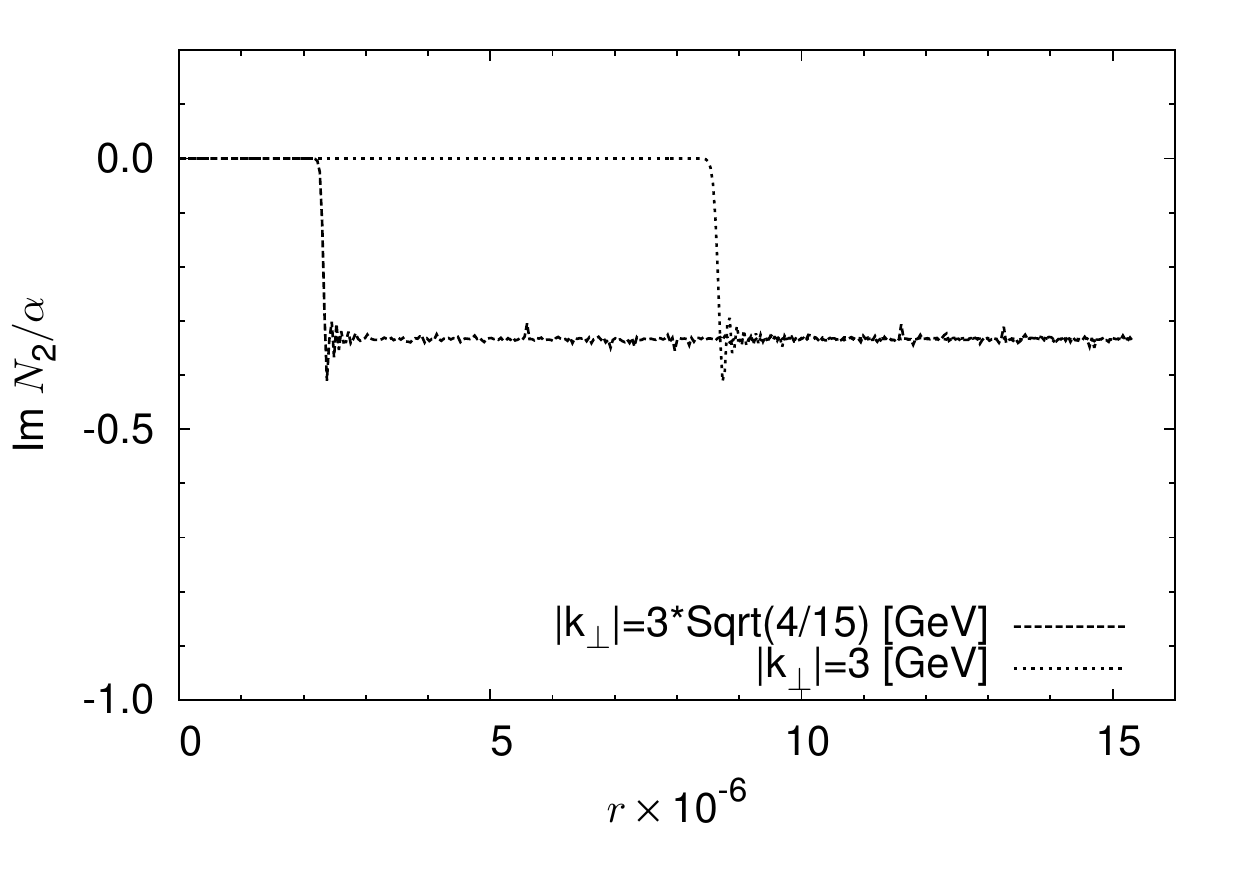}
}
\vspace*{-4pt}
\caption{Form factor $N_2$ for electrons (case [c-2]) with $\ell_{\mathrm{max}}=1000$
(top ($r<1$), middle (real part in $1<r$) and bottom (imaginary part in $1<r$).}
\label{fig:N2electron01mpi}
\end{figure}

\begin{figure}[ht]
\centerline{
\includegraphics[width=\figscale]{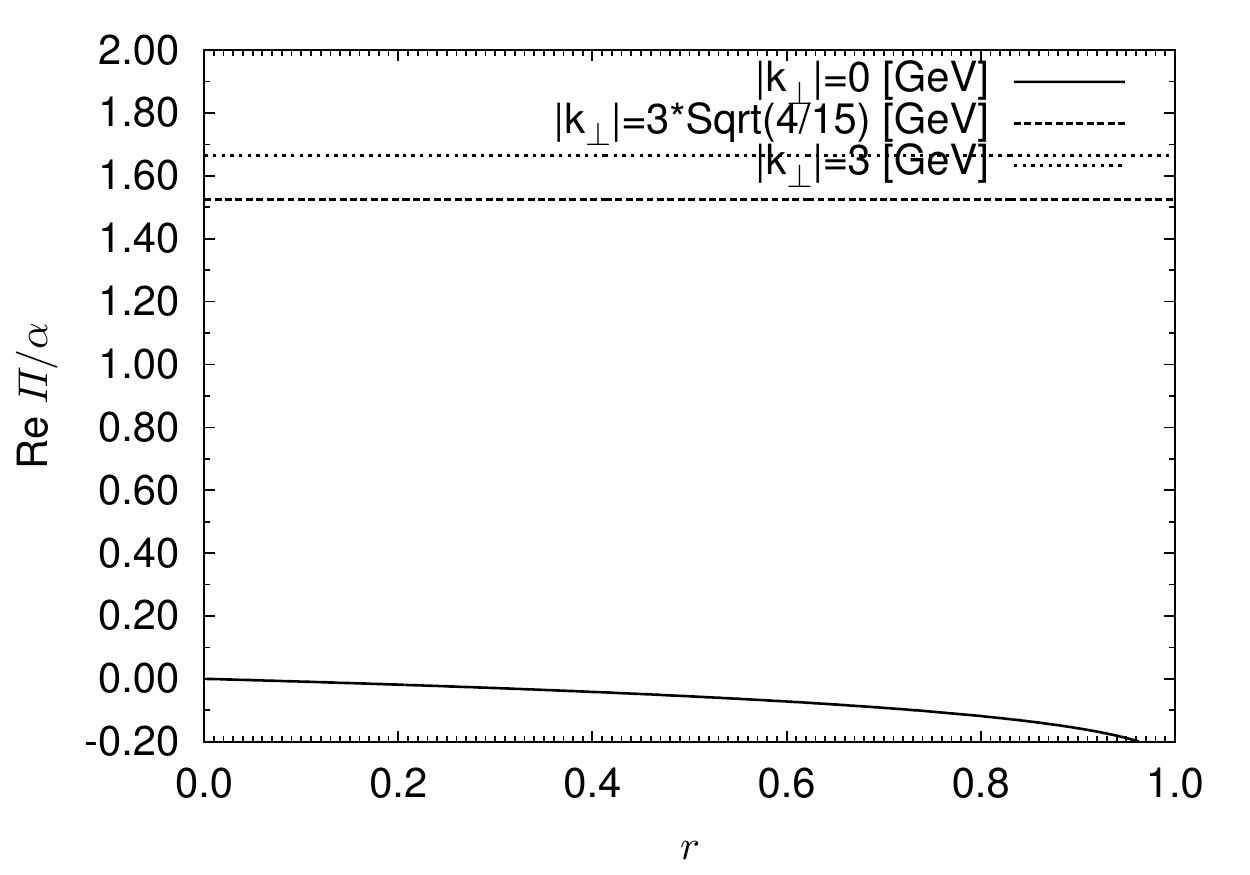}
}
\centerline{
\includegraphics[width=\figscale]{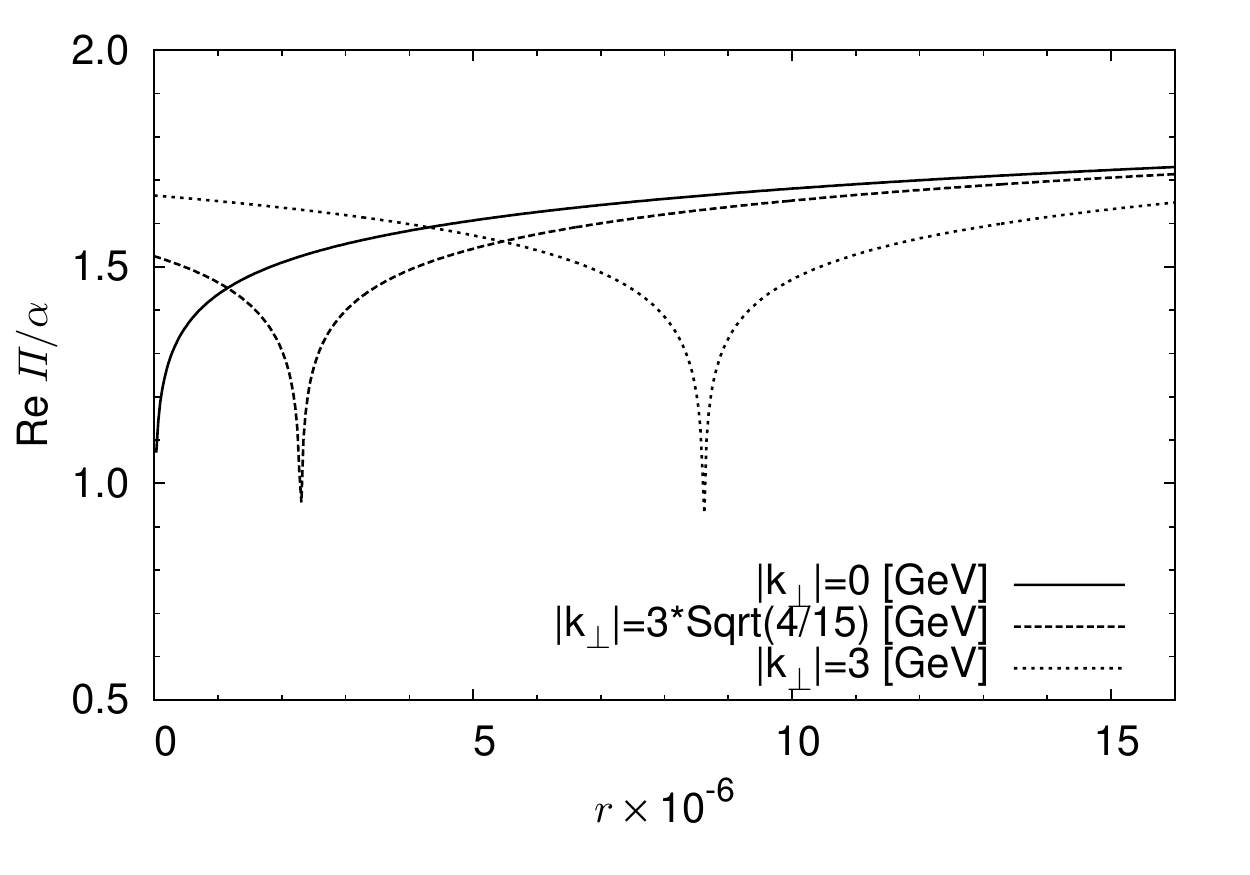}
\includegraphics[width=\figscale]{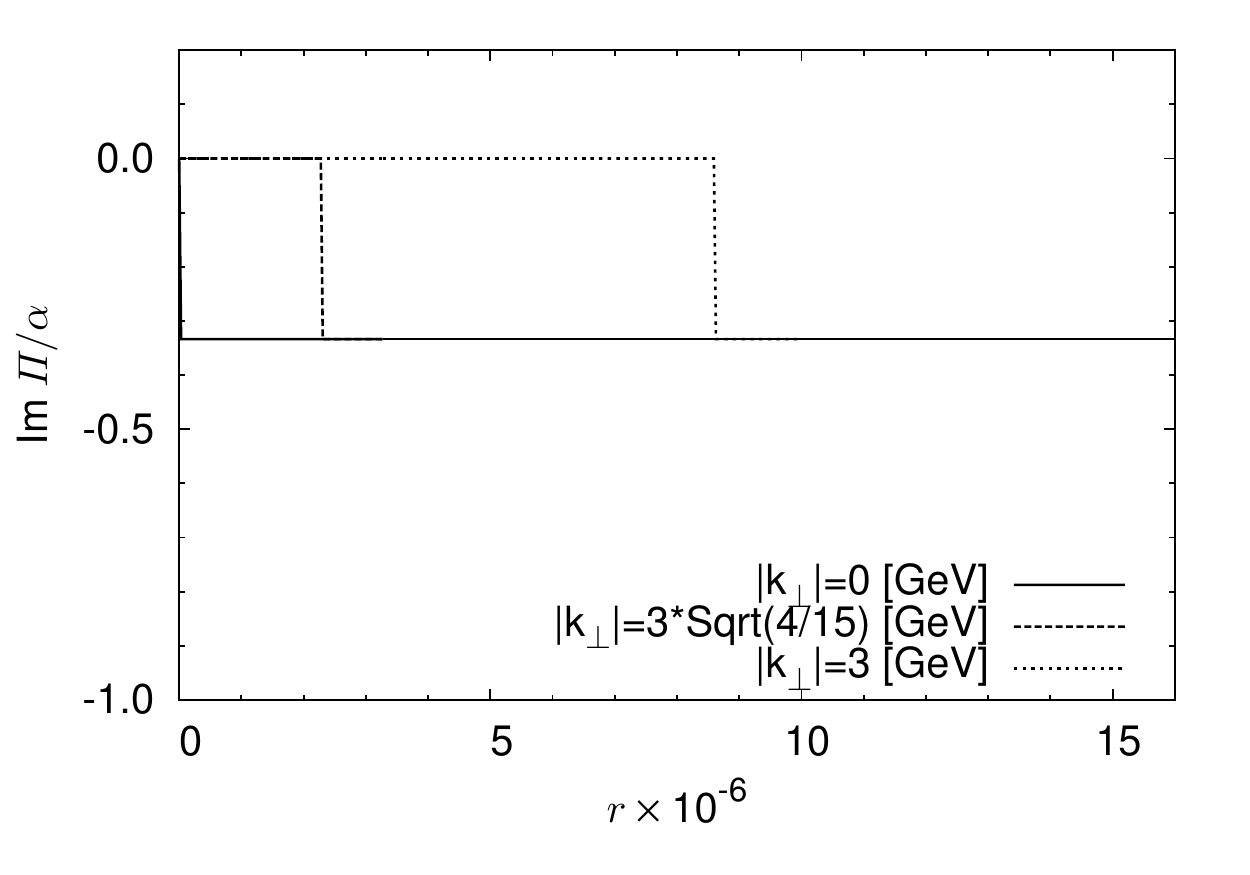}
}
\vspace*{-4pt}
\caption{Form factors $\Pi$, 
Eqs.(\ref{eq:PiBelowThZeroB}) and (\ref{eq:PiAboveThZeroB}), for electrons
(top ($r<1$), bottom left (real part in $1<r$), and bottom right (imaginary part in $1<r$).}
\label{fig:PielectronZeroB}
\end{figure}

In this appendix, we compile other figures for the form factors with weaker magnetic fields
as follows. We also include the form factor $\Pi$ of
Eqs.(\ref{eq:PiBelowThZeroB}) and (\ref{eq:PiAboveThZeroB}) for comparison.
\begin{itemize}
\item case [b-1] with $\ell_{\mathrm{max}}=1000$: Figs.~\ref{fig:N0N1muonmpi} and \ref{fig:N2muonmpi}.
\item case [c-1] with $\ell_{\mathrm{max}}=1000$: Figs.~\ref{fig:N0N1muon01mpi} and \ref{fig:N2muon01mpi}.
\item Eqs.(\ref{eq:PiBelowThZeroB}) and (\ref{eq:PiAboveThZeroB}) for muons with $eB=0$:
  Fig.~\ref{fig:PimuonZeroB}.
\item case [b-2] with $\ell_{\mathrm{max}}=1000$: Figs.~\ref{fig:N0N1electronmpi} and \ref{fig:N2electronmpi}.
\item case [c-2] with $\ell_{\mathrm{max}}=1000$: Figs.~\ref{fig:N0N1electron01mpi} and \ref{fig:N2electron01mpi}.
\item Eqs.(\ref{eq:PiBelowThZeroB}) and (\ref{eq:PiAboveThZeroB}) for electrons with $eB=0$:
  Fig.~\ref{fig:PielectronZeroB}.

\end{itemize}


\begin{thebibliography}{0}
\bibitem{Adler:1971wn} 
  S.~L.~Adler,
  Annals Phys.\  {\bf 67}, 599 (1971).

\bibitem{Shabad:1972rg}
  A.~E.~Shabad,
  Lett.\ Nuovo Cim.\  {\bf 3S2}, 457 (1972)
   [Lett.\ Nuovo Cim.\  {\bf 3}, 457 (1972)].

\bibitem{Shabad:1975ik}
  A.~E.~Shabad,
  Annals Phys.\  {\bf 90}, 166 (1975).

\bibitem{Shabad:1984ASTRO}
  A.~E.~Shabad and  V.~V.~Usov,
  Astrophys.\ Space Sci.\ {\bf 102}, 327 (1984).

\bibitem{Tsai:1974fa} 
  W.~Tsai and T.~Erber,
  Phys.\ Rev.\ D {\bf 10}, 492 (1974).

\bibitem{Tsai:1974}
  W.~Tsai, 
  Phys.\ Rev.\ D {\bf 10}, 2699 (1974).

\bibitem{Tsai:1975iz}
  W.~-y.~Tsai and T.~Erber,
  Phys.\ Rev.\ D {\bf 12}, 1132 (1975).

\bibitem{Urrutia:1977xb} 
  L.~F.~Urrutia,
  Phys.\ Rev.\ D {\bf 17}, 1977 (1978).

\bibitem{Melrose:1976dr} 
  D.~B.~Melrose and R.~J.~Stoneham,
  Nuovo Cim.\ A {\bf 32}, 435 (1976).

\bibitem{Melrose:1977} 
  D.~B.~Melrose and R.~J.~Stoneham,
  J. Phys. A: Math. Gen. {\bf 10}, 1211 (1977).

\bibitem{Ritus:1985}
  V.~I.~Ritus,
  J.~Sov.~Laser~Res. {\bf 6}, 497 (1985).

\bibitem{Dittrich:1998gt} 
  W.~Dittrich and H.~Gies,
  In {\it Frontier Tests of QED and Physics of the Vacuum}, Edited by E. Zavattini,
  D. Bakalov, and C. Rizzo, Heron Press (Sofia, Hungary), 1998., p.29
  [hep-ph/9806417].

\bibitem{Weise:2006jw} 
  J.~I.~Weise and D.~B.~Melrose,
  Phys.\ Rev.\ D {\bf 73}, 045005 (2006).

\bibitem{Baier:2009it} 
  V.~N.~Baier and V.~M.~Katkov,
  Phys.\ Lett.\ A {\bf 374}, 2201 (2010)
  [arXiv:0912.5250 [hep-ph]].

\bibitem{Baier:2007dw}
  V.~N.~Baier and V.~M.~Katkov,
  Phys.\ Rev.\ D {\bf 75}, 073009 (2007)
  [hep-ph/0701119].

\bibitem{Dobrich:2012sw}
  B.~Dobrich, H.~Gies, N.~Neitz and F.~Karbstein,
  Phys.\ Rev.\ Lett.\  {\bf 109}, 131802 (2012)
  [arXiv:1203.2533 [hep-ph]].

\bibitem{Dobrich:2012jd}
  B.~Dobrich, H.~Gies, N.~Neitz and F.~Karbstein,
  Phys.\ Rev.\ D {\bf 87}, 025022 (2013)
  [arXiv:1203.4986 [hep-ph]].

\bibitem{Schubert:2000yt} 
  C.~Schubert,
  Nucl.\ Phys.\ B {\bf 585}, 407 (2000)
  [hep-ph/0001288].


\bibitem{Schubert:2000kf} 
  C.~Schubert,
  Nucl.\ Phys.\ B {\bf 585}, 429 (2000)
  [hep-ph/0002276].

\bibitem{Gies:2001zm} 
  H.~Gies and C.~Schubert,
  Nucl.\ Phys.\ B {\bf 609}, 313 (2001)
  [hep-ph/0104077].


\bibitem{Hattori:2012je} 
  K.~Hattori and K.~Itakura,
  Annals Phys.\  {\bf 330}, 23 (2013)
  [arXiv:1209.2663 [hep-ph]].

\bibitem{Karbstein:2011ja} 
  F.~Karbstein, L.~Roessler, B.~Dobrich and H.~Gies,
  Int.\ J.\ Mod.\ Phys.\ Conf.\ Ser.\  {\bf 14}, 403 (2012)
  [arXiv:1111.5984 [hep-ph]].

\bibitem{Kohri:2001wx} 
  K.~Kohri and S.~Yamada,
  Phys.\ Rev.\ D {\bf 65}, 043006 (2002)
  [astro-ph/0102225].

\bibitem{Gies:2011he} 
  H.~Gies and L.~Roessler,
  Phys.\ Rev.\ D {\bf 84}, 065035 (2011)
  [arXiv:1107.0286 [hep-ph]].


\bibitem{Gies:2005bz} 
  H.~Gies and K.~Klingmuller,
  Phys.\ Rev.\ D {\bf 72}, 065001 (2005)
  [hep-ph/0505099].


\bibitem{Fukushima:2011nu}
  K.~Fukushima,
  Phys.\ Rev.\ D {\bf 83}, 111501 (2011)
  [arXiv:1103.4430 [hep-ph]].

\bibitem{Skokov:2009qp}
  V.~Skokov, A.~Yu.~Illarionov, V.~Toneev,
  Int.\ J.\ Mod.\ Phys.\ A {\bf 24}, 5925 (2009)
  [arXiv:0907.1396 [nucl-th]].

\bibitem{Voskresensky:1980}
  D.~N.~Voskresensky and N.~Y.~Anisimov, Sov. Phys. JETEP {\bf 51}, 13(1980) [Zh. Eksp. Teor. Fiz. {\bf 78}, 28 (1980)].

\bibitem{CHIRALMAGREV} 
  For a review, see
  K.~Fukushima,
  Lect.\ Notes Phys.\  {\bf 871}, 241 (2013)
  [arXiv:1209.5064 [hep-ph]].

\bibitem{CHIRALMAG}
  K.~Fukushima, D.~E.~Kharzeev, H.~J.~Warringa,
  Phys.\ Rev.\ D {\bf 78}, 074033 (2008)
  [arXiv:0808.3382 [hep-ph]].

\bibitem{Yee:2013qma}
  H.~-U.~Yee,
  arXiv:1303.3571 [nucl-th].
\bibitem{Tuchin:2010gx}
  K.~Tuchin,
  Phys.\ Rev.\ C {\bf 83}, 017901 (2011)
  [arXiv:1008.1604 [nucl-th]].
\bibitem{Tuchin:2010vs}
  K.~Tuchin,
  Phys.\ Rev.\ C {\bf 82}, 034904 (2010)
   [Erratum-ibid.\ C {\bf 83}, 039903 (2011)]
  [arXiv:1006.3051 [nucl-th]].
\bibitem{Turbide:2005bz}
  S.~Turbide, C.~Gale and R.~J.~Fries,
  Phys.\ Rev.\ Lett.\  {\bf 96}, 032303 (2006)
  [hep-ph/0508201].
\bibitem{Chatterjee:2005de}
  R.~Chatterjee, E.~S.~Frodermann, U.~W.~Heinz and D.~K.~Srivastava,
  Phys.\ Rev.\ Lett.\  {\bf 96}, 202302 (2006)
  [nucl-th/0511079].
\bibitem{Layek:2006um}
  B.~Layek, R.~Chatterjee and D.~K.~Srivastava,
  Phys.\ Rev.\ C {\bf 74}, 044901 (2006)
  [nucl-th/0605019].
\bibitem{Kopeliovich:2007fv}
  B.~Z.~Kopeliovich, H.~J.~Pirner, A.~H.~Rezaeian and I.~Schmidt,
  Phys.\ Rev.\ D {\bf 77}, 034011 (2008)
  [arXiv:0711.3010 [hep-ph]].
\bibitem{Kopeliovich:2007sd}
  B.~Z.~Kopeliovich, A.~H.~Rezaeian and I.~Schmidt,
  Nucl.\ Phys.\ A {\bf 807}, 61 (2008)
  [arXiv:0712.2829 [hep-ph]].

\bibitem{Adare:2009qk}
  A.~Adare {\it et al.}  [PHENIX Collaboration],
  Phys.\ Rev.\ C {\bf 81}, 034911 (2010)
  [arXiv:0912.0244 [nucl-ex]].

\bibitem{Alessandro:2006yt}
  B.~Alessandro {\it et al.}  [ALICE Collaboration],
  J.\ Phys.\ G {\bf 32}, 1295 (2006).

\bibitem{Inagaki:2003yi}  
  T.~Inagaki, D.~Kimura and T.~Murata, 
  Prog.\ Theor.\ Phys.\  {\bf 111}, 371 (2004)  [hep-ph/0312005]. 
 
\bibitem{Inagaki:2003ac}  
  T.~Inagaki, D.~Kimura and T.~Murata, 
  Int.\ J.\ Mod.\ Phys.\ A {\bf 20}, 4995 (2005)  [hep-ph/0307289]. 

\bibitem{OscillatoryRegion}
N. M. Temme,
J. Appl. Math. Phys. (ZAMP) 41, 114 (1990).

\bibitem{StrongRatioAsymptotic}
A.~Dea\~{n}o, E.~J.~Huertas, F.~Marcell\'{a}n,
J.\ Math.\ Anal.\ Appl. {\bf 403}, 477 (2013) [arXiv:1301.4266 [math.CA]].

\bibitem{DEQUADRATURE}
T. Ooura, Numerical Automatic Integrator for Improper Integral program
using Double-Exponential Quadrature formula, 
\url{http://www.kurims.kyoto-u.ac.jp/~ooura/index.html}


\end{thebibliography}
\end{document}